\documentclass[aip, amsmath, amssymb, reprint]{revtex4-1}

\usepackage{graphicx}
\usepackage{dcolumn}
\usepackage{bm}
\usepackage{braket}
\usepackage[utf8]{inputenc}
\usepackage[T1]{fontenc}

\DeclareMathOperator{\Dsquare}{\text{$D_{\scalebox{0.6}{$\square$}}$}}

\usepackage{multirow}

\usepackage{float}
\usepackage{latexsym}
\usepackage{dcolumn}
\usepackage{epsf}
\usepackage{hyperref}

\begin{document}

\preprint{AIP/123-QED}

\title[Quantum simulation and computing with Rydberg-interacting qubits]{Quantum simulation and computing with Rydberg-interacting qubits}

\author{M. Morgado}

\author{S. Whitlock}
\email{whitlock@unistra.fr}
\affiliation{ 
Institut de Science et d'Ingénierie Supramoléculaires (ISIS, UMR7006), University of Strasbourg and CNRS}

\date{\today}

\begin{abstract}
Arrays of optically trapped atoms excited to Rydberg states have recently emerged as a competitive physical platform for quantum simulation and computing, where high-fidelity state preparation and readout, quantum logic gates and controlled quantum dynamics of more than 100 qubits have all been demonstrated. These systems are now approaching the point where reliable quantum computations with hundreds of qubits and realistically thousands of multiqubit gates with low error rates should be within reach for the first time. In this article we give an overview of the Rydberg quantum toolbox, emphasizing the high degree of flexibility for encoding qubits, performing quantum operations and engineering quantum many-body Hamiltonians. We then review the state-of-the-art concerning high-fidelity quantum operations and logic gates as well as quantum simulations in many-body regimes. Finally, we discuss computing schemes that are particularly suited to the Rydberg platform and some of the remaining challenges on the road to general purpose quantum simulators and quantum computers.
\end{abstract}

\maketitle 

\section{\label{sec:level1}Introduction:}
\begin{figure}[!ht]
  \includegraphics[width=0.9\linewidth]{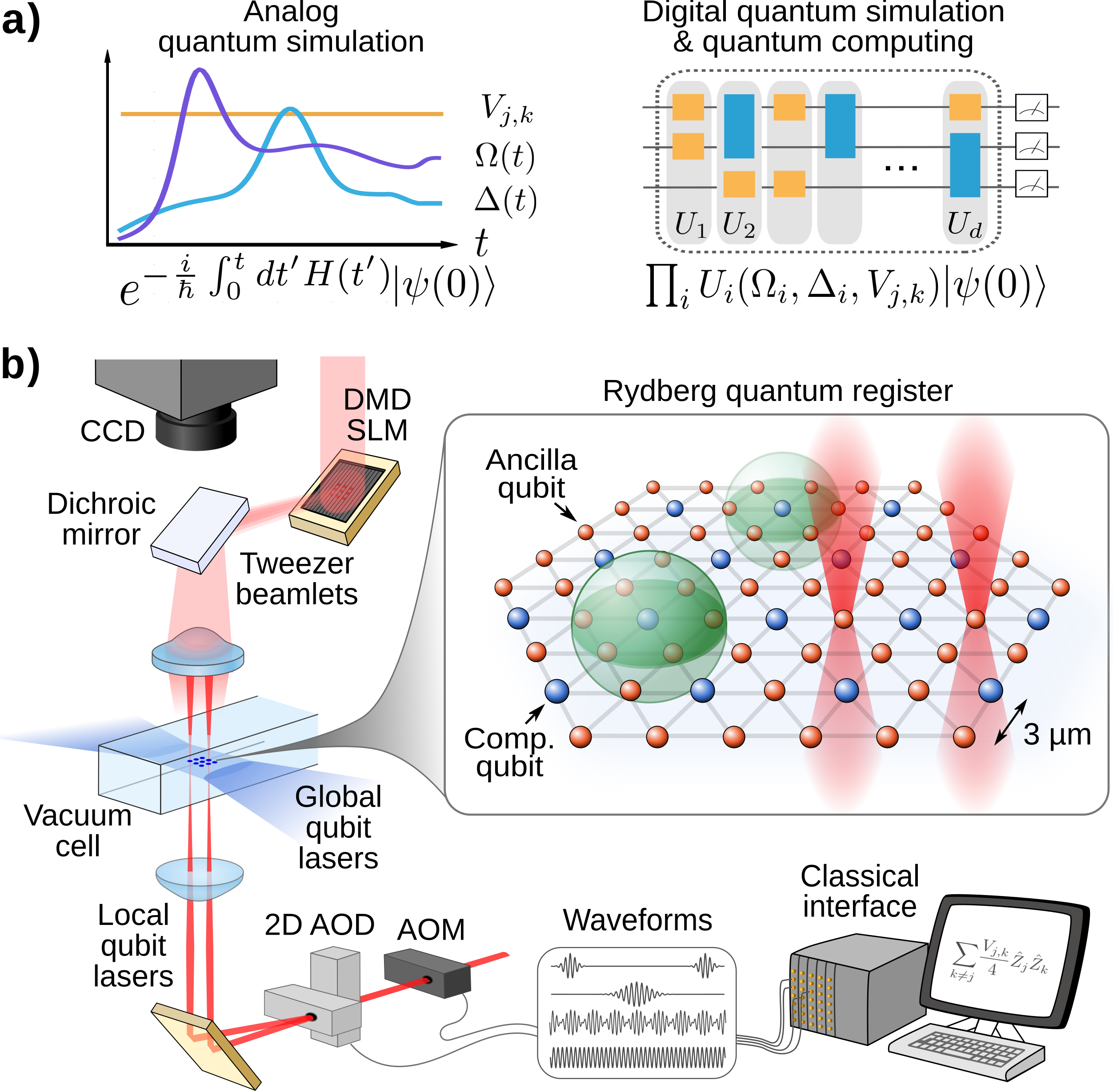}
  \caption{Depiction of a quantum processor based on Rydberg-interacting qubits. a) The Rydberg platform is uniquely suited for both analog and digital quantum simulation and quantum computing. b) Sketch of a typical setup consisting of ultracold atoms trapped in an array of optical tweezers produced by a digital micromirror device (DMD) or spatial light modulator (SLM). Qubits can be manipulated by  optical fields controlled by acousto-optical modulators (AOMs) and two-dimensional acousto-optical deflectors (AODs). The quantum register shows two species of atomic qubits (blue and orange spheres). Semi-transparent green spheres depict the Rydberg-Rydberg interactions (blockade spheres). Red shaded areas depict the addressing lasers for implementing single and multi-qubit operations.}
  \label{fig:register}
\end{figure}

Quantum simulation and quantum computing represent revolutionary ways to process information and to learn about our world\cite{feynman1982simulating,nielsen2002quantum}. The basic idea is to prepare a set of quantum objects (e.g., atoms, ions, electrons, photons) in a well defined quantum state and to transform this state by making them interact in controlled ways (as depicted in Fig.~\ref{fig:register}a for quantum evolution according to a Hamiltonian involving several time-dependent parameters or a set of discrete unitary operations). In this way one can overcome the intrinsic limitations of classical devices when representing entangled multi-particle quantum states and their evolution.

An \textit{analog quantum simulator} is a physical system which mimics another quantum system of interest, or a specific class of models, by reproducing its Hamiltonian as close as possible\cite{buluta2009quantum,cirac2012goals,georgescu2014quantum}. By performing controlled experiments and measurements on the quantum simulator we can then learn something about the target system which is hard, if not impossible, using classical computers. A \textit{digital quantum simulator} performs a similar task by encoding the quantum state of the target system in a quantum register, i.e., an array of quantum bits (\textit{qubits}) as shown in Fig.~\ref{fig:register}b, and simulating its time evolution by successively applying a programmable sequence of quantum gates\cite{lloyd1996universal,buluta2009quantum,weimer2011digital,georgescu2014quantum}. If these gates constitute a universal set then in principle a digital quantum simulator can simulate any local Hamiltonian\cite{lloyd1996universal}, including those with terms that are not natively realized by the physical system. A \textit{quantum computer} takes this a step further, making use of a set of fully addressable qubits and universal set of quantum logic gates (possibly including error correction) to implement quantum algorithms for efficiently solving classically intractable computational problems\cite{nielsen2002quantum,preskill2018quantum}. 

Today’s relatively noisy, intermediate-scale quantum processors, and larger-scale quantum simulators with limited programmability, are already starting to provide the means to solve problems, at or beyond, the limits of classical computing techniques\cite{choi2016exploring,preskill2018quantum,arute2019quantum,kokail2019self}. And as this technology continues to improve, the distinction between quantum simulators and quantum computers will become increasing blurred. Quantum computers are set to become indispensable tools for scientific discovery, e.g., for simulating quantum Hamiltonians in many-body physics, chemistry and material science\cite{alexeev2019quantum}. At the same time, sufficiently programmable quantum simulators may well be used to solve computational problems with importance well outside of physics. 

There are a variety of different physical systems that can be used for quantum simulation and quantum computing, each having their own unique advantages and challenges\cite{alexeev2019quantum,altman2019quantum}. Superconducting circuits\cite{wendin2017quantum,kjaergaard2020superconducting} (SC) and trapped ions\cite{bruzewicz2019trapped} are presently among the most advanced platforms for quantum computing, having demonstrated a very high level of programmable control over individual qubits, including high-fidelity quantum logic operations [with two-qubit gate fidelities $F>0.997$ (SC\cite{kjaergaard2020superconducting}), $F>0.999$ (ions\cite{ballance2016high,gaebler2016high, bruzewicz2019trapped})].
However, scaling up these systems much beyond $50-100$ fully-controlled qubits (the approximate threshold where the full quantum state can no longer be represented on classical computers) presents a formidable challenge due to increasing gate errors with system size\cite{erhard2019characterizing} or the need to shuttle around the ions\cite{kielpinski2002architecture,kaushal2020shuttling} and the increasing complexity of control circuits\cite{kjaergaard2020superconducting}. 

Another very promising approach to quantum information processing based on trapped neutral atoms was first proposed in $2000-2001$, in two seminal papers by \textcite{jaksch2000fast} and \textcite{lukin2001dipole}. The basic idea is to encode quantum information in the internal states of single atoms or (collective) excitations of atomic ensembles, with interactions mediated via their electronically highly-excited Rydberg states (Rydberg-interacting qubits, or Rydberg qubits for short). Neutral atoms in particular can be well isolated from the environment and prepared in relatively large systems of hundreds or thousands of particles with different geometries\cite{bloch2008manybody,bloch2012quantum} using laser cooling and trapping techniques. This includes top-down approaches, for example by transferring an evaporatively cooled gas to an optical lattice to create nearly defect-free arrays via a superfluid-Mott insulator transition, and bottom-up approaches such as atom-by-atom assembly where single atoms are trapped and individually positioned in desired geometries using optical tweezers. By combining this with coherent laser excitation to Rydberg states it is possible to introduce strong and widely-tunable interactions that can extend over micrometer distances. Accordingly, ultracold Rydberg quantum systems have proven very successful for many-body physics and analog quantum simulation, and are now emerging as a competitive platform for digital quantum simulation and scalable quantum computing.

In this article we summarize the key features of the Rydberg platform as it stands today, emphasizing its versatility for engineering quantum logic gates and many-body Hamiltonians. We will then review recent progress in realizing highly-controllable quantum systems of Rydberg qubits, paying particular attention to the impressive advances made in the last few years. While Rydberg quantum systems can take different forms, we will focus mainly on programmable quantum processors built from arrays of individually addressable atoms as qubits (see Fig.~\ref{fig:register}b for a specific vision of such a device). We refer to earlier reviews for a more comprehensive overview of the Rydberg platform\cite{saffman2010quantum,saffman2016quantum,browaeys2016experimental,adams2019rydberg} and applications to many-body physics\cite{comparat2010dipole,low2012experimental,hofmann2014experimental,schauss2018quantum,browaeys2020many}. 

The last few years have seen remarkable progress, including: the assembly of tweezer arrays with on the order of $100$ sites deterministically filled with single atoms~\cite{endres2016atom,barredo2016atom,barredo2018synthetic,mello2019defect} or small atomic ensembles\cite{wang2020preparation} that can serve as Rydberg qubits; high-fidelity state resolved readout\cite{covey20192000,madjarov2020high}; high-fidelity entangling operations with $F>0.991$\cite{madjarov2020high}; native multi-qubit gates\cite{levine2019parallel}; the generation of highly-entangled GHZ states involving $\sim 20$ qubits\cite{omran2019generation}; and controlled quantum evolution in atomic arrays featuring strong interactions\cite{schauss2015crystallization,bernien2017probing,zeiher2017coherent,guardado2018probing,lienhard2018observing,leseleuc2019observation}, now reaching system sizes well beyond 100 qubits in programmable geometries\cite{ebadi2020quantum,bluvstein2020controlling,scholl2020programmable}. These achievements reflect a growing community effort working toward the realization of highly-programmable quantum systems based on Rydberg qubit technology, which now also includes several new commercial efforts. For a recent overview of the technology being developed by the company Pasqal, we refer to \textcite{henriet2020quantum}.

The review is structured as follows. In section~\ref{sec:chtoolbox} we introduce the different ways to encode quantum information in atomic qubits, to control their interactions using Rydberg states and present the different ways to realize quantum logic gates and to engineer many-body Hamiltonians. In section~\ref{sec:demonstrations} we summarize some of the recent technological achievements, in particular focusing on the implementation of high-fidelity quantum logic gates and controlled quantum dynamics using arrays of ultracold Rydberg atoms. In section~\ref{sec:proposals} we will discuss a few novel approaches to quantum simulation and quantum computation that take advantage of the unique features of the Rydberg platform and thus might be particularly useful while we navigate the ``noisy intermediate-scale quantum'' (NISQ) era\cite{preskill2018quantum}. Finally, in section~\ref{sec:outlook} we will discuss some of the future opportunities and outstanding challenges on the road toward the more powerful quantum simulators and quantum computers of the future.

\section{Rydberg quantum toolbox}\label{sec:chtoolbox}

In this section we present the key features of the Rydberg platform that make it particularly suitable for quantum simulation and quantum computing. This starts with a description of the physical system, guided by established criteria for assessing the viability of quantum computers and quantum simulators, followed by a classification of the different types of Rydberg qubits. A distinguishing feature of Rydberg qubits is their versatile interaction properties. Therefore, after explaining the physical origin of these interactions we spend a considerable part of this section to present the different entangling gates and many-body Hamiltonians that can be considered native to the Rydberg platform.

\begin{table*}[!t]\caption{Criteria for quantum simulators and quantum computers}\label{tab:criteria}
\begin{tabular}{l|p{65mm}|p{85mm}}
\textbf{Criteria} & \textbf{Quantum computers\cite{divincenzo2000physical}}                           & \multicolumn{1}{l}{\textbf{Quantum simulators\cite{cirac2012goals}}}     \\ \hline  \hline    
Quantum system    & A scalable physical system with well characterized qubits                                      & A system of quantum particles (bosons, fermions, pseudo-spins) confined in space and collectively possessing a large number of degrees of freedom                                                             \\ \hline    
Initialization    & The ability to initialize the state of the qubits to a simple fiducial state, such as $|000\ldots\rangle$ & The ability to prepare (approximately) a known quantum state (typically a pure state)                                                                                                \\ \hline    
Coherence         & Long relevant decoherence times, much longer than the gate operation time                      & 
\\ \hline    
Interactions      & A “universal” set of quantum gates                                                             & An adjustable set of interactions used to engineer Hamiltonians/quantum master equations including some that cannot be efficiently simulated classically 
\\ \hline    
Measurement       & A qubit-specific measurement capability                                                        & The ability to perform measurements on the system; either individual particles or collective properties                                                                     \\ \hline    
Verification      &                                                                                                & A way to verify the results of the simulation are correct \\   \hline                                                                                      
\end{tabular}
\end{table*}

\subsection{Characteristics of the Rydberg platform}\label{sec:criteria}
In 2000, \textcite{divincenzo2000physical} outlined a set of criteria for assessing the viability of physical platforms for quantum information processing. In 2012, \textcite{cirac2012goals} put forward a similar set of criteria for quantum simulators (summarized in Table~\ref{tab:criteria}). Rydberg-interacting atoms fulfil both sets of criteria, which makes them a very attractive platform for general purpose programmable quantum simulation and scalable quantum computing. In the following we will outline the key features of the Rydberg platform as they relate to these criteria.\\

\noindent{\textit{1. Quantum system}}\\
Figure~\ref{fig:register}b depicts an archetypal Rydberg quantum processor, consisting of an array of atoms precisely positioned in space with separations of a few micrometers. Most experiments to date are based on alkali atoms (e.g., Li, K, Rb, Cs), but recently we have also seen important experimental demonstrations with alkaline-earth atoms\cite{madjarov2020high} and trapped ions\cite{zhang2020submicrosecond}. This opens up interesting new possibilities, e.g., combining ultra-coherent atomic clock qubits with Rydberg mediated interactions or hybrid approaches to quantum information processing which leverage some of the best features of the trapped-ion and (neutral atom) Rydberg platforms\cite{MOKHBERI2020233}. Indeed any quantum system featuring strongly-interacting excited states could be used as Rydberg qubits, possibly even solid-state systems\cite{crane2020rydberg}. In the near future we expect that quantum processors based on multiple different atomic species\cite{beterov2015Rydberg} or even fundamentally different technologies could offer unique advantages for state preparation, manipulation, storage and readout\cite{xiang2013hybrid,Kurizki3866}.
 
\begin{figure}
  \includegraphics[width=\linewidth]{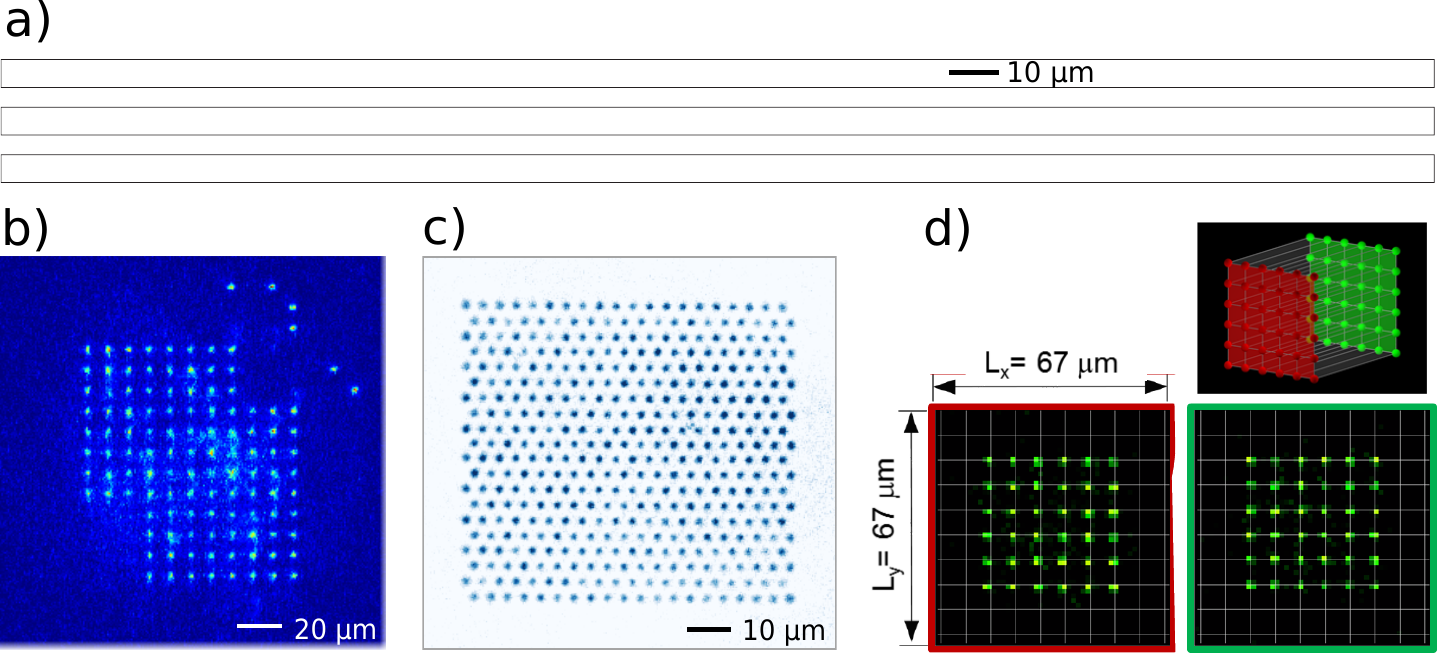}
  \caption{\small Examples of atomic quantum registers of optically trapped neutral atoms in programmable tweezer arrays. a) Fluorescence images of deterministically loaded one-dimensional arrays containing more than 50 atoms in periodic, dimerized and cluster geometries. From \textcite{endres2016atom} Science, 354, 1024-1027 (2016). Reprinted with permission from AAAS. b) Fluorescence image of a deterministically loaded two-dimensional square lattice with 111 atoms. Reprinted figure with permission from \textcite{mello2019defect} Phys. Rev. Lett. 122, 203601 (2019), Copyright 2019 by the American Physical Society. c) Absorption image of a triangular lattice of microscopic atomic ensembles containing more than 400 sites with high filling factors. Reproduced from \textcite{wang2020preparation} npj Quantum Information, 6, 54 (2020), licensed under CC-BY 4.0. d) Fluorescence image of a deterministically loaded bilayer lattice of 72 sites. Reprinted by permission from Springer Nature: \textcite{barredo2018synthetic} Nature 561, 79 (2018), Copyright 2018.}
  \label{fig:100atoms}
\end{figure}

Currently one of the best ways to engineer Rydberg quantum processors is by laser cooling the atoms and then trapping them in optical microtraps (e.g., tweezers) generated by a spatial light modulator or digital micromirror device (Fig.~\ref{fig:register}b). Each trap site can be filled with precisely one atom by exploiting light-assisted collisions combined with a rearrangement procedure to fill empty sites in a process called ``atom assembly''\cite{endres2016atom,barredo2016atom,kim2016situ,brown2019gray,mello2019defect}. However in this review we cover demonstrations involving both bottom-up assembly and top-down approaches involving optical tweezer arrays and small optical lattices. These techniques have been used to realize deterministically loaded and almost defect-free quantum registers in different spatial geometries including 1D\cite{endres2016atom,madjarov2020high}, 2D\cite{schauss2012observation,schauss2015crystallization,zeiher2016many,barredo2016atom,lienhard2018observing,guardado2018probing,mello2019defect,wang2020preparation} and 3D\cite{barredo2018synthetic,schlosser2019large} arrays (see Fig.~\ref{fig:100atoms} for a selection of experimental images showing different geometries). Each atom possesses discrete quantum states that can be used to encode a qubit and to mediate interactions. This could involve either long-lived ground states or highly-excited Rydberg states or combinations of both, and, particularly for neutral particles, they can be isolated almost completely from their environment and from each other when desired. These qualities are favourable for realizing large systems of hundreds or potentially thousands of qubits using current technology, without the need for advanced nanofabrication or cryogenic systems.\\ 

\noindent{\textit{2. Initialization}}\\
Trapped atoms have well defined energy levels and all atoms of the same species are identical. They can also be initialized in a desired quantum state (e.g., $\ket{0}^{\otimes N}$) with high efficiency using dissipative optical pumping techniques (see Refs.\cite{antoni2017generation,levine2019parallel} for recent demonstrations of initializing neutral atoms in magnetically-insensitive states with $>95\%$ purity) and coherent transfer pulses to other ground or electronically-excited states. Once initialized, these states are quite stable. The qubit energy splittings $\omega_q/2\pi\gtrsim\textrm{500 MHz}$ are large compared to the typical couplings by external fields and the energy scales associated to thermal motion $k_BT/h\lesssim 1\,$MHz. The typical lifetime of Rydberg states is $\sim 100\,\mu$s, much longer than the typical timescales of Rydberg-Rydberg interactions or the typical gate times of $\tau_g=(0.05-6)\,\mu$s (see Table~\ref{tab:qubitgates} for an overview), while the lifetime of ground-state atoms in a trap can be many seconds. \\
\vskip -1em

\noindent{\textit{3. Coherence}}\\
Depending on the type of atom and atomic states involved, typical coherence times in experiments employing Rydberg states to mediate interactions vary from several microseconds (for Rydberg state encoding) to tens or hundreds of milliseconds (for ground state encoding). However this is far from the fundamental limit, as coherence times in the seconds range or longer have already been demonstrated using magic-intensity or magic-wavelength tweezers (without Rydberg excitation)\cite{sheng2018high,norcia2019seconds,young2020tweezer}. However, this does not typically include extra decoherence stemming from the interaction with laser fields or the interactions between particles, which can depend on the number of qubits, noise characteristics, and the types of interactions or circuits that one wishes to simulate or compute. 

A better performance indicator that can be used to assess Rydberg quantum processors today is the depth of the largest random circuit (consisting of the parallel application of two-qubit operations to all qubits) that can be reliably executed before an error is likely to occur. This can be defined as $\Dsquare = \lfloor \epsilon^{-1/2}\rfloor$, where $\epsilon$ is the average error probability per two-qubit gate or half interaction cycle (see Appendix~\ref{app:dsquare} for more details). This also implicitly assumes the availability of at least $\Dsquare$ qubits. 

Ideally $\Dsquare$ would be measured using benchmark protocols performed on the real quantum hardware\cite{cross2019validating,blume2019volumetric,derbyshire2020randomized}, which has not yet been done for the Rydberg platform. Instead, for the purpose of this review, we try to estimate $\Dsquare$ based on available data. Therefore we assume $\epsilon \sim 1-F$ for digital circuits and $\epsilon\sim \pi/(V_\text{max}T_\text{coh})$ for analog quantum simulations, where $V_\text{max}$ is the maximum effective interaction strength (usually between nearest neighbors) and $T_\text{coh}$ is the characteristic coherence time for the simulation. However, used in this way, $\Dsquare$ does not account for other important criteria such as the dependence of gate errors on circuit size, circuit connectivity, and the availability of different types of quantum gates. Recent demonstrations of two-qubit entangling operations and quantum simulations with many Rydberg qubits are compatible with $\Dsquare\approx 10$ (Sec~\ref{sec:demonstrations}), similar to the latest reported demonstrations with SC quantum circuits\cite{arute2019quantum,jurcevic2020demonstration} and trapped ions\cite{erhard2019characterizing,wright2019benchmarking,pino2020demonstration,egan2020fault}. In view of the current rate of progress, and assuming realistic improvements to the existing technology (explored further in Sec.~\ref{sec:outlook}), we anticipate that Rydberg quantum systems with $\Dsquare \gtrsim 40$ and several hundred qubits will soon be within reach.  \\

\noindent{\textit{4. Interactions}}\\
The main way to manipulate Rydberg qubits is using laser fields or combined laser and microwave fields, which is advantageous for local qubit addressing with low crosstalk (to be discussed in more detail in Sec.~\ref{sec:singlequbit}). Multi-qubit interactions originating from electric dipole-dipole couplings between Rydberg states are orders of magnitude stronger than the relatively weak or short-range interactions typical of ground-state neutral atoms, thanks to the exaggerated size of Rydberg states (scaling with the square of the principal quantum number $n$\cite{comparat2010dipole,saffman2010quantum}). These interactions can easily extend over several micrometers (beyond nearest neighbors) and take different forms depending on the specific atomic states used, which offers advantages for implementing quantum algorithms that benefit from high qubit connectivity and gate expressivity\cite{Linke3305}. A prominent example is the Rydberg blockade interaction, whereby one excited atom shifts the Rydberg states of neighboring atoms out of resonance (depicted by green spheres in Fig.~\ref{fig:register}b). Other examples include the resonant electric dipole interaction which allows for the coherent exchange of energy between nearby atoms prepared in different Rydberg states, or the tunable Rydberg-dressed interaction obtained using stroboscopic or off-resonant laser couplings to map the Rydberg state properties to long-lived ground states\cite{santos2000bose,pupillo2010strongly,henkel2010three}. These interactions are particularly useful for realizing fast and robust multi-qubit quantum gates as well as for implementing a relatively wide range of interesting models for quantum simulation (discussed in Sec.~\ref{sec:interactions}).\\ 

\noindent{\textit{5. Measurement}}\\
Currently the main method to read out Rydberg qubits is via single-atom sensitive fluorescence imaging from the ground states. Rydberg excited atoms can be detected either by first transferring them to a suitable ground state, or by removing them from the trap prior to imaging in which case they show up as the absence of a signal. Rydberg state detection efficiencies $\gtrsim 0.95$ are routinely achieved within $\gtrsim 10\,$ms for parallel readout of the entire array\cite{labuhn2016tunable,levine2018highfidelity,picken2018entanglement}, with the best results reported so far of $>0.996$\cite{madjarov2020high}. While this type of detection is usually destructive, high-fidelity lossless readout schemes for ground state qubits have also been demonstrated using state-selective fluorescence in free space\cite{kwon2017parallel,martinez2017fast,norcia2018microscopic,covey20192000}, using cavity enhancement~\cite{bochmann2010lossless,gehr2010cavity}, or using state-dependent potentials\cite{Boll1257,robens2017atomic,wu2019stern}. This would enable repeated measurements on qubits and to act on measurement outcomes, e.g., for quantum feedback\cite{ZHANG20171} and quantum error correction protocols\cite{Devitt_2013}.\\

\noindent{\textit{6. Verification}}\\
An important issue that applies to both quantum computers and quantum simulators in any platform is whether one can trust that they are producing correct results\cite{gheorghiu2019verification,hauke2012can}. For small quantum systems (comprised of a few qubits) it is possible to verify quantum operations using tomographic methods\cite{cramer2010efficient}. Otherwise, for system sizes of $\lesssim 50$ qubits (or systems with limited entanglement) it is possible to benchmark quantum simulations and computations against exact numerical calculations on high performance classical computers. Beyond this, the high degree of tunability of the Rydberg system and the possibility to spatially reconfigure the system could allow for benchmarking against highly optimized numerical methods, such as effectively exact matrix product state calculations\cite{vidal2004efficient,daley2004timedependent} or integrable models, which are especially suitable for one-dimensional problems, before extending to regimes where these methods fail. 

Verifying quantum simulations and quantum computations in classically intractable regimes is still a largely open theoretical question\cite{fitzsimons2017unconditionally,gheorghiu2019verification}. However recent ideas, such as self-verifying quantum simulations\cite{kokail2019self} and randomized benchmarking protocols\cite{brydges2019probing,derbyshire2020randomized} would be interesting to implement Rydberg quantum systems (particularly as we approach regimes that cannot be simulated on classical computers). Similar ideas could provide a setting to achieve certifiable quantum speedups over classical computers, even with limited control over individual qubits\cite{bermejovega2018architectures,eisert2020quantum}. Ultimately the results of a quantum simulation or computation could be verified by another quantum simulator or quantum computer, highlighting the importance of developing quantum hardware based on different technologies\cite{Linke3305,elben2020cross}.

\subsection{Types of Rydberg qubits}\label{sec:toolbox}

\begin{figure}
  \includegraphics[width=\linewidth]{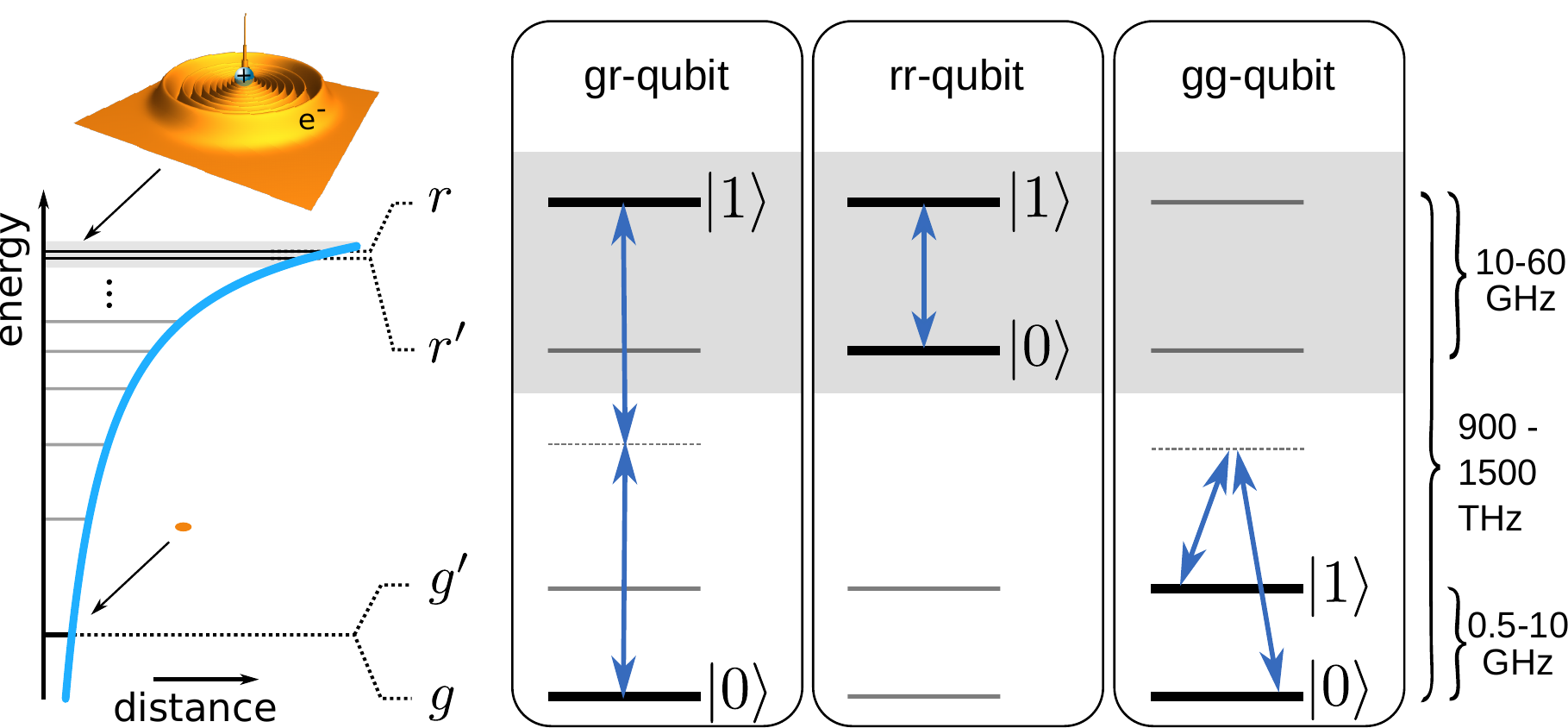}
  \caption{\small Rydberg qubits can be classified according to three main types depending on the number of ground and Rydberg states that make up the qubit. Each type of qubit involves different characteristic energy scales and different ways to manipulate the qubit states using either microwave or optical fields (blue arrows).}
  \label{fig:qubits}
\end{figure}

The relatively rich energy level structure of Rydberg atoms and techniques for coherently manipulating the internal states of atoms, including different combinations of ground and Rydberg states, offers many possibilities for storing and manipulating quantum information. In the following we classify Rydberg qubits according to three main types, distinguished by the number of ground or Rydberg states that make up the qubit (depicted in Fig.~\ref{fig:qubits}).

\subsubsection*{Ground-Rydberg (gr) qubits}
The simplest class of Rydberg qubits are composed of one weakly-interacting state $\ket{g}\equiv\ket{0}$ and one strongly-interacting Rydberg state $\ket{r}\equiv\ket{1}$ (gr-qubits). The latter is normally specified in terms of a principal quantum number, typically in the range of $n=60-80$ and an azimuthal quantum number $\ell$. In most (but not all) experiments so far, $\ell=0$, corresponding to $S$-states for their convenient interaction properties (discussed further in Sec.~\ref{sec:interactions}). $\ket{g}$ is typically the electronic ground state, or sometimes an excited metastable state (effective ground state) in the case of alkaline-earth atoms\cite{cooper2018alkaline,norcia2019seconds,madjarov2020high,wilson2019trapped} or ions\cite{higgins2017coherent,zhang2020submicrosecond}. 
The typical energy splitting of gr-qubits in frequency units is $(900-\nobreak 1500)\,$THz, depending on the atomic species and the Rydberg states used. Therefore qubit manipulations usually involve an ultraviolet laser or a combination of visible and infrared lasers in a ladder configuration. Lifetimes of gr-qubits are typically limited by decay of the Rydberg state to around ${\sim}100\,\mu$s due to spontaneous emission and stimulated emission by black-body radiation. Coherence times for gr-qubits presently fall in the range $T_2^*\sim (2 - 20)\,\mu$s\cite{picken2018entanglement,levine2018highfidelity,madjarov2020high,wilson2019trapped}, limited primarily by phase noise on the excitation lasers and dephasing due to thermal motion and anti-trapping of the Rydberg states\cite{saffman2011rydberg,leseleuc2018analysis}. The physical interactions between gr-qubits are typically of the form $-\frac{C_6}{R^6}\ket{11}\bra{11}$ and are generally always on (Sec.~\ref{sec:interactions}), making it hard to act on individual qubits in isolation from other qubits. However gr-qubits are relatively easy to initialize, manipulate and measure. For these reasons they are arguably state-of-the-art right now for Rydberg qubits for fast $\lesssim 100\,$ns and high-fidelity entangling operations\cite{picken2018entanglement,levine2018highfidelity,madjarov2020high}, as well as for many-body quantum state engineering and quantum simulation \cite{schauss2015crystallization,bernien2017probing,zeiher2017coherent,guardado2018probing,lienhard2018observing,leseleuc2019observation,ebadi2020quantum,bluvstein2020controlling,scholl2020programmable}. 

While gr-qubits are usually encoded using single atoms, they could also be realized using collective states of atomic ensembles confined smaller than the blockade volume (collective qubits)\cite{lukin2001dipole,ebert2015coherence,spong2020robustness}. This can be advantageous for realizing larger arrays without complex single atom assembly protocols\cite{saffman2008scaling,wang2020preparation}. Collective qubits can also benefit from a $\sqrt{N}$ collective enhancement of the atom-light coupling and could provide more options for achieving fast single qubit readout using the ensemble atoms as an amplifier\cite{guenter2012interaction}. But they also come with additional challenges associated to intrinsic atom number fluctuations and short-range interactions between ensemble atoms, so they are not widely used at present.

\subsubsection*{Rydberg-Rydberg (rr) qubits}
Qubits encoded using two different Rydberg states (rr-qubits, $\ket{r'}\equiv\ket{0},\ket{r}\equiv\ket{1}$) offer a high degree of flexibility for engineering Rydberg-Rydberg interactions, including longer range dipolar-exchange interactions of the form $\frac{C_3}{R^3}\left(\ket{10}\bra{01}+\ket{01}\bra{10}\right)$. Interactions between Rydberg states are generally strongest for similar principal quantum numbers, e.g., $n,n-1$ and similar azimuthal quantum numbers $\delta\ell=|\ell-\ell'|=0,1,2$. This corresponds to typical rr-qubit energy splittings in the $(10-60)\,$GHz range (scaling approximately with $n^{-3}$).

To initialize rr-qubits one can use an optical excitation pulse from the ground state or use robust stimulated Raman adiabatic passage (STIRAP)\cite{unanyan1998robust,deiglmayr2006quantum,higgins2017coherent,leseleuc2019observation} and composite pulse sequences\cite{beterov2013quantum}. State manipulation is typically via microwave coupling fields which is advantageous due to the availability of low noise sources for driving microwave qubit transitions and negligible Doppler dephasing~\cite{signoles2019observation,leseleuc2017optical,wilson2019trapped}. However this makes it harder to address individual qubits. As for gr-qubits, coherence times are limited by the difficulty to trap Rydberg states and the finite Rydberg state lifetime $\sim 100\,\mu$s. Recent experiments with alkaline-earth Rydberg atoms trapped in optical tweezers have observed coherent single atom dynamics with coherence times of $T_2^*=22\,\mu$s\cite{wilson2019trapped}, and coherent dynamics in small arrays of alkali atoms over approximately $\sim (4-7)\,\mu$s with typical nearest-neighbor interaction strengths $(1-2.5)\,$MHz\cite{barredo2015coherent,leseleuc2017optical,leseleuc2019observation}. A route to much longer lifetimes is possible using qubits formed by two circular Rydberg states with large principal quantum numbers and maximal azimuthal and magnetic quantum numbers in cryogenic environments \cite{nguyen2018towards,cortinas2020laser}. These states were used in pioneering experiments on atom-photon entanglement and cavity quantum electrodynamics\cite{Raimond2001manipulating}. However, combining laser trapping with cryogenic environments comes at the cost of significantly increased technical complexity.

\subsubsection*{Ground-ground (gg) qubits}
Qubits encoded in two long-lived low-lying atomic states $\ket{g}\equiv\ket{0},\ket{g'}\equiv\ket{1}$ (gg-qubits) offer the best performance in terms of coherence times combined with switchable interactions, making them good candidates for universal quantum computing. This can involve two (usually magnetically insensitive) hyperfine sublevels of the electronic ground state\cite{xia2015randomized,jau2016entangling,zeng2017entangling,picken2018entanglement,graham2019rydberg,levine2019parallel}, or it can be the electronic ground state and a metastable excited state (e.g., in the case of alkaline-earth species). Qubit energy splittings for hyperfine qubits fall in the $(1-10)\,$GHz range while metastable qubits involve optical frequencies. 

In addition to their long lifetimes, gg-qubits can be trapped with similar trapping potentials for both qubit states. Accordingly, single qubit dephasing times of $T_2^*=(1 - 20)\,$ms are typical without spin echo pulses\cite{xia2015randomized,picken2018entanglement,graham2019rydberg} and can be boosted to the seconds range using spin echo pulses or magic trapping techniques, which is a significant advantage of gg-qubits over gr- and rr-qubits.

Single qubit gates have been demonstrated with very high fidelities of $F>0.9999$ for global manipulation\cite{sheng2018high} and $F=0.992$ for local addressing with low crosstalk\cite{xia2015randomized}. Compared to gr- and rr-qubits, gg-qubits are weakly interacting. Therefore interactions must be mediated by momentarily exciting and de-exciting them via Rydberg states using precisely timed or shaped optical fields. This has been used to demonstrate two-qubit quantum logic gates\cite{isenhower2010demonstration,levine2019parallel,graham2019rydberg} with fidelities as high as $F=0.974$\cite{levine2019parallel}, and $F=0.89$\cite{graham2019rydberg} with individual qubit addressing. Alternatively, gg-qubits can be made to interact by weakly admixing some Rydberg state character to the ground states using an off-resonant laser coupling (Rydberg dressed qubits). This offers additional possibilities for realizing new many-body phases with engineered long-range interactions\cite{santos2000bose,pupillo2010strongly,henkel2010three,balewski2014rydberg,glaetzle2015designing, bijnen2015quantum,zeiher2016many,borish2020transverse,guardado2020quench} and to 
spatially and temporally control the interactions for realizing gate operations by modulating the dressing laser fields\cite{petrosyan2014binding,jau2016entangling,arias2019realization,mitra2020robust}.

\subsection{Single qubit manipulation}\label{sec:singlequbit}

\subsubsection*{Atom-light interactions}
Rydberg qubits can differ greatly in terms of typical energy splittings, lifetimes and other details. However single qubit manipulations are generally realized by nearly monochromatic optical or microwave fields that realize the Hamiltonian (using units where $\hbar=1$ and neglecting atomic motion for the moment)

\begin{equation}\label{eq:HLM}
    \hat H^{ab}_j=\left(\frac{\Omega_j(t)}{2}e^{i\varphi_j(t)}\ket{a}_j\!\bra{b}+\textrm{h.c.}\right )-\Delta_j(t)\ket{b}_j\!\bra{b}.
\end{equation}
The states $\ket{a},\ket{b}$ could refer to the qubit states or auxiliary states used to mediate the interactions. The Hamiltonian~\eqref{eq:HLM} can be derived by treating the atom-light interactions semi-classically and by applying the rotating wave approximation to drop rapidly oscillating phase factors. $\Omega_j(t)$ characterizes the strength of the coupling at the position of atom $j$, $\varphi_j(t)$ the local phase, and $\Delta_j(t)=\omega_j-\omega_0$ the local detuning of the coupling field frequency $\omega_j$ from the atomic resonance frequency $\omega_0$. 

\begin{figure}[!t]
    \centering
    \includegraphics[width=\columnwidth]{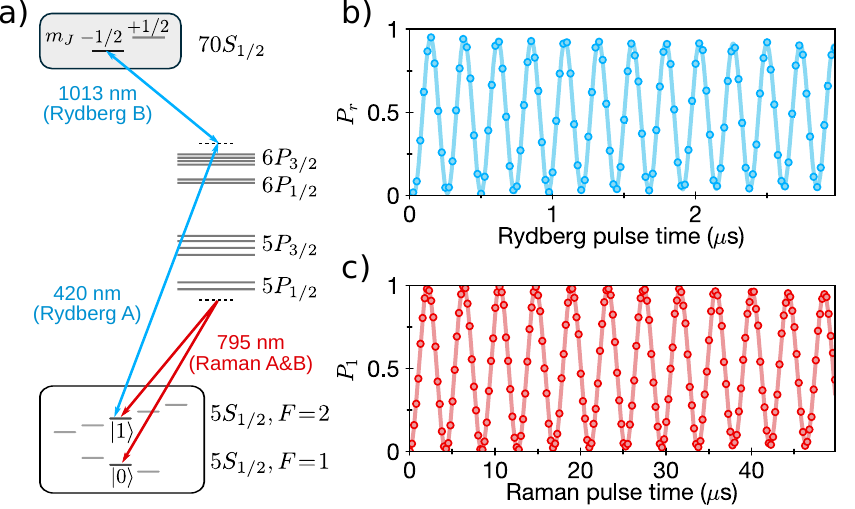}
    \caption{High fidelity manipulation of $^{87}$Rb  Rydberg qubits used in \textcite{levine2019parallel}. a) Single qubit operations are realized using a two-photon Raman transition (red arrows). Multi-qubit interactions are implemented using a two-photon excitation to one Zeeman sublevel of the $70S_{1/2}$ state (blue arrows). b) and c) show experimental data of coherent Rabi oscillations for ground-Rydberg and ground-ground transitions respectively. Reprinted figure with permission from \textcite{levine2019parallel} Phys. Rev. Lett. 123, 170503 (2019), Copyright 2019 by the American Physical Society. }
    \label{fig:rabioscillationslevine}
\end{figure}

In many situations the transition $\ket{a}\leftrightarrow\ket{b}$ is electric dipole forbidden, in which case Eq.~\eqref{eq:HLM} can be realized via two phase-coherent fields with a relative phase $\varphi_j$ and a large detuning $\delta_e$ from some (dipole-coupled) intermediate state(s) $\ket{e}$ (Fig.~\ref{fig:rabioscillationslevine}a). Omitting the subindex $j$, adiabatic elimination of $\ket{e}$ yields $\Omega\approx \frac{\Omega_A\Omega_B}{2\delta_e}$ and $\Delta=\Delta_{2ph}+(\Omega_A^2-\Omega_B^2)/(4\delta_e)$, where $\Omega_A, \Omega_B$ are the coupling strengths for the two fields and $\Delta_{2ph}=\omega_A+\omega_B-\omega_0$. Off-resonant scattering from the intermediate state leads to extra decay (and dephasing) of the atomic states during state manipulations: $\gamma_a\approx \Omega_A^2\Gamma/(4\delta_e^2),\gamma_b\approx\nobreak \Omega_B ^2\Gamma/(4\delta_e^2)$, for $\ket{a}$ and $\ket{b}$ respectively, where $\Gamma$ is the bare decay rate of the intermediate state. 

In practice, additional thermal motion of the atoms can lead to random frequency shifts (and dephasing) of the qubits\cite{saffman2011rydberg,leseleuc2018analysis} proportional to the effective wavevector of the coupling fields and the mean (1D) thermal velocity $|\vec k_A\pm\nobreak \vec k_B|v_{th}$, depending on whether the photon energies add (e.g., two-photon ground-Rydberg transitions) or subtract (e.g., Raman transitions between two ground states). This can be minimized if the two lasers have similar wavelengths and if they are aligned in a co-propagating geometry (for Raman transitions) or in a counter-propagating geometry (for two-photon ground-Rydberg transitions). Motional dephasing could also be suppressed using dynamical decoupling sequences\cite{suter2016colloquium} or by cooling the atoms close to their motional ground state\cite{kaufman2012cooling,thompson2013coherence,norcia2018microscopic,wang2019preparation,lorenz2020raman}.

Ground-Rydberg transitions are commonly driven using a two-photon optical excitation with lasers that are frequency stabilized to high-finesse cavities, yielding linewidths ${\lesssim}10\,$kHz\cite{legaie2018subkilo,leseleuc2018analysis,levine2018highfidelity}. To minimize off-resonant scattering, a large detuning is used from the typically short-lived intermediate state. Figure~\ref{fig:rabioscillationslevine}b shows exemplary data from \textcite{levine2019parallel}, demonstrating high contrast Rabi oscillations between the $5S_{1/2}$ electronic ground state and the $70S_{1/2}$ Rydberg state of $^{87}$Rb, which is a key step for realizing high-fidelity multi-qubit operations.

Rydberg-Rydberg transitions can be efficiently driven using externally controlled microwave fields. 
Direct one photon ($\delta \ell=1$)~\cite{leseleuc2017optical,orioli2018relaxation} and two-photon microwave couplings (with $\delta \ell=0,2$) via an intermediate Rydberg state have been used~\cite{signoles2019observation,wilson2019trapped}. Technology for microwave signal generation is very well developed and the large transition dipole moments of Rydberg states (scaling with $n^2$) enables fast quantum operations with modest microwave powers\cite{tommey2019resonant}, even reaching the ultrastrong coupling regime ($\Omega>\omega_0$)\cite{kockum2019ultrastrong}. 

Ground-ground transitions can be driven using either microwave fields (magnetic dipole coupling in the case of hyperfine states) or two-photon optical Raman transitions involving two phase referenced laser fields (depicted as red arrows in Fig.~\ref{fig:rabioscillationslevine}a, with exemplary Rabi oscillation data in Fig.~\ref{fig:rabioscillationslevine}c). To minimize decoherence the Raman lasers must be far detuned from any short-lived electronically excited states to suppress photon scattering. In the case of ground and meta-stable states (i.e., alkaline earth atoms/ions) the qubit transition could be driven by a single highly-stable optical clock laser.

\subsubsection*{Global versus local manipulation}

The technology for controlling laser and microwave couplings (that determine the parameters $\Omega_j,\varphi_j$ and $\Delta_j$) on sub$-\mu$s timescales is by now quite well developed, e.g., using acousto-optical modulators driven by arbitrary wave generators (AWGs). It is also possible to spatially address a few qubits at a time by passing the light fields through crossed acousto-optical deflectors (AODs) (Fig.~\ref{fig:register}b), each driven by one or more radio-frequency tones. Each tone generates a diffracted beam with an angle that depends on the radio-frequency. Passing this light through a high numerical aperture lens allows to target atoms at different positions in a two-dimensional plane. 

\begin{figure}[!t]
    \centering
    \includegraphics[width=\linewidth]{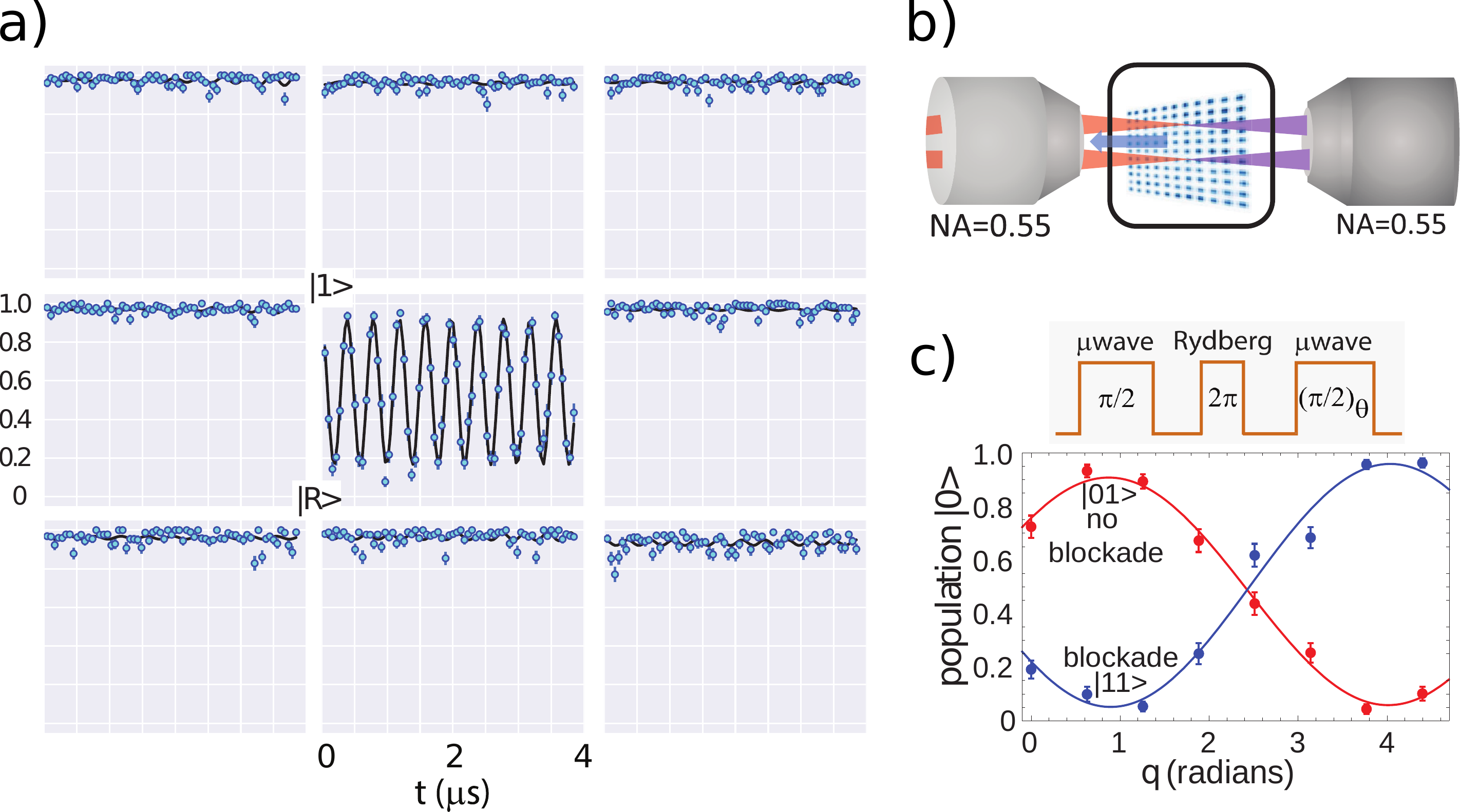}
    \caption{Demonstration of local qubit addressing and quantum gates in a two-dimensional register with low crosstalk from \textcite{graham2019rydberg}. a) Population in the atomic ground state $\ket{1}$ showing single-qubit Rabi oscillations on the central site of a 9 site neighborhood. b) Sketch of the setup used to realize two-qubit gates with individual addressing using a two-photon (red and purple) Rydberg excitation lasers in counter propagating geometry. c) "Eye" diagram used to characterize the conditional phase shift of a Rydberg blockade CZ gate. Reprinted figure with permission from \textcite{graham2019rydberg} Phys. Rev. Lett. 123, 230501 (2019), Copyright 2019 by the American Physical Society.}
    \label{fig:graham2019}
\end{figure}

Single and two qubit operations with negligible influence on neighboring qubits (low crosstalk) have been experimentally demonstrated using focused Rydberg excitation lasers or Raman coupling lasers\cite{yavuz2006fast,isenhower2010demonstration,graham2019rydberg} (one example from \textcite{graham2019rydberg} is shown in Fig.~\ref{fig:graham2019}). Another approach is to use tightly focused lasers to produce local AC Stark shifts for addressing certain qubits in combination with global microwave couplings  \cite{xia2015randomized,graham2019rydberg,leseleuc2017optical}. This method can even be applied to address individual atomic qubits in 3D quantum registers with high fidelity and low crosstalk using pairs of intersecting addressing lasers\cite{wang2015coherent,wang2016single}.

This technology for Rydberg qubit manipulation is rapidly improving, but has not yet reached the sophistication to allow for fully independent manipulations on much more than two qubits at a time. Therefore, to date, most demonstrations on larger systems have focused on sequential operations acting on one or two qubits at a time or parallel operations acting similarly on multiple qubits at the same time (Sec.~\ref{sec:demonstrations}). Global control fields (that simultaneously act on all sites in the array) are often sufficient for quantum simulation applications and some models for quantum computation based on the evolution of quantum Hamiltonians consisting of homogeneous couplings and local interactions\cite{lloyd1993potentially,benjamin2000schemes,wintermantel2020unitary} (see, for example, Sec.~\ref{sec:proposals}\,B for a proposed realization of quantum cellular automata using Rydberg qubits).  

\subsubsection*{Single qubit gates}

The physical interaction described by Eq.~\eqref{eq:HLM} can be used to realize a set of discrete single qubit quantum operations or quantum gates. For simplicity we assume a single qubit is coupled with a constant Rabi frequency $\Omega_j(t)=\Omega$ and detuning $\Delta_j(t)=\Delta$ for a fixed gate time $\tau_g$. In this case the time evolution operator can be represented by the following single qubit unitary operation
\begin{align}
\textrm{U} &=\textrm{exp}(-i\hat H^{01}_j\tau_g) \nonumber\\
&=\exp\left(\frac{i\Delta\tau_g}{2}\right)\times \exp\left(-\frac{i\tilde \Omega \tau_g}{2}\vec v\cdot\vec \sigma\right)
\end{align}
where $\tilde \Omega = \sqrt{\Omega^2+\Delta^2}$ is the generalized Rabi frequency, $\vec v = \tilde\Omega^{-1}\{\Omega\cos\varphi,\,-\Omega\sin\varphi,\,\Delta\}$ and $\vec{\sigma}=\{\hat X,\hat Y,\hat Z\}$ denotes the three-component vector of Pauli operators. This can be understood as a rotation of the Bloch vector by an angle $\tilde \Omega\tau_g$ around the axis defined by $\vec v$. This highlights the importance of the physical parameters $\Omega, \Delta$ and $\varphi$ (which depend on the excitation lasers and the position of the atom in the field) for achieving high gate fidelities, even if shaped or composite pulses can give better performance in the presence of noise. 

For $\Delta=0$ the time evolution operator can be simplified to an $xy$ rotation gate:
\begin{align}\label{eq:UXY}
    \textrm{U}_{xy}(\theta,\varphi)&=\left(\begin{array}{cc} \cos(\theta/2) & -i\sin(\theta/2) e^{i\varphi} \\ -i\sin(\theta/2) e^{-i\varphi}  & \cos(\theta/2) \end{array}\right),
\end{align}
with $\theta=\Omega \tau_g$. From this gate it is possible to generate a full set of single qubit rotation gates:
\begin{align*}
    \textrm{R}_x(\theta) &= \textrm{U}_{xy}(\theta,0),\\
    \textrm{R}_y(\theta) &= \textrm{U}_{xy}(\theta,-\pi/2),\\
    \textrm{R}_z(\theta) &= \textrm{U}_{xy}(\pi/2,\pi/2) \textrm{U}_{xy}(\theta,0) \textrm{U}_{xy}(\pi/2,-\pi/2).
\end{align*}
R$_z$ rotations can alternatively be natively implemented for $\Delta\neq 0$ (without three U$_{xy}$ rotations) using focused lasers to locally AC Stark shift the qubit energy \cite{graham2019rydberg,levine2019parallel}. Any other single qubit operation can be realized using combinations of rotation operators, e.g., the Hadamard gate within a global phase factor is $\textrm{U}_{xy}(\pi,0)\textrm{U}_{xy}(\pi/2,-\pi/2)$. Together with the multi-qubit gates described in the next section, this forms a universal set from which any computation or digital quantum simulation can in principle be realized.

\subsection{Multi-qubit manipulation}\label{sec:interactions}

\begin{figure}
  \includegraphics[width=0.95\linewidth]{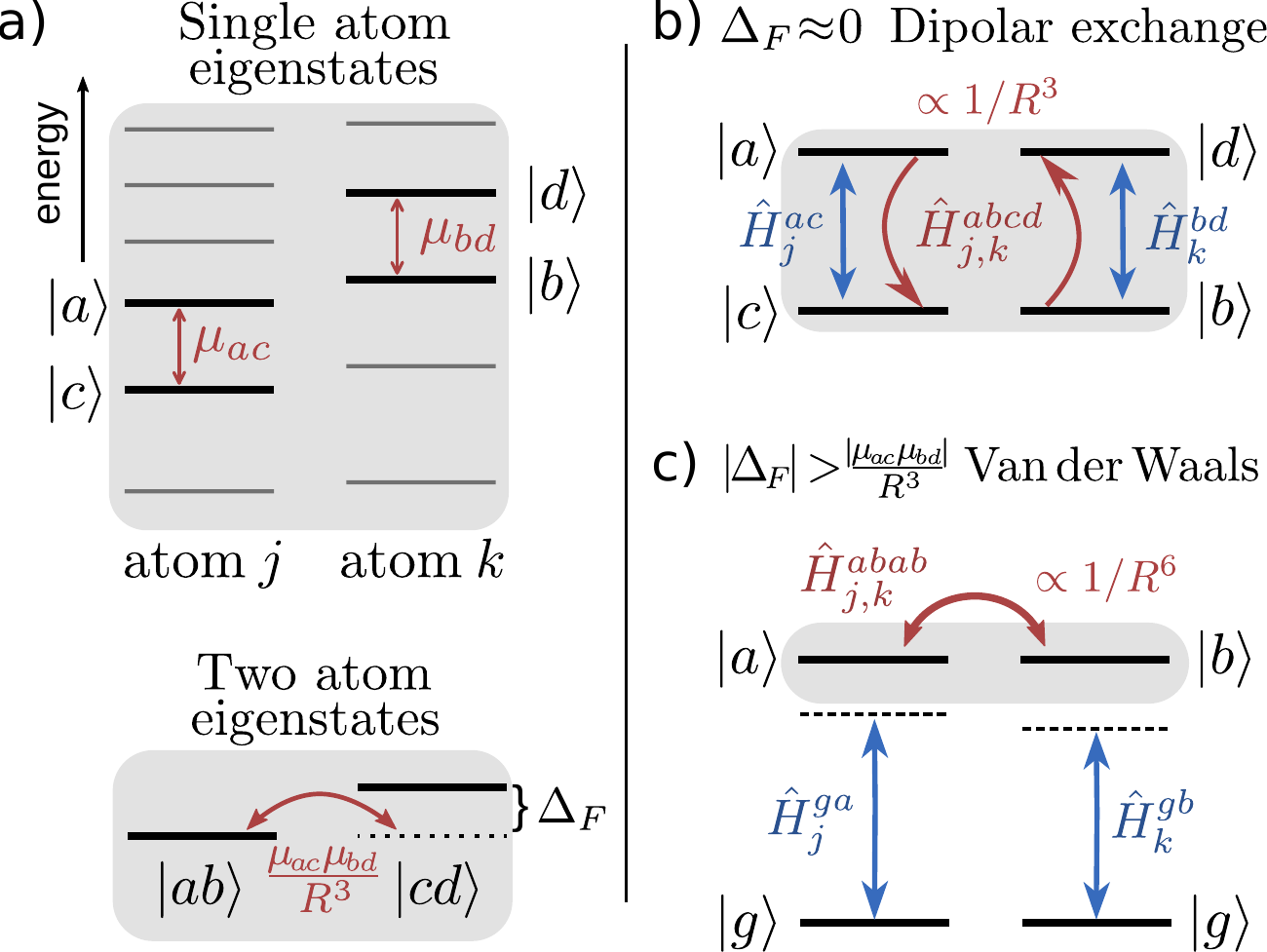}
  \caption{Physical origin of Rydberg-Rydberg interactions that can be used to construct quantum Hamiltonians and gates. a) Rydberg manifold energy level diagram for two atoms. Each atom possesses a pair of dipole coupled states $(a,c),(b,d)$ which have transition dipole moments $\mu_{ac},\mu_{bd}$ and similar energy differences. The lower panel shows the relevant pair state diagram showing the F\"orster defect $\Delta_F$. b) For $\Delta_F\approx 0$ the Rydberg-Rydberg interactions take the form of a dipolar exchange $\ket{ab}\bra{cd} + \textrm{h.c.}$ interaction with a $1/R^3$ distance dependence. c) For $|\Delta_F|>|\mu_{ac}\mu_{bd}|/R^3$ the interactions take a Van der Waals form $\ket{ab}\bra{ab}$ with a $1/R^6$ distance dependence.}
  \label{fig:interactions}
\end{figure}

\subsubsection*{Rydberg-Rydberg interactions}
Rydberg states are distinguished by their strong and tunable interactions (illustrated in Fig.~\ref{fig:interactions}) which can be used to mediate quantum gates between trapped atoms (or ions) and to engineer many-body Hamiltonians. Usually these interactions can be described by a general two-body Hamiltonian term:
\begin{align}\label{eq:Habcd}
    \hat H^{abcd}_{j,k} = \frac{V_{j,k}}{2}\ket{ab}\bra{cd}+\textrm{h.c.}\, ,
\end{align}
where $(a,c)$ and $(b,d)$ denote any two pairs of states in the Rydberg manifold of atom $j$ and $k$ respectively (depicted in Fig.~\ref{fig:interactions}a), $V_{j,k}$ is the (distance dependent) interaction strength coefficient, and $\ket{\alpha\beta}\equiv\ket{\alpha}_j\otimes\nobreak\ket{\beta}_k$. The precise form of this interaction will depend on the specific choice of states $a,b,c,d$. 

Interactions between Rydberg qubits originate from the large transition dipole moments of the Rydberg states\cite{saffman2010quantum}. Here we focus on the effect of the interactions on the internal states of the atoms (see Refs.\cite{wuster2011excitation,hague2012quantum,secker2016controlled,plodzien2018simulating,gambetta2020engineering,skannrup2020phonon} for ideas on how to exploit the effects of Rydberg interactions on motional degrees of freedom). We start from the dipole-dipole interaction operator for two atoms $j$ and $k$:
\begin{align}\label{eq:Vdd}
\hat{V}=\frac{1}{4\pi\epsilon_0}\frac{\hat{\mu}_j\cdot\hat{\mu}_k-3(\hat{\mu}_j\cdot \vec{n})(\hat{\mu}_k\cdot \vec{n})}{R_{j,k}^3}
\end{align}
where $\hat\mu_j$ is the electric dipole operator for $j$-th atom and $\vec{n}$ is a unit vector connecting the two atoms with separation $R_{j,k}$.

As a first approximation it is often sufficient to consider just a few dominant dipole-coupled states (depicted in Fig.~\ref{fig:interactions}a). For simplicity we neglect the angular dependence of the dipole-dipole interaction and assume just two dipole-coupled Rydberg states per atom $(a\leftrightarrow c)$ and $(b\leftrightarrow d)$, chosen based on their similar transition frequencies $E_a-E_c\approx E_{d}-E_{b}$. These could represent states of a single atomic species (in which case $a$ and $b$ could even be the same electronic state) or they could refer to Rydberg states of two different types of qubits (e.g., computational and ancilla qubits that could be useful for logical qubit encoding, minimally destructive readout and quantum error correction schemes). 

In the pair state basis $\{\ket{ab},\ket{cd}\}$ (Fig.~\ref{fig:interactions}a-lower) the two-atom Hamiltonian can be written
\begin{align}\label{eq:H2}
    \hat H&=\left(\frac{\mu_{ac}\mu_{bd}}{R_{j,k}^3}\ket{ab}\bra{cd}+\textrm{h.c.}\right)+\Delta_F\ket{cd}\bra{cd},
\end{align}
with $\mu_{\alpha\beta} = \langle \alpha| \hat  \mu|\beta\rangle/\sqrt{4\pi\epsilon_0}$ and $\Delta_F=E_c+E_{d}-(E_a+E_{b})$ is called the F\"orster defect. In writing Eq.~\eqref{eq:H2} we neglect the pair states that are far away in energy and we have subtracted the energy offset $E_a+E_{b}$. From this general two-body Hamiltonian it is possible to realize different types of basic interactions (depicted in Fig.~\ref{fig:interactions}b,c), depending on the magnitude of the F\"orster defect.\\

\noindent\textit{Resonant interactions ($\Delta_F\approx 0$)}\\
In this case Rydberg-Rydberg interactions take the form of a coherent state exchange process (Fig.~\ref{fig:interactions}b) with a dipolar $1/R^3$ distance dependence
\begin{align}
\hat H &= \frac{\mu_{ac}\mu_{bd}}{R_{j,k}^3}\ket{ab}\bra{cd}+\textrm{h.c.},  \\ \nonumber
&= \hat H^{abcd}_{j,k}\!\biggl(V_{j,k}=\frac{2\mu_{ac}\mu_{bd}}{R_{j,k}^3}\biggr).
\end{align}
This Hamiltonian is naturally realized for two atoms of the same atomic species prepared in two dipole-coupled Rydberg states (e.g., $a=d=nS$ and $b=c=nP$). It can also be realized by shifting other pair states into F\"orster resonance using external electric fields\cite{ryabtsev2010observation,ravets2014coherent}. This opens the possibility to modulate the effective two-body Rydberg-Rydberg interaction strength, which could be beneficial for implementing certain gate protocols\cite{huang2018robust,beterov2018fast,yu2019adiabatic}.\\

\begin{figure*}[!ht]
  \includegraphics[width=1\linewidth]{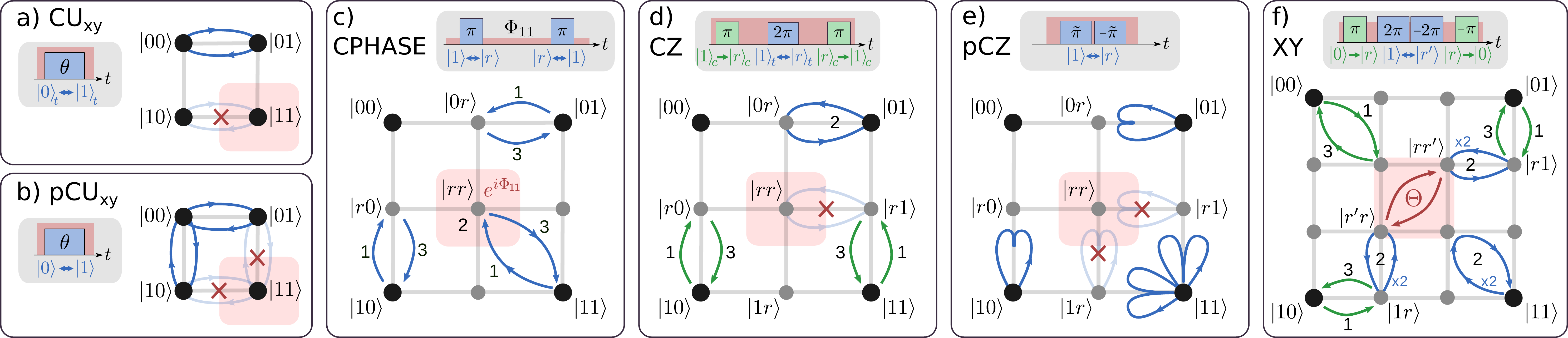}
  \caption{Overview of the native two-qubit gate protocols for Rydberg qubits and corresponding process diagrams. The computational states are depicted by black disks while auxiliary states used to mediate the interactions are depicted by gray disks. Rydberg interacting states are shaded in red and Rydberg blockaded transitions are marked with a red cross. (a) and (b) concern gr-qubits (and dressed gg-qubits) while (c-f) concern gg-qubits. A protocol consists of a sequence of coupling pulses acting on one or both qubits in parallel [depicted by blue arrows ($\pi$-pulses) and loops ($2\pi$-pulses)]. The numbers indicate the temporal order of the coupling pulses. Inconsequential couplings are omitted for clarity. Insets (in grey shaded areas) show the control pulse sequences.}
  \label{fig:operations}
\end{figure*}

\noindent\textit{Non-resonant interactions ($|\Delta_F|\gg |\mu_{ac}\mu_{bd}|/R_{j,k}^3$)}\\
In this case the dipole interaction Hamiltonian can be treated using second-order perturbation theory\cite{brion2007adiabatic}, yielding for the low energy subspace
\begin{align}
\hat H = -\frac{C_6}{R_{j,k}^6}\ket{ab}\bra{ab} = \hat H^{abab}_{j,k}\!\biggl(V_{j,k}=-\frac{C_6}{R_{j,k}^6}\biggr ),
\end{align}
with the van der Waals coefficient $C_6=\frac{|\mu_{ac}|^2|\mu_{bd}|^2}{\Delta_F}$. This corresponds to a distance dependent energy shift of the $\ket{ab}$ state by an amount $-C_6/R_{j,k}^6$ (Fig.~\ref{fig:interactions}c). This is usually encountered for two atoms prepared in the same Rydberg state, e.g., $a=b=nS$ or two different Rydberg states that are not directly dipole-coupled, e.g., $a=nS, b=n'S$.

In general, the precise form of the Rydberg-Rydberg interactions, including their dependence on interatomic separation and relative angles is determined by a sum of contributions from all dipole-coupled pair states. However as a general rule in alkali atoms, Rydberg $nS$ states exhibit approximately isotropic and repulsive van der Waals interactions ($C_6<0$) and interactions between $nS$ and $nP$ states have a relatively simple dipolar form $V_{j,k} \propto 1-3\cos^2\theta_{j,k}$. In 2018 two open source software packages were released that enable high accuracy calculations of Rydberg state properties including their crucial interaction properties\cite{sibalic2017arc,robertson2020arc,weber2017calculation}. 

\subsubsection*{Native two-qubit gates}
The first protocols for realizing two-qubit entangling gates based on controllable Rydberg interactions were put forward by \textcite{jaksch2000fast}. However the Rydberg-Rydberg interactions described by Eq.~\eqref{eq:Habcd} combined with single qubit manipulations described by Eq.~\eqref{eq:HLM} provide the basis for constructing a much broader set of fast and robust multiqubit gates that exploit the versatile interaction properties of Rydberg atoms. In the following we give an overview of these gates and protocols for their implementation (not an exhaustive list), with an emphasis on the types of gates that have been experimentally demonstrated (several examples are sketched in Fig.~\ref{fig:operations}).\\

\noindent\textit{Controlled-$U_{xy}$ rotation gate (gr- and gg-qubits)}\\
A conceptually simple entangling operation that can be natively realized with gr-qubits is the controlled-$\mathrm{U}_{xy}$ rotation (CU$_{xy}$) gate described by the unitary operator
\begin{align}\label{eq:CUxy}
\textrm{CU}_{xy} & =\left (
\begin{array}{cccc}
\cos{(\theta/2)} & -i\sin(\theta/2)e^{i\varphi} & 0 & 0\\
-i\sin(\theta/2)e^{-i\varphi} & \cos{(\theta/2)} & 0 & 0\\
0 & 0 & 1 & 0 \\
0 & 0 & 0 & 1
\end{array}
\right ),
\end{align}
in the basis $\{\ket{0}_c\ket{0}_t, \ket{0}_c\ket{1}_t, \ket{1}_c\ket{0}_t, \ket{1}_c\ket{1}_t\}$ where $c$ and $t$ refer to the control and target qubits respectively. This describes a single qubit U$_{xy}$ rotation on the target qubit, conditional on the control being in state $\ket{0}$. This corresponds to a special case of the Barenco gates \cite{barenco1995universal} and is also sometimes called a controlled rotation (CROT) gate\cite{li2003alloptical}. The CU$_{xy}$ gate can be converted to a controlled-NOT gate (CNOT) using $\theta=\pi,\varphi=0$ and preceding it with a R$_z(-\pi/2)$ rotation on the control qubit.

This gate can be obtained from the time evolution operator resulting from the physical interactions described by Eq.~\eqref{eq:HLM} and Eq.~\eqref{eq:Habcd} assuming non-resonant interactions (with $\ket{g}\equiv\ket{0}$ and $\ket{r}\equiv\ket{1}$):
\begin{align}
\hat U &=\exp[-i( \hat H^{01}_t + \hat H^{1111}_{c,t}) \tau_g].
\end{align}
The action of this pulse sequence is also sketched in Fig.~\ref{fig:operations}a. It corresponds to a Rabi oscillation $\ket{0}_t\leftrightarrow\ket{1}_t$ on the target qubit $t$ in the presence of a strong interaction described by a diagonal operator which blocks the oscillation if the control qubit $c$ is in the $\ket{1}_c$ state. Eq.~\eqref{eq:CUxy} can be derived for $\theta=\Omega_t \tau_g$, $\Delta_t=0$ and $|V_{c,t}| \gg \Omega_t$ to ensure the blockade condition is met. Technically there is an additional interaction-induced phase accumulated on the $\ket{1}_c\ket{1}_t$ state, which can be compensated by fixing $|V_{c,t}|= 2\pi m/\tau_g$ with $m$ a large integer. Though this requirement can be relaxed when the system is constrained to the $\{\ket{0}_c\ket{0}_t, \ket{0}_c\ket{1}_t, \ket{1}_c\ket{0}_t\}$ subspace. This type of gate is relatively easy to realize with gr-qubits, but could be implemented with any qubits that possess a (strong) diagonal two-body interaction. 

The $\textrm{CU}_{xy}$ Rydberg gate is robust to variations in the precise strength of the Rydberg-Rydberg interaction (aside from the phase accumulated on the $\ket{1}_c\ket{1}_t$ state, as long as the blockade condition is met), atomic center of mass motion or mechanical forces between Rydberg excited atoms. It can also be straightforwardly extended to $k-$control qubits within the blockade volume. It is important to note however, that the physical two-body interaction $\hat H^{1111}_{j,k}$ is always on and acts in the computational subspace. This means that in larger systems this gate does not act in isolation to other qubits and the system may evolve even when the control fields (via $\hat{H}^{01}_{j}$) are switched off. This could be overcome using a strong Rydberg-dressed blockade interaction or F\"orster resonance that can be switched on and off using external fields. On the other hand, always-on interactions might be advantageous for implementing multi-qubit gates and protocols that do not require individual qubit addressing\cite{benjamin2001quantum}.\\

\noindent\textit{Parallel controlled-$U_{xy}$ rotation gate (gr- and gg-qubits)}\\
A variant of the $\textrm{CU}_{xy}$ rotation gate can be applied to two qubits at the same time to give the following unitary transformation
\begin{align}\label{eq:CUxyparallel}
\textrm{pCU}_{xy} & =\left (
\begin{array}{cccc}
\cos(\theta/2) & s(\theta,\varphi)  & s(\theta,\varphi) & 0\\
s(\theta,-\varphi) & \cos^2(\theta/4) & -\sin^2(\theta/4) & 0\\
s(\theta,-\varphi) & -\sin^2(\theta/4) & \cos^2(\theta/4) &0\\
0 & 0 & 0 & 1\\
\end{array}
\right ),
\end{align}
with $s(\theta,\varphi)=-i\sin(\theta/2)e^{i\varphi}/\sqrt{2}$.
This corresponds to the pulse sequence
\begin{align}
\hat U &=\exp[-i( \hat H^{01}_c + \hat H^{01}_t + \hat H^{1111}_{c,t}) \tau_g].
\end{align}
using $\Omega = \theta/(\sqrt{2}\tau_g)$ and $\Delta=0$ for both qubits and $|V_{c,t}| = 2\pi m/\tau_g \gg \Omega$, with $m\in \mathbb{Z}$ if compensating the phase accumulated on the $\ket{11}$ state.
Since both qubits are addressed in parallel there is no distinction between control and target for this gate. It describes collective Rabi oscillations of both qubits between a state with zero Rydberg excitations to a state with precisely one Rydberg excitation shared by both qubits (Fig.~\ref{fig:operations}b). These oscillations were first observed in 2009, for two gr-qubits by \textcite{gaetan2009observation} and subsequently used to realize entangling operations for gr-\cite{wilk2010entanglement,levine2018highfidelity,picken2018entanglement,madjarov2020high} and dressed gg-qubits\cite{jau2016entangling}. Generally this is achieved starting from the $\ket{00}$ state and choosing $\Omega_c=\Omega_t=\Omega, \Delta_c=\Delta_t=0$, $|V_{c,t}|\gg \Omega$ and $\theta=\pi$ which results in a maximally entangled Bell state of two qubits, i.e., $1/\sqrt{2}\left(\ket{01}+\ket{10}\right )$. \\

\noindent\textit{Controlled phase gate family (gg-qubits)}\\
The problem of the always-on interactions can be overcome by encoding qubits in two ground states and transiently exciting to an auxiliary Rydberg state. This can be used to produce the controlled phase gate:
\begin{align}\label{eq:Cphase}
\textrm{CPHASE} & =\left (
\begin{array}{cccc}
e^{i\Phi_{00}} & 0 & 0 & 0\\
0 & e^{i\Phi_{01}} & 0 & 0\\
0 & 0 & e^{i\Phi_{10}} & 0\\
0 & 0 & 0 & e^{i\Phi_{11}}
\end{array}
\right ).
\end{align}
This is an entangling gate as long as $(\Phi_{00}+\Phi_{11})-(\Phi_{10}+\nobreak\Phi_{01})$ is not a integer multiple of $2\pi$. A special case is when $\Phi_{00}=\Phi_{10}=\Phi_{01}=0$ and $\Phi_{11}=\pi$ when the controlled phase gate is equivalent to the canonical controlled-Z (CZ) gate. 

\begin{figure}[!t]
    \centering
    \includegraphics[width=0.4\textwidth]{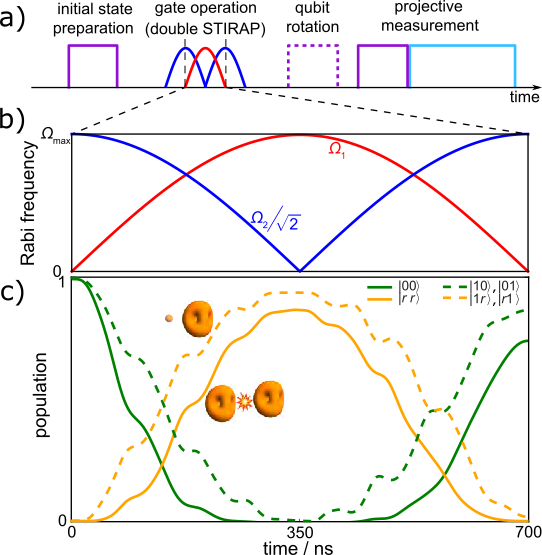}
    \caption{Protocol for a Rydberg mediated CPHASE gate for trapped ions. a) Pulse sequence used in \textcite{zhang2020submicrosecond} to implement the gate including, b) a double STIRAP pulse used to transfer population to strongly interacting Rydberg states. c) Simulation of the populations in the computational and Rydberg manifold during gate operation. In this figure the $\ket{00}$ state acquires an additional phase proportional to the area under the $\ket{rr}$ curve. Reprinted by permission from Springer Nature: \textcite{zhang2020submicrosecond} Nature 580, 345-349 (2020), Copyright 2020.}
    \label{fig:zhang2020}
\end{figure}

A specific physical implementation of $\textrm{CPHASE}$ (originally proposed as `Model A' by \textcite{jaksch2000fast}) can be expressed by the pulse sequence
\begin{align}\hat U &= \exp[-i(\hat H^{r1}_c\!+\hat H^{r1}_t)\tau_1]\\\nonumber
&\quad\times\exp[-i \hat H^{rrrr}_{c,t} \tau_2]\\\nonumber
&\quad\times\exp[-i (\hat H^{r1}_c\!+\hat H^{r1}_t)\tau_1],
\end{align}
with $\Delta_{c}=\Delta_t=0,\Omega_c=\Omega_t=\Omega$ and $\tau_1=\pi/\Omega$. This sequence is depicted in Fig.~\ref{fig:operations}c and describes a Rabi $\pi$-pulse on both qubits from $\ket{1}$ to $\ket{r}$ (labelled $1$) followed by a waiting period $\tau_2$ in the presence of Rydberg-Rydberg interactions (labelled $2$) and then a subsequent Rabi $\pi$-pulse back to the $\ket{1}$ state (labelled $3$). Depending on the waiting period the state $\ket{11}$ will acquire a different phase than the $\ket{01},\ket{10}$ states. This yields $\textrm{CPHASE}$ with $\Phi_{00}=0$, $\Phi_{01}=\Phi_{10}=\pi$ and $\Phi_{11}=-V_{c,t}\tau_2$. Assuming van der Waals interactions this corresponds to a two-qubit phase shift of $\Phi_{11} = (C_6/R_{c,t}^6)\tau_2$. A specific implementation of this type of gate for trapped ions, realized by \textcite{zhang2020submicrosecond}, is shown in Fig.~\ref{fig:zhang2020}.

This particular implementation of a CPHASE gate lacks the robustness of the $\textrm{CU}_{xy}$ gate or the Rydberg blockade gate below because of its sensitivity to the interaction strength $V_{c,t}$ which can cause errors in the accumulated phase (depending on the relative positions of the atoms) or the efficiency of population transfer between the computational states and the Rydberg states that mediate the interaction. Robustness of the CPHASE gate can be improved using composite pulses or stimulated Raman adiabatic passage pulses instead of the Rabi $\pi$-pulses\cite{zhang2020submicrosecond}, and using additional fields to control the interaction strength during the gate operation.

A more sophisticated version of the controlled phase gate was recently proposed by \textcite{Shi2019fast}. The protocol involves a single off-resonant coupling pulse on both qubits in parallel or individually to realize gates in the general form of Eq.~\eqref{eq:Cphase}. Gate errors due to motional dephasing and variations of the interaction strength can be minimized through careful choice of detunings, Rabi couplings and pulse durations. 

Errors induced by Doppler dephasing and fluctuating light shifts could be circumvented using Rydberg dressing and adiabatically varying two-atom light shifts \cite{keating2015robust} interleaved with spin echo sequences\cite{mitra2020robust}.\\

\begin{figure}[!t]
    \centering
    \includegraphics[width=0.80\columnwidth]{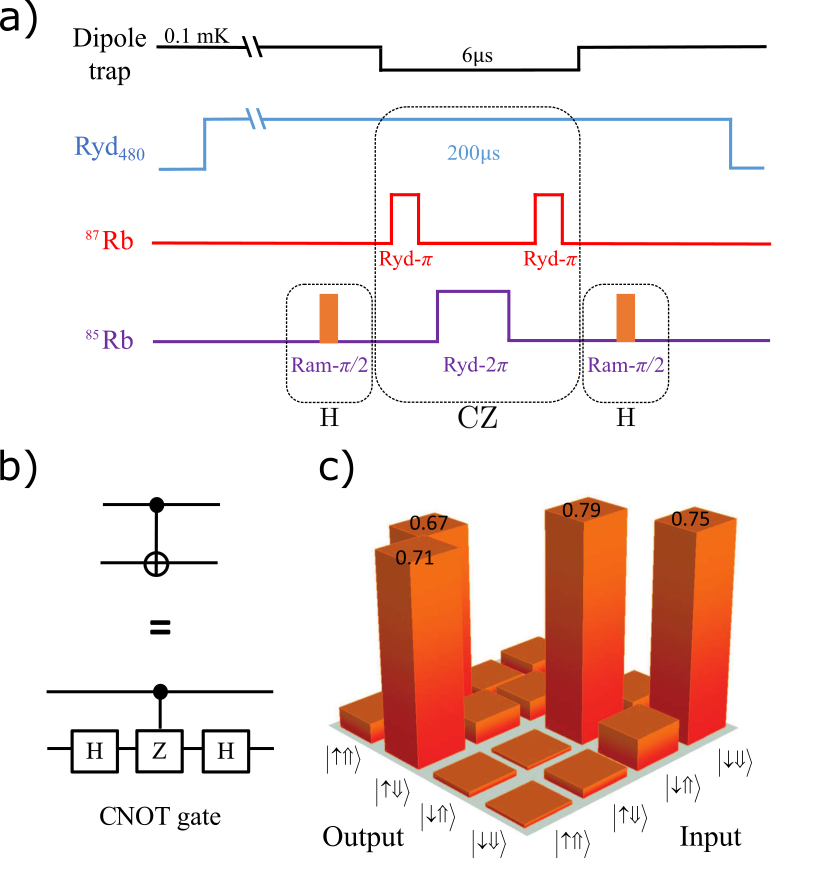}
    \caption{Implementation of a heteronuclear ($^{85}\text{Rb}$ and $^{87}\text{Rb}$) CZ and CNOT quantum gate with gg-qubit encoding. a) Pulse sequence of the gate implementation which involves $\textrm{R}_x(\pi/2)$ rotations (Raman pulses) on the target qubit before and after the Rydberg blockade CZ gate. b) Corresponding elementary circuit diagram. c) Measured CNOT gate matrix showing a fidelity $F=0.73(1)$ (uncorrected). Reprinted with permission from \textcite{zeng2017entangling} Phys. Rev. Lett. 119, 160502 (2017), Copyright 2017 by the American Physical Society.}
    \label{fig:heteronuclearCNOT}
\end{figure}

\noindent\textit{Rydberg blockade (CZ) gate (gg-qubits)}\\
An elegant way to overcome the sensitivity of Rydberg gates to the Rydberg interaction strength was also proposed by \textcite{jaksch2000fast} as `Model B', and first experimentally realized by \textcite{urban2009observation} and improved further in \textcite{graham2019rydberg}. An experimental demonstration of a hetronuclear CZ gate by \textcite{zeng2017entangling} (and its conversion to a CNOT gate using 5 pulses in total) is depicted in Fig.~\ref{fig:heteronuclearCNOT}a. The basic idea is that the presence of a Rydberg excitation of the control qubit induces a large level shift of the nearby target-qubit Rydberg-state which prevents its laser excitation. Since at most one atom is excited to the Rydberg state at a time, the gate also minimizes possible dephasing due to interatomic forces.

The Rydberg blockade (CZ) gate realizes the unitary Eq.\,\eqref{eq:Cphase} with $\Phi_{00}=0,\Phi_{01}=\Phi_{10}=\Phi_{11}=\pi$.
This is equivalent to the canonical controlled-Z (CZ gate) within a global phase factor and a choice of basis (or qubit state definitions). It can be implemented using a three pulse sequence sketched in Fig.~\ref{fig:operations}d and is described by the time evolution operator
\begin{align}
\hat U &= \exp[-i \hat H^{r1}_c \tau_1]\\\nonumber
&\quad\times\exp[-i (\hat H^{r1}_t+\hat H^{rrrr}_{c,t}) \tau_2]\\\nonumber
&\quad\times\exp[-i \hat H^{r1}_c \tau_1]
\end{align}
with $\Delta_c=\Delta_t=0,  \Omega_{c}=\Omega_t=\Omega$ and $\tau_2=2\tau_1=2\pi/\Omega$ and $|V_{c,t}|\gg \Omega$. This describes a first $\pi$-pulse on the control qubit from $\ket{1}\rightarrow\ket{r}$ (labelled $1$ in Fig.~\ref{fig:operations}d) then a $2\pi$-pulse on the target qubit from $\ket{1}\rightarrow\ket{r}$ (labelled $2$) followed by a final $\pi$-pulse on the control qubit back to $\ket{1}$ (labelled $3$). The $\ket{0}$ state is decoupled from the lasers. The strong Rydberg blockade interaction $|V_{c,t}|\gg \Omega$ can be present at all times but is only effective during the second coupling pulse. In the ideal case it has no effect on the first or last pulse, however it prevents the $2\pi$ rotation if both qubits start in the state $\ket{1}$. As a result, each computational state returns to its original state after the sequence is complete but acquires a phase shift of $\pi$ for each full Rabi oscillation it underwent. The use of optimally shaped pulses can further reduce the gate time and improve the fidelity (potentially $F>0.9999$ within $50\,$ns) by suppressing leakage to unwanted Rydberg states\cite{theis2016high}. Like the $\textrm{CU}_{xy}$ operation which is also based on the Rydberg blockade, this gate is insensitive to the precise interaction strength, motion of the atoms and uncertainties in the atomic positions, but without the complications associated with always-on interactions. \\

\noindent\textit{Parallel Rydberg blockade (pCZ) gate (gg-qubits)}\\
The original Rydberg blockade gate protocol requires independent manipulation of the control and target qubits, and generally it is found that multiple $\pi$-pulses applied in sequence lead to increased loss and dephasing as compared to a single continuous Rabi pulse that minimizes time spent in the Rydberg state\cite{levine2019parallel}. To overcome these difficulties, a variant was recently proposed and experimentally demonstrated by \textcite{levine2019parallel}. Several related protocols based on optimized adiabatic pulses were also recently proposed\cite{saffman2020symmetric}. Here we describe the \textcite{levine2019parallel} protocol which can be described by a sequence of two pulses applied to both qubits simultaneously:
\begin{align}
\hat U &= \exp[-i \bigl(\hat H^{rrrr}_{c,t}+\sum_\alpha\hat H^{r1}_\alpha(\varphi_\alpha=\varphi)\bigr) \frac{\tau_g}{2}]\\ \nonumber
&\quad\times\exp[-i \bigl (\hat H^{rrrr}_{c,t}+\sum_\alpha \hat H^{r1}_\alpha(\varphi_\alpha=0)\bigr) \frac{\tau_g}{2}],
\end{align}
with $\alpha=\{c,t\}$. This corresponds to the coupling pulse sequence depicted in Fig.~\ref{fig:operations}e where the first and second pulses have a relative phase difference $\varphi$. The gate works in the blockade regime $|V_{c,t}|\gg \Omega_\alpha,|\Delta_\alpha|$ and to ensure proper operation the pulse parameters must be carefully chosen to ensure that the $\ket{01},\ket{10}$ and $\ket{11}$ states perform a complete (detuned) Rabi oscillation back to their original states. However they each acquire a phase that depends on the geometric surface enclosed by their trajectory on the Bloch sphere, which is in general different for the $\ket{11}$ state (due to the blockade condition and collective enhancement). A set of parameters that achieves this is: $\Omega_{\alpha}=\Omega$, $\Delta_{\alpha}=0.377\Omega$, $\varphi=3.90242$ and the total gate time $\tau_g=2.7328\pi/\Omega$. The resulting gate is described by Eq.\,\eqref{eq:Cphase} with $\Phi_{00}=0$, $\Phi_{01}=\Phi_{10}=\phi_1$ and $\Phi_{11}=2\phi_1-\pi$, which is equivalent to a CZ gate within a defined single qubit phase $\phi_1$.  
This implementation is slightly faster than the original Rydberg blockade gate and simplifies the practical implementation since the coupling pulses are applied in parallel to both qubits without requiring individual control of the single qubit couplings.\\

\noindent\textit{Dark state adiabatic (CZ) gate (gg-qubits)}\\
\textcite{petrosyan2017high} propose a version of the Rydberg blockade gate that replaces the Rydberg blockade interaction by a resonant dipolar exchange interaction and adiabatic following of a two-atom dark state. Multi-qubit variants of this concept were proposed by \textcite{khazali2020fast} and will be discussed further in the next section. Related protocols using Stark tuned F\"orster resonances were proposed by \textcite{beterov2016} and a general introduction to adiabatic passage techniques for Rydberg quantum gates, especially applied to atomic ensembles, can be found in \textcite{beterov2020application}. The pulse sequence for the the dark state adiabatic gate is similar to the original Rydberg blockade gate except the second $2\pi$-pulse on the target qubit is replaced with a smooth pulse in the presence of a strong dipolar exchange interaction with an auxiliary Rydberg pair state. In addition to its robustness against uncertainties in the interaction strength and interatomic forces, it is also robust against phase errors induced by, e.g., residual Rydberg excitation under imperfect blockade conditions. However it is slower than other Rydberg blockade gates due to the adiabatic following condition and therefore may be more sensitive to Rydberg state decay. This could be partially overcome using short-cuts to adiabaticity\cite{kai2019geometric}.\\

\noindent\textit{Rydberg blockade CNOT gate (gg-qubits)}\\
The Rydberg blockade gate in the original `Model B' protocol can be transformed into a CNOT gate  by replacing the middle $2\pi$-pulse with three $\pi$-pulses on the target qubit ($\ket{1}_t\rightarrow\ket{r}_t,\ket{r}_t\leftrightarrow\ket{0}_t,\ket{r}_t\rightarrow\ket{1}_t$) that swaps the target state as long it is not blockaded\cite{isenhower2011multibit}. Alternatively a CNOT can be realized by preceding and following the CZ gate with Hadamard gates on one of the qubits as shown in Fig.~\ref{fig:heteronuclearCNOT}, or a $\mathrm{R}_x(\pi/2)$ gate on alternate qubits before and after the CZ\cite{isenhower2010demonstration}.\\

\noindent\textit{XY gate family (gg-qubits or rr-qubits)}\\
A final type of gate which can be naturally implemented with Rydberg qubits is the XY family of gates\cite{shuch2003natural,abrams2019implementation,ni2018dipolar}. This can be thought of as a coherent rotation by an angle $\Theta$ between the $\ket{01}$ and $\ket{10}$ states, with XY$(\Theta=3\pi)$ being equivalent to the iSWAP gate. 

The $\textrm{XY}$ gate can be represented as
\begin{align}\label{eq:XYgate}
\textrm{XY}(\Theta) & =\left(\begin{array}{cccc} 1 & 0 & 0 & 0 \\ 0 & \cos(\Theta/2) & -i\sin(\Theta/2) & 0 \\ 0 & -i\sin(\Theta/2) &  \cos(\Theta/2) & 0 \\ 0 & 0 & 0 &1\end{array}\right).
\end{align}
In the following we introduce a new four pulse protocol that exploits the strong dipolar exchange interactions between two different Rydberg states and does not require individual addressing. A different protocol for implementing a Rydberg SWAP gate was proposed by \textcite{huaizhi2012quantum} using strong state-dependent blockade interactions. Our protocol consists of the pulse sequence
\begin{align}\label{eq:XYgateprotocol}
\hat U &= \exp[-i \sum_\alpha\hat H^{r0}_\alpha(\Omega_\alpha=\Omega_1,\varphi=\pi)\tau_1]\\\nonumber
&\quad\times\exp[-i (\hat H^{rr'r'r}_{c,t}\!+\!\sum_\alpha\hat H^{r'1}_\alpha(\Omega_\alpha\!=\!\Omega_2,\varphi\!=\!\pi)) \tau_2]\\\nonumber
&\quad\times\exp[-i (\hat H^{rr'r'r}_{c,t}\!+\!\sum_\alpha\hat H^{r'1}_\alpha(\Omega_\alpha\!=\!\Omega_2,\varphi\!=\!0)) \tau_2]\\\nonumber
&\quad\times\exp[-i \sum_\alpha\hat H^{r0}_\alpha(\Omega_\alpha=\Omega_1,\varphi=0)\tau_1],
\end{align}
with $\alpha=\{c,t\}$, $\Delta_\alpha=0$, $\tau_1\Omega_1=\pi$, $\tau_2\Omega_2= 2\pi$, $\tau_2=\Theta/V_{c,t}$. The states $r,r'$ denote two different dipole-coupled Rydberg states (e.g., $nS$ and $nP$ Rydberg states). This pulse sequence is depicted in Fig.~\ref{fig:operations}f and describes a $\pi$-pulse of both qubits from $\ket{0}\rightarrow\ket{r}$ (labelled $1$), followed by two out-of-phase $2\pi$ rotations on the $\ket{1}\rightarrow\ket{r'}$ transition in the presence of a resonant dipolar exchange interaction $H^{rr'r'r}_{c,t}= (V_{c,t}/2)\ket{rr'}\bra{r'r} + \text{h.c.}$ (labelled $2$). The duration of these pulses determines the XY rotation angle, while the strength of the Rabi coupling must be set accordingly to ensure two full $2\pi$ rotations. A final $\pi$-pulse on both qubits from $\ket{r}\rightarrow\ket{0}$ returns the system to the computational subspace (labelled $3$). According to this procedure the $\ket{00},\ket{11}$ components undergo a full rotation back to their original states. A simple comparison of Eq.~\eqref{eq:XYgateprotocol} with the ideal gate Eq.~\eqref{eq:XYgate} including the effect of the always-on dipole-dipole interactions (but neglecting other sources of decoherence) suggests that this protocol can yield high fidelities comparable to other proposed protocols with $F>0.99$~\cite{huaizhi2012quantum,ni2018dipolar}. This native XY gate benefits from $1/R^3$ interactions that could be advantageous for efficiently routing quantum information across a quantum register or for digital quantum simulation of spin models and efficient quantum algorithms requiring the minimum number of gate operations~\cite{kivlichan2018quantum,foxen2020demonstrating}. 

\subsection*{Multi-qubit (more than two-qubit) Rydberg gates}

\begin{figure}[!t]
  \includegraphics[width=1\columnwidth]{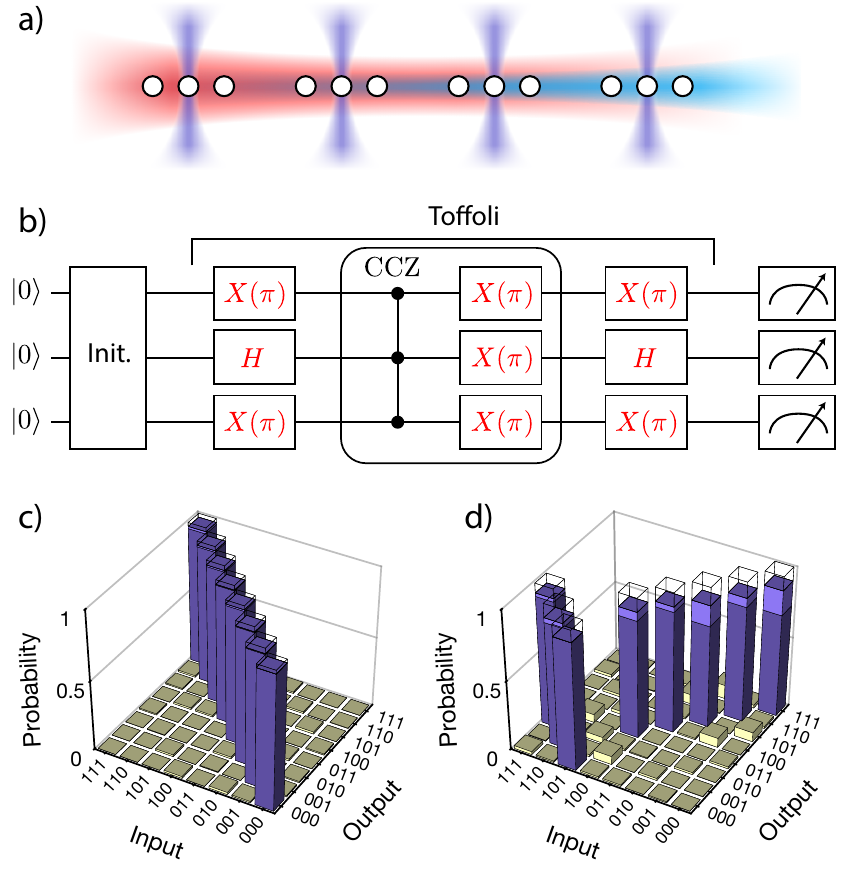}
  \caption{Demonstration of a three-qubit Toffoli (C$_2$NOT) gate with $^{87}$Rb Rydberg gg-qubits. a) Experimental setup consisting of a trimerized linear tweezer array with global state manipulation and local addressing on the central qubits. b) Circuit diagram used to realize the Toffoli gate from native Rydberg gates ($\textrm{X}(\pi)$ refers to the $\textrm{R}_x(\pi)$ rotation gate). c) Initialization of the system in each of the 8 basis states. d) Measured truth table for the C$_2$NOT gate with fidelity $F>0.870(4)$ corrected for state preparation and measurement errors. Reprinted figure with permission from \textcite{levine2019parallel} Phys. Rev. Lett. 123, 170503 (2019), Copyright 2019 by the American Physical Society.}
  \label{fig:toffoli}
\end{figure}

Rydberg qubits, with their strong and long-range dipole-mediated interactions are also naturally suited for implementing quantum operations acting on more than two qubits simultaneously\cite{lukin2001dipole,unanyan2002efficient,weimer2011digital,isenhower2011multibit}. While these gates can generally be decomposed in terms of one- and two-qubit gates, native $k-$qubit gates can be advantageous for constructing efficient quantum algorithms. For example the three-qubit Toffoli gate ($\equiv$C$_2$NOT) may be implemented as a sequence of six two-qubit CNOT gates and nine single qubit gates\cite{shende2009on,gulliksen2015characterization}. Alternatively, by exploiting the Rydberg blockade interaction with Rydberg qubits, the same gate could be implemented with a sequence of $3-5$ pulses.\\

\noindent\textit{Asymmetric blockade (C$_k$Z) gate}\\
A native multi-qubit gate for Rydberg qubits is the C$_k$Z blockade gate\cite{brion2007conditional,wu2010implementation,molmer2011efficient}, which can be understood as a generalization of the Rydberg blockade CZ gate to $k$ control qubits:
\begin{align}\label{eq:CkZ}
\textrm{C$_k$Z} & =\left (
\begin{array}{cc}
\mathbb{I}_{k} & 0 \\
0 & \hat{\textrm{Z}}\\
\end{array}
\right ).
\end{align}
where $\mathbb{I}_k$ is the $(2^{k+1}-2)$ dimensional identity operator and $\hat{\textrm{Z}}$ is the Pauli-Z operator acting on the target qubit. To implement this gate one can make use of resonant control pulses and asymmetric interactions between the Rydberg states of the control and target atoms to avoid unwanted interactions between control qubits. This is the case for two different Rydberg states with a strong interstate interaction or if there is a large separation between the control qubits as compared to the control-target separation (as is the case for a 1D chain with one control qubit on each side of the target qubit). Another way to achieve control-target interactions that are much stronger than control-control (and target-target) interactions is using microwave dressing of Rydberg states to enhance the interactions and induce an asymmetric Rydberg blockade\cite{muller2008trapped,tanasittikosol2011microwave,Sevin_li_2014,petrosyan2014binding,young2020asymmetric}. The gate can then be described by the unitary operation%
\begin{align}
\hat U &= \exp[-i \sum_k \hat H^{r0}_k(\Omega_k=\Omega_c,\varphi_k=\pi) \tau_1]\\\nonumber
&\quad\times\exp[-i \bigr(\hat H^{r'1}_t(\Omega_t,\varphi_t\!=\!0)\!+\!\sum_k\hat H^{rr'rr'}_{k,t}\bigl) \tau_2]\\\nonumber
&\quad\times\exp[-i \sum_k \hat H^{r0}_k(\Omega_k=\Omega_c,\varphi_k=0) \tau_1].
\end{align}
where $k$ is an index for all control atoms and $t$ refers to the target atom, $\Delta_c=\Delta_k=0$, $\tau_1\Omega_c=\pi$, $\tau_2\Omega_t=2\pi$, and $|V_{k,t}|\gg \Omega_t$. This describes a simultaneous $\pi$-pulse to all control atoms from the $\ket{0}$ state to a mutually non-interacting Rydberg state $\ket{r}$, then a $2\pi$-pulse to the target qubit to drive the transition $\ket{1}\rightarrow\ket{r'}\rightarrow \ket{1}$ while the $\ket{r'}$ state interacts with $\ket{r}$ via a strong diagonal interaction $\hat H^{rr'rr'}_{k,t}$. Accordingly the target qubit acquires a $\pi$ phase shift only if all of the control qubits were initially in the $\ket{1}$ (otherwise the target qubit rotation is blocked). The control qubits are then returned to $\ket{0}$ with a final $\pi$-pulse that is ideally out of phase with the first pulse. 

A parallel variant of the C$_2$Z gate has been experimentally demonstrated by \textcite{levine2019parallel}, using a global amplitude and frequency modulated laser pulse, analogous to the parallel CZ gate implementation described in the previous section. In this case however the required pulse shape is quite complex and was found by numerically optimization using an optimal control algorithm.

Generalizations of the C$_k$Z gate can be made to include multiple target qubits exploiting shortcuts to adiabaticity\cite{shen2019construction} and exploiting the coupling to an optical cavity mode\cite{liang2015shortcuts}.  \textcite{shi2018deutsch} proposed a generalization of the C$_2$Z gate to a three qubit Deutsch gate (as well as CNOT and Toffoli gate protocols that require only three pulses to realize) which replaces $\hat{\textrm{Z}}$ on the target qubit by something similar to a U$_{xy}$ operation. An alternative approach to generating multi-qubit phase gates is to make use of detuned laser fields and the antiblockade condition\cite{su2017fast,su2018one}.\\

\noindent\textit{Multi-qubit Toffoli ($\textrm{C}_k\textrm{NOT}$) and fan-out ($\textrm{CNOT}^k$) gates}\\
The $\textrm{C}_k\textrm{NOT}$ and $\textrm{CNOT}^k$ gates are generalizations of the Rydberg blockade gate whereby multiple control qubits condition a state change on the target qubit ($\textrm{C}_k\textrm{NOT}$) or a single control qubit conditions a state change on multiple target qubits ($\textrm{CNOT}^k$). 

In 2009, \textcite{muller2009mesoscopic} proposed a parallel version of a $\textrm{CNOT}^k$ gate using an electromagnetically-induced transparency resonance that can be used to entangle multiple atoms with a single control atom. In 2011, \textcite{isenhower2011multibit} proposed a $\textrm{C}_k\textrm{NOT}$ gate which replaces the $2\pi$-pulse in the C$_k$Z gate by three pulses which realize a state swap on the target qubit (with only 5 Rydberg $\pi$-pulses in total, independent of $k$). A $\textrm{C}_2\textrm{NOT}$ (Toffoli) gate was experimentally realized by \textcite{levine2019parallel} by applying a local Hadamard on one of the atoms before and after the C$_2$Z gate (Fig.~\ref{fig:toffoli}). 

A proposal for a three-qubit Toffoli gate was made by \textcite{beterov2018fast}, using Stark tuned three-body F\"orster resonant interactions. $\textrm{C}_k\textrm{NOT}$ gates can also be extended to multiple target qubits in a single step as long as the target-target interactions are weak compared to the control-target interactions. \textcite{khazali2020fast} recently proposed a set of $\textrm{C}_k\textrm{NOT}$ and $\textrm{CNOT}^k$ gates making use of dark-state adiabatic evolution and resonant exchange interactions (generalization of the dark state adiabatic gate), which could also be implemented in superconducting circuit qubits. 

A single step implementation of a $k-$qubit Toffoli gate was recently proposed which uses a similar protocol to the CU$_{xy}$ gate \cite{rasmussen2020single}.

\subsection{Many-body Hamiltonians}\label{sec:hamiltonians}
In this section we will discuss some of the many-body Hamiltonians that are naturally realized with Rydberg qubits. To date, experimental demonstrations with more than a few interacting Rydberg qubits have focused mainly on quantum spin-1/2 Hamiltonians. These models are of central interest in condensed matter and non-equilibrium physics, in particular for studying quantum dynamics, quantum phase transitions, and quasi-adiabatic and non-adiabatic quantum state preparation protocols involving different types of interactions and spatial configurations of the qubits (several examples will be discussed in Sec.~\ref{sec:demonstrations}). In the near future it should become possible to further generalize these models using, e.g., Hamiltonian engineering protocols\cite{ajoy2013quantum,hayes2014programmable,choi2020robust} and digital quantum simulation\cite{weimer2010rydberg,weimer2011digital}, which would potentially open up important applications in high-energy physics\cite{weimer2010rydberg,kokail2019self,celi2020emerging}, mathematics\cite{Smilansky_2007,anantharaman2015}, quantum chemistry\cite{babbush2014adiabatic,moll2018quantum,kuhn2019accuracy}, biophysics\cite{gunter2013observing,schonleber2015quantum,plodzien2018simulating,yang2019quantum} and even in some areas of machine learning\cite{biamonte2017quantum}, finance\cite{ORUS2019100028} and logistics\cite{neukart2017traffic,ding2019logistic}, among others. In the following we summarize a few of the paradigmatic models that have been explored so far.

\subsubsection*{Quantum Ising model (gr- and gg-qubits)}
One of the earliest and most extensively studied models to be realized experimentally with Rydberg qubits is the quantum Ising model in transverse and longitudinal fields\cite{schauss2015crystallization,labuhn2016tunable,takei2016direct,bernien2017probing,leseleuc2018,schauss2018quantum,lienhard2018observing,guardado2018probing}. This model is naturally implemented by an array of qubits with diagonal two-body Rydberg-Rydberg interactions\cite{robicheaux2005manybody}:

\begin{align}
\hat H = \sum_j \biggl[\hat H_j^{gr}(\varphi_j=0)+\frac{1}{2}\sum_{k\neq j}\hat H^{rrrr}_{j,k}\biggr].
\end{align}
Identifying $\ket{g}\equiv\ket{0}$ and $\ket{r}\equiv\ket{1}$ we can write this system as a spin Hamiltonian:
\begin{align}\label{eq:ising}
    \hat{H}&=\frac{1}{2}\sum_j\biggl [ \Omega_j(t)\hat X_j +  \bigl(\Delta_j(t)\!-\!\mathcal{I}_j\bigr)\hat Z_j + \sum_ {k\neq j}\!\frac{V_{j,k}}{4}\hat Z_j \hat Z_k\biggr ].
\end{align}
Here $\hat{X}_j,\hat{Z}_j$ are the Pauli-X and Pauli-Z operators acting on site $j$ (in the $\{\ket{0},\ket{1}\}$ basis) and we have used $\ket{1}_j\!\bra{1}=(1-\hat Z_j)/2$ and subtracted a global energy offset. The first two terms represent transverse and longitudinal fields controlled by the excitation lasers, that are usually globally applied to all the qubits, i.e., $\Omega_j=\Omega, \Delta_j=\Delta$. The third term is the Ising interaction. Two differences from the usual quantum Ising model are: (i) the site-dependent energy shift $\mathcal{I}_j=\sum_{k,(j\neq k)}V_{j,k}/2$. However in homogeneous systems and sufficiently far from the edges this equates to an inconsequential energy offset; (ii) the finite range interactions, typically scaling like $V_{j,k}=-C_p/R_{j,k}^p$ ($p=6$ for van der Waals interactions) that can be controlled by choosing different Rydberg states (Sec.~\ref{sec:interactions}) and by varying the atomic separations.

\begin{figure}[!t]
    \centering
    \includegraphics[width=0.9\columnwidth]{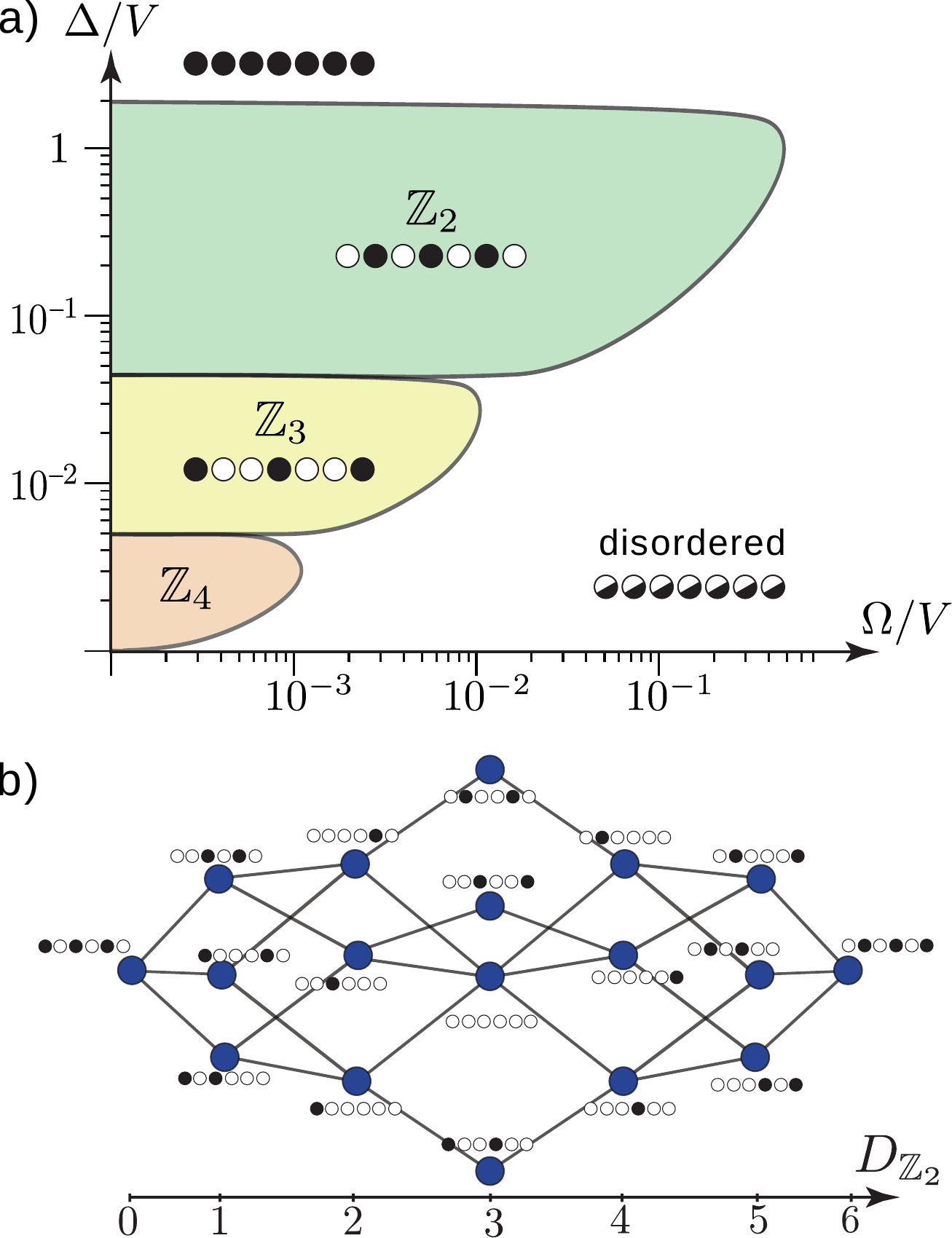}
    \caption{Quantum phases of a one-dimensional array of Rydberg qubits with Ising interactions. a) Ground state phase diagram of Eq.~\eqref{eq:ising} as a function of the transverse and longitudinal fields $\Omega, \Delta$ and the nearest neighbor interaction strength $V$ showing a series of crystalline phases. Figure is a sketch based on calculations in \textcite{rader2019floating}. b) Graph representation of the PXP model. Reprinted figure with permission from \textcite{turner2018quantum}, Phys. Rev. B, 98, 155134 (2018), Copyright 2018 by the American Physical Society.}
    \label{fig:phasediagram}
\end{figure}

For lattice geometries with period $a$ we denote the nearest neighbor interaction strength $V=-C_p/a^p$. For $|\Omega/V|\lesssim 1$ the influence of nearest and beyond nearest neighbor interactions in Eq.\eqref{eq:ising} gives rise to rich ground state phase diagram exhibiting multiple crystalline phases depending on the ratio $\Delta/V$ (Fig.~\ref{fig:phasediagram}a). In 1D these can be labelled according to their broken translational symmetry
(e.g., $\mathbb{Z}_{q=2}$ for $\ket{0101010\ldots}$, where $1/q$ is the fractional filling of the excitation crystal structure). Focusing on $\Omega\rightarrow 0$, the boundaries between phases $\mathbb{Z}_q, \mathbb{Z}_{q+1}$ occur at detunings
$$\delta_q =V\zeta(p) \left(\frac{q+1}{q^p}-\frac{q}{(q+1)^p}\right ),$$
with $\zeta(p)$ the Riemann Zeta function and $\zeta(6)\approx 1.017$.
Accordingly the ground state of the system has symmetry $\mathbb{Z}_q$ if  $\delta_q\!>\!\Delta\!\geq\! \delta_{q-1}$ and the characteristic energy scale associated with the interactions is $V/(qa)^p\approx \Delta$. \textcite{rader2019floating} performed numerical simulations of the phase structure and identified the presence of additional floating crystal phases surrounding the ordered phases. Samajdar \textit{et al.} calculated the corresponding quantum phase diagram for a two-dimensional square lattice\cite{samajdar2020complex} and the kagome lattice\cite{samajdar2020quantum}, showing intricate competing ordered phases and exotic phase transitions. Experimental protocols to prepare such ordered phases using quasi-adiabatic sweeps are discussed in Sec.~\ref{sec:demonstrations}.

Another important parameter is the blockade radius $R_b=|C_p/\Omega|^{1/p}=a|\Omega/V|^{-1/p}$, i.e., the distance over which interatomic interactions prevent the simultaneous excitation of two atoms. If $R_b\gg a$ the size of the relevant Hilbert space is dramatically reduced due to the blockade constraint which prevents Rydberg excitations on nearby lattice sites and the underlying lattice structure hardly matters. If $R_b\ll a$ then the interactions are relatively weak and the ground state of the system is generally uncorrelated. For van der Waals interactions and $R_b \sim a$, the system can be well approximated by a nearest-neighbor model due to the rapid $1/R^6$ fall-off of the interaction strength. 

\subsubsection*{PXP model (gr-qubits)}
An interesting special case of the quantum Ising model arises when $\Delta_j\approx 0$ and $a<R_b<2a$ (nearest-neighbor blockade) and $V_{j,k}\approx 0$ for everything beyond nearest neighbors. Such a situation was experimentally realized in a 1D chain of Rydberg atoms by \textcite{bernien2017probing}. In this case one can derive an effective Hamiltonian for the low-energy subspace\cite{turner2018quantum} which amounts to neglecting configurations with two adjacent excitations. In 1D the resulting Hamiltonian takes the form of a PXP model\cite{fendley2004competing}:
\begin{align}
    \hat H= \frac{1}{2}\sum_j \Omega_j(t) \hat P^0_{j-1}\hat{X}_j\hat P^0_{j+1},
\end{align}
where $P^0_j=|0 \rangle_j\langle 0|$ is a projection operator acting on site $j$. This Hamiltonian describes Rabi rotations on site $j$ under the constraint that all neighboring qubits are in the $|0\rangle$ state. It can also be thought of as a many-body version of the CU$_{xy}$ operation applied to three-qubit neighborhoods that can be represented by a quantum graph shown in Fig.~\ref{fig:phasediagram}b. The nodes of the graph correspond to classical (product state) configurations which grow in number exponentially  according to the Fibonacci sequence with the number of qubits, while the links connect those configurations under the action of the Hamiltonian. The eigenstates are found to converge toward a Wigner-Dyson level distribution with a small number of special eigenstates that have nearly equally spaced eigenvalues\cite{turner2018quantum,choi2019emergent} which gives rise to interesting non-ergodic behavior. This model and its variants have been theoretically studied in 1D\cite{lesanovsky2012liquid,lesanovsky2012interacting,turner2018weak,turner2018quantum} and more recently 2D on different lattices\cite{lin2020quantum,michailidis2020stabilizing} and found to exhibit liquid like ground states\cite{lesanovsky2012liquid}, interacting Fibonacci anyons\cite{lesanovsky2012interacting} and quantum scars (particular states that do not thermalize)\cite{turner2018weak,turner2018quantum,choi2019emergent} and synchronization effects\cite{michailidis2020stabilizing,lin2020quantum}.

\subsubsection*{Quantum XY model (rr-qubits)}
The quantum XY model is another paradigmatic quantum spin model, featuring in this case spin-exchange  ``flip-flop'' interactions. It can be mapped to a system of hardcore bosons hopping on a graph (or non-interacting fermions via the Jordan Wigner transformation). The XY model might also find applications for universal quantum simulation and non-gate based quantum computation\cite{Cubitt9497}. 

A spin-1/2 XY model can be realized using rr-qubits involving two different Rydberg states with orbital angular momentum difference $\delta\ell=1$. For illustration we consider two Rydberg states $\ket{s}=nS_{1/2}$ and $\ket{p}=nP_{1/2}$. In this case the resonant dipolar exchange interactions take the form $V_{j,k}=C_3(\theta_{j,k})/R_{j,k}^3$ and are generally much stronger than the second-order van der Waals interactions. These states can additionally be coupled by an external microwave field to realize single qubit rotations. The many-body Hamiltonian can then be written as
\begin{align}
\hat H = \sum_j \biggl [\hat H_j^{sp}(\varphi_j=0)+\frac{1}{2}\sum_{k\neq j}\hat H^{spps}_{j,k}\biggr ].
\end{align}

Identifying $\ket{s}\equiv \ket{0}$ and $\ket{p}\equiv \ket{1}$ then this can be written in the form of a quantum XY-Hamiltonian (or more precisely an XX-Hamiltonian where the XX and YY terms have equal weights):
\begin{align}
    \hat{H}&=\frac{1}{2}\sum_j\biggl [ \Omega_j(t)\hat X_j + \Delta_j(t)\hat Z_j \\\nonumber
    &\qquad+ \sum_ {k\neq j}\!\frac{V_{j,k}}{4}\bigr(\hat X_j \hat X_k
\!+\!\hat Y_j \hat Y_k\bigr)\biggr ].
\end{align}
This model has been experimentally realized in a many-body regime in disordered three-dimensional systems\cite{orioli2018relaxation,lippe2020experimental}, and in a small 1D chain of 3 atoms\cite{barredo2015coherent}. Recently this Hamiltonian was used to simulate the Su-Schrieffer-Heeger (SSH) model in a 1D chain with alternating weak and strong links which supports topological phases\cite{leseleuc2019observation}.

\subsubsection*{XXZ and other spin models (rr-qubits)}
Another class of models that can be naturally realized using rr-qubits is the XXZ model. This occurs for example for two Rydberg states with $\delta\ell=0,2$, when the van der Waals interactions are mediated via a common intermediate state with a large F\"orster defect (e.g., $\ket{r}=nS\leftrightarrow n'P\leftrightarrow (n+1)S=\ket{r'}$). In this case the two-body interactions include a combination of exchange and direct terms:
\begin{align}
\hat H &= \sum_j \biggl[\hat H_j^{rr'}(\varphi=0)\\ \nonumber
&\qquad +\frac{1}{2}\sum_{k\neq j}\hat H^{rr'r'r}_{j,k}(V_{j,k}=2J_{j,k})\\ \nonumber
&\qquad +\frac{1}{2}\sum_{k\neq j}\hat H^{rr'rr'}_{j,k}(V_{j,k}=2\delta J_{j,k})\biggr].
\end{align}
Identifying $\ket{r}\equiv\ket{0}$ and $\ket{r'}\equiv\ket{1}$ we arrive at the XXZ-Hamiltonian
\begin{align}
    \hat{H}=&\frac{1}{2}\sum_j \biggl[ \Omega_j(t)\hat X_j + (\Delta_j(t)-\mathcal{I}_j)\hat Z_j \\\nonumber
    &+ \frac{1}{2}\sum_{k\neq j} J_{j,k}\left(\hat X_j \hat X_k+\hat Y_j \hat Y_k +\delta \hat Z_j \hat Z_k\right )\biggr],
\end{align}
with $\mathcal{I}_j=\sum_{k,j\neq k} \delta J_{j,k}$
This is an anisotropic spin-1/2 XXZ model with interaction strength $J_{j,k}=-C_6/R_{j,k}^6$ and anisotropy parameter $\delta$ that both depend on the chosen Rydberg states. This model has been experimentally studied in a disordered 3D gas\cite{signoles2019observation}. It also describes the interactions between Rydberg qubits encoded in circular Rydberg states\cite{nguyen2018towards}.

Related models with additional spin non-conserving terms were proposed for Rydberg qubits involving two sublevels of an $nP$ Rydberg state\cite{glaetzle2015designing,whitlock2017simulating}, and experimentally and theoretically studied in the context of spin-orbit coupled Rydberg systems\cite{weber2018topologically,lienhard2020realization}. 

\section{Experimental demonstrations}\label{sec:demonstrations}
Recent experimental demonstrations of quantum logic operations on small collections of qubits, or the dynamics of larger quantum systems with limited control over individual qubits, have shown the potential of Rydberg qubits as a competitive technology for programmable quantum simulation and quantum computing. In this section we give an overview of a few key examples focusing on the last $5$ years, which we can loosely categorize according to: high-fidelity entangling operations; gate implementations and elementary quantum circuits; and many-body quantum state engineering. We distinguish entangling operations and gate implementations depending on whether the corresponding operation is verified for a single initial state or applied to the entire computational state space.

\subsection{High-fidelity entangling operations}\label{sec:entanglingops}

The first experiments to realize entangling operations for a pair of Rydberg qubits were performed at the University of Wisconsin\cite{urban2009observation} (gg-qubits) and at the Institut d'Optique\cite{gaetan2009observation} (gr-qubits), with both sets of results published in 2009. In the Wisconsin experiments they implemented a Rydberg blockade CZ operation with individual addressing for control and target qubits. At the Institut d'Optique they observed the collective Rabi oscillations (parallel CU$_{xy}$) that signal the generation of a highly entangled Bell state, e.g., $\ket{\Psi^+}=(\ket{01}+\ket{10})/\sqrt{2}$. 

An established way to characterize the performance of an entangling operation is via the Bell state entanglement fidelity $F_\textrm{Bell}=\braket{\Psi^+|\rho|\Psi^+}$ (for a two qubit state described by the density matrix $\rho$) measured by parity oscillations\cite{turchette1998deterministic,sackett2000experimental}. This involves the application of a global U$_{xy}(\theta=\Omega\tau_g,\varphi=0)$ rotation to both qubits after the entangling operation and then reading out the parity signal $\Pi(\theta)=P_{00}(\theta)+P_{11}(\theta)-P_{01}(\theta)-P_{10}(\theta)$ (where $P_{\alpha\beta}(\theta)$ is the population in the state $\ket{\alpha\beta}$ after application of the analysis pulse). For the maximally entangled state $\ket{\Psi^+}$ the parity should oscillate between $\pm 1$ as a function of $\theta$ with a period of $\pi$, while in practice the amplitude of these oscillations will be $|C|<1$\cite{maller2015rydberg,wilk2010entanglement,gaetan2010analysis,picken2018entanglement}.

An estimate for $F_\textrm{Bell}$ close to the Bell state $\ket{\Psi^+}$ can be calculated from the amplitude of the parity oscillations as $F_\textrm{Bell}= (P_{00}(0)+P_{11}(0))/2+|C|$ and if $F_\textrm{Bell}>0.5$ the state is entangled. \textcite{wilk2010entanglement} detailed a method to correct for loss of qubits during measurement. 

\renewcommand{\arraystretch}{1.2}
\begin{table*}[!ht]
\begin{tabular}{llllll}\hline
{\textbf{Year, reference}}  & 
{\textbf{Qubit\qquad}} & 
{\textbf{Operation\qquad}} & 
{\textbf{Fidelity\qquad}} & 
{$\bm{T_2^*}$\qquad\quad} & 
{$\bm{\tau_g}$} \\
\hline\hline

2016, \textcite{jau2016entangling} & ${}^{133}$Cs (gg) &  pCU$_{xy}(\pi)$ & $\geq 0.81(2) {}^{\text{i,iii}}$ &  & 2$\,\mu$s \\

2017, \textcite{zeng2017entangling} & ${}^{87\!-\!85}$Rb (gg) & \begin{tabular}[t]{@{}l@{}} Heteronuclear\\ CZ $\rightarrow$ CNOT \end{tabular} &0.73(1)${}^{\text{ii}}$ &   & 6$\,\mu$s \\

2018, \textcite{levine2018highfidelity} & $^{87}$Rb (gr) & pCU$_{xy}(\pi)$  & $\geq$ 0.97(3)${}^{\text{i,iii}}$   & $4.5(1)\,\mu$s & 177\,ns  \\

2018, \textcite{picken2018entanglement} &${}^{133}$Cs (gg) & pCU$_{xy}(\pi)$ & $\geq 0.81(5){}^{\text{i,iii}}$ & 10(2)\,ms & 1.85$\,\mu$s  \\

2019, \textcite{graham2019rydberg} &${}^{133}$Cs (gg)  & CZ $\rightarrow$ CNOT & 0.89${}^{\text{i,iii}}$ & $\leq 1.6\,$ms & 1.12$\,\mu$s  \\ 

2019, \textcite{levine2019parallel} &{${}^{87}$Rb (gg)} & \begin{tabular}[t]{@{}l@{}}pCZ \\ pCZ $\rightarrow$ CNOT\\
C$_2$Z $\rightarrow$ Toffoli \end{tabular}
& 
\begin{tabular}[t]{@{}l@{}}
$\geq 0.974(3)^{\text{i,iii}}$ \\ 
$\geq 0.965(3){}^{\text{i,ii}}$\\
$\geq 0.870(4){}^{\text{i,ii}}$ \end{tabular}
& 
\begin{tabular}[t]{@{}l@{}}
  \\ 
\\
 \end{tabular}
& 
\begin{tabular}[t]{@{}l@{}}
$0.4\,\mu$s \\ 
\\
$1.2\,\mu$s \end{tabular}
\\

2020, \textcite{zhang2020submicrosecond} & ${}^{88}$Sr${}^{\!+}$ (gg)  & CPHASE$\rightarrow$ CZ & 0.78(3)${}^{\text{iii}}$ &   & 700\,ns \\

2020, \textcite{madjarov2020high} & $^{88}$Sr (gr)   & pCU$_{xy}(\pi)$  & $\geq$ 0.991(4)${}^{\text{i,iii}}$  & $\approx 2\,\mu\textrm{s}$ & 51\,ns \\ \hline

\end{tabular}
\caption{Reported quantum operations and quantum gates realized with Rydberg qubits and performance parameters. $T_2^*$ refers to the qubit coherence time measured via Ramsey interferometry without spin echo pulses (coherence times with echo pulses are typically an order of magnitude longer). $\tau_g$ is the operation time. For results prior to 2016, see the review by  \textcite{saffman2016quantum}. $^{\text{i}}$Fidelity corrected for SPAM errors, $^{\text{ii}}$gate fidelity and $^{\text{iii}}$entanglement (Bell) fidelity. For reference, a fidelity of $F=0.99$ corresponds to an estimated achievable circuit depth of $\Dsquare=10$. }\label{tab:qubitgates}
\end{table*}

In 2010, the Wisconsin and the Institut d'Optique groups both demonstrated entanglement between Rydberg qubits with (atom loss corrected) fidelities of $F=0.58(7), F=0.71(5)$  (Wisconsin\cite{isenhower2010demonstration,zhang2010deterministic}) and $F=0.75(7)$ (Institut d'Optique\cite{wilk2010entanglement}). In 2016, \textcite{jau2016entangling} demonstrated a Bell state entanglement fidelity of $F_\textrm{Bell}\geq 0.81(2)$ using a pair of gg-qubits using a parallel CU$_{xy}$ operation directly on the ground state with Rydberg dressed interactions\cite{jau2016entangling}. Since these first experiments a number of groups have demonstrated higher fidelity entangling operations between Rydberg qubits. These demonstrations have made use of a variety of different types of atoms, qubit encodings and control methods, pushing the fidelities of entangling operations to levels that compare well with other types of qubits (see Table \ref{tab:qubitgates}).

The first demonstration of entangling operations involving Rydberg qubits with fidelities significantly above $0.90$ was in 2018, by \textcite{levine2018highfidelity}. They demonstrated high fidelity manipulation and Bell state entanglement of gr-qubits with a fidelity of $0.97(3)$, after correcting for detection errors. Their system consisted of a chain of $^{87}$Rb gr-qubits initialized in the state $\ket{0}=\ket{5S_{1/2},F=2,m_F=-2}$. Coherent coupling to the  $\ket{1}=\ket{70S,J=1/2,m_J=-1/2}$ state was realized with a global two-photon coherent laser excitation via the intermediate $6P_{3/2}$ state similar to that shown in Fig.~\ref{fig:rabioscillationslevine}a. A key improvement made in this experiment was to suppress laser amplitude and phase noise on the Rydberg excitation lasers, the latter by using the transmission through an optical cavity as a spectral filter to suppress residual servo bumps. To entangle two qubits they positioned them with a separation of $5.7\mu$m (within the blockade radius) and then drove collective Rabi oscillations which implemented a pCU$_{xy}(\pi)$ operation: $\ket{00}\rightarrow \ket{\Psi^{+}}$. To determine $F_\textrm{Bell}$ they measured the off-diagonal elements of the density matrix by applying a local phase shift ($\textrm{R}_z(\theta)$ operation) to one of the qubits using a focused off-resonant addressing laser and embedded the whole sequence in a spin echo protocol to partially cancel random Doppler shifts. The time to implement a pCU$_{xy}(\pi)$ operation was $\tau_g=177\,$ns which can be compared with the measured dephasing timescale of $T_2^{*}=4.5(1)\,\mu$s (without spin echo pulses) and $T_2=32(6)\,\mu$s (including spin echo). 

Around a similar time, \textcite{picken2018entanglement} reported Bell state entanglement between a pair of gg-qubits formed by two magnetically insensitive hyperfine ground states of $^{133}$Cs atoms. The measured entanglement fidelity was $F_\textrm{Bell}=0.81(5)$ after correction for particle loss (limited by phase noise on the Rydberg excitation laser). Their protocol consisted of a collective Rabi oscillation from $\ket{11}\rightarrow (\ket{1r}+\ket{r1})/\sqrt{2}$ which was subsequently mapped to $\ket{\Psi^+}$ in the ground state computational subspace via a $\pi$-pulse on the $\ket{r}\rightarrow\ket{0}$ transition. The measured coherence time of the ground-state encoded qubit states of $T_2^*=10(1)\,$ms is much longer than the gate time of $\tau_g=1.85\,\mu$s, which in principle would allow thousands of quantum operations assuming the fidelities can be further improved.

\begin{figure}[!t]
    \centering
    \includegraphics[width=0.95\columnwidth]{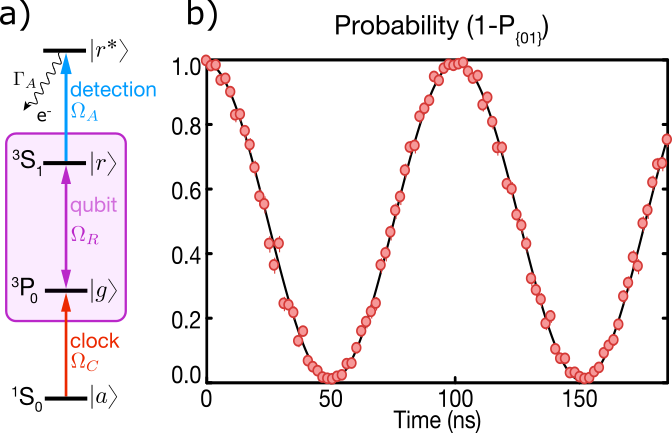}
    \caption{High-fidelity manipulation and entanglement of $^{88}$Sr Rydberg qubits. a) Level scheme used to prepare, encode and detect the qubits. b) High-fidelity collective Rabi oscillations for pairs of gr-qubits that realizes a pCU$_{xy}(\theta)$ entangling operation. A two qubit Bell state $(\ket{gr}+\ket{rg})/\sqrt{2}$ is realized by a $\pi$ Rabi rotation corresponding to a $51\,$ns gate time. Reprinted by permission from Springer Nature: \textcite{madjarov2020high}, Nat. Phys. 16, 857-861 (2020), Copyright 2020.}
    \label{fig:rabioscillationsendres}
\end{figure}

In 2020, \textcite{madjarov2020high} demonstrated the highest Bell state fidelities to date of $F>0.991(4)$, as well as comparably high-fidelity single qubit rotations. Their system consists of a dimerized 1D array of $^{88}$Sr alkaline-earth atoms for which the Rydberg blockade is effective between nearest pairs. The qubit is encoded in the metastable $5s5p ^3P_0\equiv \ket{0}$ state and the $5s61s ^3S_1 \equiv \ket{1}$ state (gr-qubit) shown in Fig.~\ref{fig:rabioscillationsendres}a. Two-qubit entangling operations are realized using pCU$_{xy}(\theta)$ operations. A lower bound for the Bell state fidelity was obtained from the measured populations after a pCU$_{xy}(\pi)$ operation (Fig.~\ref{fig:rabioscillationsendres}b) and an estimate of the purity of the resulting two-qubit state. This work represents a first step towards combining high-fidelity Rydberg quantum gates with ultra-coherent optical clock qubits. Furthermore the very high-fidelity for these operations combined with the fast gate times of $\tau_g=51\,$ns, is very encouraging for the realization of deep quantum circuits with many qubits ($\Dsquare \gtrsim 10$).

\subsection{Gate implementations and elementary quantum circuits}

The first demonstration and characterization of a Rydberg quantum gate was a CNOT gate in 2010 by the Wisconsin group\cite{isenhower2010demonstration}, using a pair of gg-qubits encoded in the hyperfine ground states of $^{133}$Cs atoms. The use of CNOT as opposed to native CZ gates has the advantage that the truth table can be verified by population measurements alone. This requires additional pulses to prepare qubits in each of the computational states and population measurements of each output state to obtain the process matrix $U_{exp}$. The gate fidelity can be estimated by $F=\mathrm{Tr}[|U_{ideal}
^T|U_{exp}]/4$. To further verify that the gate preserves (two-qubit) coherence, one can apply single qubit R$_x(\pi/2)$ rotations on the control qubit before the CNOT gate. This generates Bell states which can then be characterized via parity oscillations.

The first heteronuclear Rydberg quantum gate was demonstrated by \textcite{zeng2017entangling} in 2017. The system consisted of one $^{87}$Rb atom and one $^{85}$Rb atom confined in two optical traps separated by $3.8\,\mu$m and each prepared in a combination of two magnetically insensitive hyperfine ground states (gg-qubits). They implemented a CNOT operation using the Rydberg blockade protocol shown in Fig.~\ref{fig:heteronuclearCNOT}. The raw gate fidelity was $F=0.73(1)$ (including state preparation errors) and the Bell state entanglement fidelity was $F_\textrm{Bell}=0.59(3)$, limited mainly by power fluctuations of the Rydberg excitation lasers and Doppler dephasing. This shows that Rydberg gates acting on two species can in principle perform as well as single species demonstrations, thus opening the possibility for architectures that rely on two different species, e.g., serving as computational and ancilla qubits.

Building on the first experiments to demonstrate Rydberg excitation of trapped ions\cite{feldker2015rydberg,bachor2016addressing,higgins2017coherent,higgins2017single}, \textcite{zhang2020submicrosecond} recently experimentally demonstrated a quantum gate between trapped ions mediated by strong Rydberg-Rydberg interactions, speeding up quantum operations by several orders of magnitude. For this they realized a CPHASE gate acting on two gg-qubits encoded in the ground and metastable $4D_{5/2}$ states of $^{88}$Sr$^+$ ions in a linear Paul trap. The second-order van der Waals interaction between Rydberg ions is relatively weak, so they induced a strong first order interaction using a microwave dressing field between the $46S_{1/2}$ and $46P_{1/2}$ Rydberg states. The gate sequence is shown in Fig.~\ref{fig:zhang2020}. It started with the initial state $(\ket{00}+\ket{01}+\ket{10}+\ket{11})/2$ followed by a two-photon double stimulated Raman adiabatic passage (STIRAP) pulse sequence to transfer both ions to the dressed Rydberg state $\ket{r}$, and back. Consequently the state $\ket{00}$ acquires a phase $\Phi_{00}=V\int_0^T\braket{rr|\rho(t)|rr}\approx \pi$ for a specific choice of the final time $T$. They characterize the gate by rotating the phase shifted two-qubit state to a Bell state and then performed parity oscillations. This yields $F_\textrm{Bell}=0.78(3)$, limited mainly by microwave power fluctuations and the finite coherence of the Rydberg laser coupling. The speed of this Rydberg gate and insensitivity to ionic motion opens a promising new route to implement multi-qubit gates in large ion crystals without the need for phonon-mediated interactions.

Most gate demonstrations to date have involved isolated pairs of qubits or global addressing pulses. However to scale up quantum processors it is important to show good gate performance in larger qubit arrays, i.e., using tightly focused addressing beams to minimize crosstalk. This was demonstrated in 2019, by \textcite{graham2019rydberg} using a 2D array of 121 sites with an average filling of $0.55$, generated by diffractive optical elements and blue-detuned light. Qubits were encoded in two hyperfine ground states of $^{133}$Cs atoms (gg-qubits). The two-photon Rydberg excitation lasers were focused to waists of $3\,\mu$m and pointed at individual sites using two crossed AODs per laser. Combined with local AC Stark shifted microwave pulses\cite{xia2015randomized} they demonstrate site specific single and two-qubit operations. Starting from the input state $(\ket{00}+\ket{01}+\ket{10}+\ket{11})/2$ they apply a Rydberg blockade CZ gate, convert it to a CNOT gate, and characterize the Bell state fidelity $F_\mathrm{Bell}=0.86(2)$ (0.89 correcting for single qubit and state preparation and measurement SPAM errors). The effect of crosstalk on neighboring sites was estimated to be $<4\%$, with the dominant errors associated to the finite temperature and atomic position variations in the traps.

In 2019, \textcite{levine2019parallel} demonstrated high-fidelity multi-qubit gates in a 1D array of $^{87}$Rb atoms serving as gg-qubits encoded in two magnetically insensitive hyperfine ground states. Ground-ground and ground-Rydberg transitions were driven by coherent laser fields as shown in Fig.~\ref{fig:rabioscillationslevine}. A major innovation in this experiment was the design and implementation of a new protocol for realizing controlled-Z gates that do not require individual addressing (introduced as the pCZ gate in Sec.~\ref{sec:interactions}). They embedded this gate in a sequence which produces a Bell state and characterized its performance via parity oscillations. The resulting raw fidelity was $0.959(2)$ and $>0.974(3)$ after correcting for SPAM errors. They separately characterized the gate by converting it to a CNOT and measuring the truth table fidelity $F>0.965(3)$ (SPAM corrected). 

In the same work, \textcite{levine2019parallel} demonstrated and characterised the first three-qubit Rydberg gates (C$_2$Z and Toffoli). To implement these gates they generated a trimerized 1D array of qubits as shown in Fig.~\ref{fig:toffoli}a. The nearest neighbors were strongly blockaded while the two outer atoms of each trimer were not, thus providing the required asymmetric interactions. Compared to the two-qubit pCZ protocol, the three qubit C$_2$Z pulse sequence is relatively complicated. Therefore the sequence was optimized using the remote version of the dCRAB optimal control algorithm (RedCRAB)\cite{rach2015dressing,heck2018remote}. To characterize the gate they apply a local Hadamard gate to the central qubits before and after the C$_2Z$ which converts it to a Toffoli gate. The measured truth table fidelity was $F\geq 0.870(4)$ (SPAM corrected), with comparable results obtained via limited process tomography. 
These results compare quite well with Toffoli gate implementations with trapped ions ($F=0.71$\cite{monz2009realization} and $F=0.896$\cite{figgatt2017complete}) and superconducting circuits ($F=0.685$\cite{fedorov2012implementation} and $F=0.78$\cite{reed2012realization}). It also opens the possibility of realizing more efficient quantum circuits that exploit highly optimized multi-qubit gates without the need for fully independent control over all qubits. 

\begin{figure}[!t]
    \centering
    \includegraphics[width=\columnwidth]{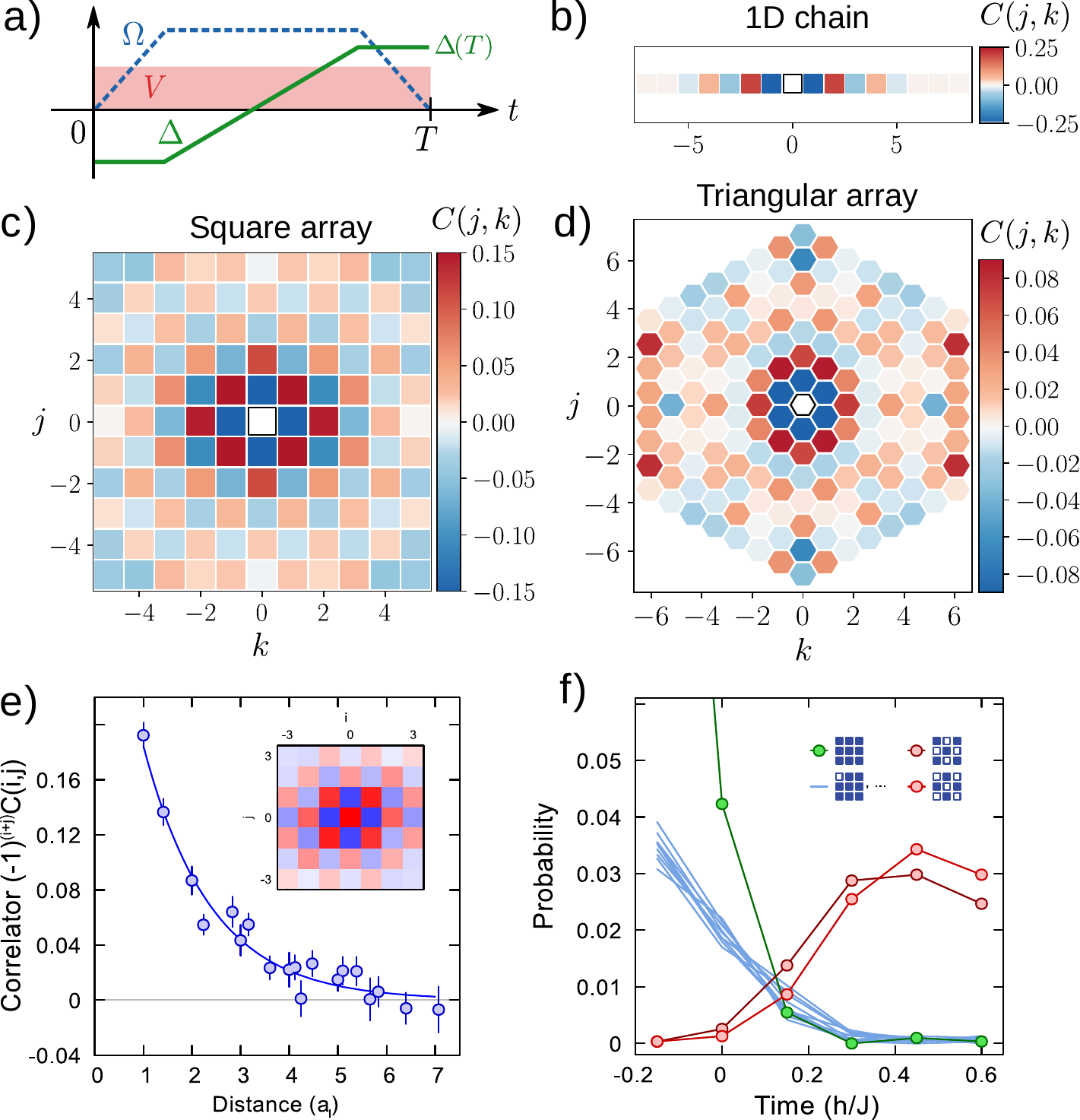}
    \caption{Build up of magnetic correlations in qubit arrays with different spatial configurations. a) Protocol for realizing antiferromagnetically ordered states for the Ising model by ramping the external control parameters $\Omega,\Delta$. b-d) Experimentally measured spin-spin correlation functions for three different configurations. Reproduced from \textcite{lienhard2018observing}, Phys. Rev. X, 8, 021070 (2018), licensed under CC-BY 4.0. e) Correlation length measurement for a square array and f) probability for finding different spin configurations in $3\times 3$ sublattices. Reproduced from \textcite{guardado2018probing}, Phys. Rev. X, 8, 021069 (2018), licensed under CC-BY 4.0. }
    \label{fig:afm}
\end{figure}

\subsection{Many-body quantum state engineering}\label{sec:demonstrations_qsim}

Due to their strong state-dependent interactions (Sec.~\ref{sec:interactions}), Rydberg qubits are naturally suited for the simulation of quantum spin models. To date, experimental demonstrations have been analog quantum simulations focusing on low energy states of the model and dynamics following quantum quenches, usually involving different spatial configurations and time-dependent (quasi-homogeneous) coupling fields ($\Delta_j(t)\approx \Delta(t)$, $\Omega_j(t)\approx \Omega(t)$). A protocol for producing strongly-correlated states via the Ising model\cite{pohl2010dynamical,Schachenmayer_2010,vanBijnen2011adiabatic,Vermersch_2015,schauss2015crystallization,petrosyan2016on} (Eq.~\eqref{eq:ising}) is sketched in Fig.~\ref{fig:afm}a. It starts with $\Omega=0,\Delta/V\ll -1$ and the initial state $\ket{0000\ldots}$, which for $V>0$ coincides with the ground state of the system. The coupling $\Omega$ is then slowly ramped up, followed by a sweep of the detuning to $\Delta/V \gg 1$ which couples states with different numbers of Rydberg excitations. Finally $\Omega$ is ramped back down to zero. For sufficiently slow ramps the system should adiabatically follow the ground state of the system at all times, ending in the strongly-correlated ground state (such as those depicted in Fig.~\ref{fig:phasediagram}a for $\Omega\rightarrow 0,\Delta/V>0$). 

In practice the speed of the ramp and strength of the resulting correlations are limited by Rydberg state decay and dephasing of the ground-Rydberg transition. However the characteristic coherence time $T_\text{coh}$ can be difficult to quantify in many-body settings. Instead, commonly used quantities to characterize coherence in current experiments are the Ramsey coherence time $T_2$ or the $1/e$ damping time of single-atom Rabi oscillations $T_\mathrm{Rabi}$ measured on widely separated qubits (the latter can be related to the $\pi-$pulse infidelity $\epsilon_{\pi}\approx \pi/(2\Omega T_\mathrm{Rabi})$ assuming exponential damping). These quantities can be loosely related by $T_2=2/\gamma \approx T_\mathrm{Rabi}/2$ assuming the system is modelled by a quantum master equation in Lindblad form that includes irreversible decay of the off-diagonal elements of the density matrix with rate $\gamma/2$ (pure dephasing). However we note that in general the relevant coherence time could be very different from $T_2$ or $T_\mathrm{Rabi}$ depending on the model being simulated, the noise spectrum\cite{graham2019rydberg} and many-body effects, e.g., mechanical forces between the atoms.

The first experiments to explore the quantum Ising model with individually resolved Rydberg qubits were by \textcite{schauss2012observation} in 2012 and again in 2015 under even more controlled conditions\cite{schauss2015crystallization}. The system consisted of a 2D array of $^{87}$Rb atoms prepared in an optical lattice. The system was driven from the ground state to a Rydberg state by a two-photon coupling, realizing the Ising-like Hamiltonian of Eq.~\eqref{eq:ising} with global couplings. In these experiments the blockade radius was much larger than the lattice spacing ($R_b/a\gg 1$) such that the effective Hilbert space was strongly constrained. The relevant interaction strength is therefore given by the final detuning $V_\text{max}\approx \Delta(T)\lesssim 2\pi\times  (800\,\textrm{kHz}$). This can be compared with the measured Rabi damping time of $T_\mathrm{Rabi}\approx 3\,\mu$s\cite{ediss18152}. Although this corresponds to a relatively small effective circuit depth, by sweeping $\Omega(t)$ and $\Delta(t)$ along a predefined trajectory similar to Fig.~\ref{fig:afm}a, it was possible to observe small spatially correlated excitation clusters expected for the low energy states of this model. 

In 2017, \textcite{bernien2017probing} realized the same type of Hamiltonian with tunable interactions and system sizes of up to 51 Rydberg-ground qubits. The system consisted of a deterministically prepared 1D array of $^{87}$Rb atoms with variable spacings realized by a dynamically adjustable array of optical tweezers\cite{endres2016atom}. This made it possible to vary $R_b/a$ from $\sim 0.4$ (weakly interacting) to $3.1$ (strongly blockaded). For $2a>R_b\geq a$ the system is described by an effective PXP model. Therefore the maximum interaction strength is given by $\Omega/2\pi\lesssim 2\,$MHz. This compared favorably with the observed damping time for single-atom Rabi oscillations of $T_\mathrm{Rabi}=6\,\mu$s.

By performing slow detuning sweeps from $\Delta<0$ to different final values of $\Delta>0$ they could prepare the system in different ordered phases. The ground state preparation fidelity for the $\mathbb{Z}_2$ crystalline ground state was $0.77(6)$ for $N=7$ and $0.009(2)$ for $N=51$. Although this appears low, the crystalline ground state is the most probable final state in view of the effective Hilbert space dimension, which scales approximately with $([1+\sqrt{5}]/2)^N$ accounting for the Rydberg blockade constraint (Fig.~\ref{fig:phasediagram}b). This extends earlier results in trapped ion chains for 14 qubits\cite{richerme2013experimental} with improved scaling of the fidelity with system size. This confirmed that classical ground states of relatively-large, programmable, Ising-like quantum systems can be prepared with statistical significance, promising for e.g., quantum annealing protocols\cite{glaetzle2017coherent} and quantum approximate optimization algorithms\cite{farhi2016quantum} (discussed in Sec.~\ref{sec:proposals}). 

In a final experiment, they performed a quench from a crystalline state to the disordered phase and observed unusually long-lived oscillations in the many-body quantum dynamics. These oscillations were later associated with weak ergodicity breaking and many-body scars\cite{turner2018weak}. In subsequent experiments studied the quantum Kibble-Zurek mechanism and the growth of correlations by sweeping through the quantum phase transition\cite{keesling2019quantum} and stabilization of revivals by periodic driving in different 2D geometries\cite{bluvstein2020controlling}.

Also in 2017, \textcite{zeiher2017coherent} realized an Ising model in a 10 site 1D array of $^{87}$Rb Rydberg-dressed atoms in an optical lattice with an average filling of $0.87$. Qubits were prepared in a superposition of two hyperfine ground states (gg-qubits) where one state was continuously and off-resonantly coupled to a Rydberg state in order to mediate the interactions. In this case the Ising interaction $V_{j,k}$ takes a soft-core shape\cite{zeiher2016many} that extends over several qubits, with $V_\text{max}/2\pi=13\,$kHz. They inferred the dominant decoherence mechanism to be population decay of the Rydberg-dressed state with a time constant of $1.21(3)\,$ms. In this way it was possible to observe long-lived coherent dynamics and partial revivals for $\approx 12$ interaction cycles, comparable to the state-of-the-art at that time for implementing spin Hamiltonians in trapped-ion chains\cite{zhang2017observation,zhang2017observation53}. In 2020, \textcite{guardado2020quench} realized quantum simulations of a fermionic lattice model with nearest-neighbor Rydberg-dressed interactions and hopping.

In 2018, \textcite{lienhard2018observing} and \textcite{guardado2018probing} performed studies of the quantum Ising model in two-dimensional lattices. Both experiments implemented ramps of $\Delta(t)$ and $\Omega(t)$ in order to bring the system into antiferromagnetically ordered states, evidenced by the growth of magnetic correlations quantified by the spin-spin correlation function
$$C(j,k)=\langle \hat Z_j\hat Z_k\rangle - \langle \hat Z_j\rangle \langle \hat Z_k\rangle.$$

\textcite{guardado2018probing} prepared highly excited states of the ferromagnetic Ising model with attractive interactions, which is equivalent to the low-energy states of the antiferromagnetic Ising model. Their system consisted of an array of $^{6}$Li atoms in an optical lattice, focusing on an annular region of $\sim 150$ sites with approximately 96\% occupancy. The atoms are driven from the electronic ground state to the $23P$ Rydberg state using a single-photon ultraviolet laser coupling. The nearest-neighbor interaction strength was $V_\mathrm{max}/2\pi=-6.0\,\textrm{MHz}$ and the Rabi oscillation decay time was $T_\mathrm{Rabi}=1.5\,\mu$s. Using a sudden quench from a paramagnetic state they probed the growth of antiferromagnetic correlations with a maximum range of 1.9 sites using a near adiabatic sweeps (shown in Fig.~\ref{fig:afm}e). They also observe the growth of correlations in the probability to end up in different configurations of small $3\times 3$ sub-arrays, with a combined probability of $\approx 0.06$ to end up in one of the two antiferromagnetically ordered configurations (Fig.~\ref{fig:afm}f). For short-time quench dynamics they obtain good agreement with numerical simulation up to about half an interaction cycle without taking into account decoherence, while longer times showed the influence of decoherence beyond single-particle decay and dephasing.

\textcite{lienhard2018observing}, realized up to 36 gr-qubits in 1D chains and 2D square and triangular lattice geometries using atom assembly\cite{barredo2016atom}. The qubits were encoded in the $5S_{1/2}$ electronic ground state and the $64D_{3/2}$ states of $^{87}$Rb atoms and coupled by a global two-photon laser field. The lattice spacing was chosen such that $R_b\approx a$ (effective nearest-neighbor model) with effectively isotropic interactions $V_\text{max}/2\pi=2.7\,$MHz. The damping time for single-atom Rabi oscillations was $T_\textrm{Rabi}\approx 1.2\,\mu$s, attributed to Doppler dephasing, spontaneous emission from the intermediate state and laser phase noise, among other smaller effects\cite{leseleuc2018analysis}. By performing controlled parameter ramps shown in Fig.~\ref{fig:afm}a, they could measure the growth and spreading of antiferromagnetic correlations for different array configurations shown in Fig.~\ref{fig:afm}b-d. Antiferromagnetic correlations could be observed over the whole lattice, but with a correlation length of around $1.4$ sites. In 2020 similar experiments were performed in 2D arrays with more than 100 qubits in different geometries\cite{scholl2020programmable,ebadi2020quantum}. The use of improved Rydberg excitation lasers made it possible to increase the coherence length to $7$ sites\cite{scholl2020programmable} and $>11$ sites respectively\cite{ebadi2020quantum}, demonstrating controlled (adiabatic) quantum evolution for systems reaching the limits of state-of-the-art numerical methods.

\begin{figure}[!t]
    \centering
    \includegraphics[width=0.5\textwidth]{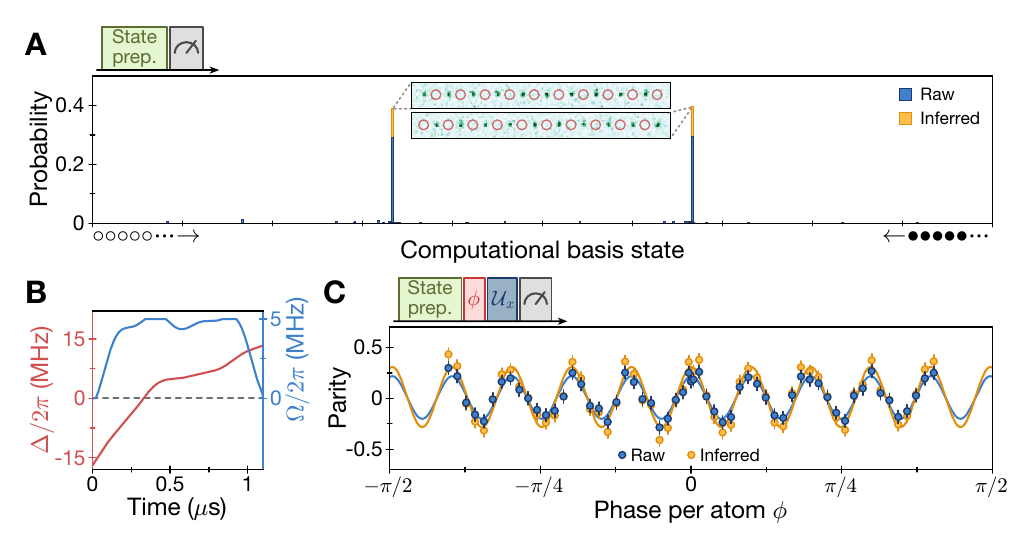}
    \caption{Preparation and characterization of a 20-atom GHZ state. a) Probability of observing different patterns, showing a large probability for the two antiferromagnetically ordered configurations (shown in the inset). b) Optimal control pulse used to generate the GHZ state. c) Parity oscillations produced by acquiring a relative phase between the GHZ components which indicates a lower bound on the 20-atom GHZ fidelity of $F\geq 0.54(2)$. From \textcite{omran2019generation}, Science, 365, 570-574 (2019). Reprinted with permission from AAAS.}
    \label{fig:omran2019}
\end{figure}

In 2019, \textcite{omran2019generation} demonstrated the generation and manipulation of multiparticle entangled Greenberger-Horne-Zeilinger (GHZ) states of up to 20 Rydberg qubits. The system consisted of a 1D array of gr-qubits (similar to \textcite{bernien2017probing}) which realizes the PXP model, with additional addressing beams to introduce local energy shifts on specific sites. The effective interaction strength was $\Omega/2\pi=5\,\textrm{MHz}$, the measured Rabi $\pi-$pulse infidelity was $0.006(3)$ and they estimate fluctuating Doppler shifts of order $\sim 43\,\textrm{kHz}$. These parameters are compatible with $\Dsquare\approx 7-9$ which, along with Refs.~\cite{scholl2020programmable,bluvstein2020controlling,ebadi2020quantum}, is among the highest values for a many-body interacting array of Rydberg qubits to date. They applied sweeps experimentally optimized by the RedCRAB optimal control algorithm via a cloud server\cite{rach2015dressing,heck2018remote} (the sweep for $N=20$ is shown in Fig.~\ref{fig:omran2019}b). This brings the system from the initial state $\ket{000\dots}$ to a target $N$-particle GHZ state:  $$\ket{\textrm{GHZ}^N}=\frac{1}{\sqrt{2}}\left ( \ket{0101\ldots} + \ket{1010\ldots}\right ),$$
within just $5-6$ interaction cycles. 

For $N=20$, the measured probability to observe each of the $2^{20}=1,048,576$ possible states of the computational basis shows two large peaks corresponding to the perfectly antiferromagnetically ordered configurations as shown in Fig.~\ref{fig:omran2019}a. To confirm that the system is in the entangled GHZ state they measure the GHZ fidelity using parity oscillations (similar to the two qubit protocol introduced in Sec.~\ref{sec:entanglingops}) induced by applying a $(-1)^jZ_j$ rotation to each site (using local light shifts) to control the phase of the GHZ state (Fig.~\ref{fig:omran2019}c). From this they extract a lower bound for the 20 qubit GHZ state fidelity of $F\geq 0.54(2)$. This clearly demonstrates that  Rydberg atom arrays constitute a competitive platform for quantum state engineering and for the realization of quantum circuits supporting many-particle entangled states.

\begin{figure}[!t]
    \centering
    \includegraphics[width=0.95\columnwidth]{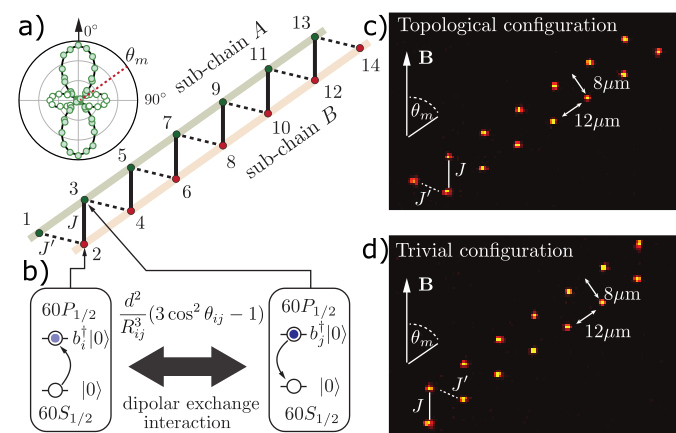}
    \caption{Experimental realization of a bosonic Su-Schrieffer-Heeger model with dipolar interacting Rydberg qubits. a) system of 14 qubits in a staggered chain. The chain is tilted by the angle $\theta_m=54.7^\circ$ to cancel couplings between qubits in the same sublattice (shown by the inset). b) Qubit states and relevant interactions. c) and d) Fluorescence images of the atoms prepared in two configurations which in c) gives rise to a symmetry protected topological phase. From \textcite{leseleuc2019observation} Science, 365, 775-780 (2019). Reprinted with permission from AAAS.}
    \label{fig:léséleuc2019}
\end{figure}

Also in 2019, \textcite{leseleuc2019observation}, implemented the XY model in a tailored geometry and realized a symmetry protected topological phase of Rydberg rr-qubits. The system consisted of a staggered chain of 14 sites coupled by an angular dependent exchange interaction shown in Fig.~\ref{fig:léséleuc2019}. This realizes an effective 1D chain with alternating weak and strong links along which an excitation can hop. This maps onto the Su-Schrieffer-Heeger model for hardcore bosons (originally formulated for fermionic particles hopping on a dimerized lattice as a model for long-chain polymers)\cite{su1979solitons}. The maximum interaction strength was $V_\text{max}/2\pi=2.42\,\textrm{MHz}$ (strong links) and they observed coherent dynamics up to $\approx 6\,\mu$s. Using an additional microwave field they could study the single particle energy spectrum of this model and prepare the many-body ground state at half-filling using a quasi adiabatic sweep, showing that it exhibits robustness with respect to perturbations characteristic of a symmetry protected topological phase. This shows that robust topological states can be prepared with Rydberg qubits exploiting tailored anisotropic interactions.

\section{Toward more programmable quantum simulations and quantum computations}\label{sec:proposals}

\begin{figure*}[!t]
  \includegraphics[width=1\textwidth]{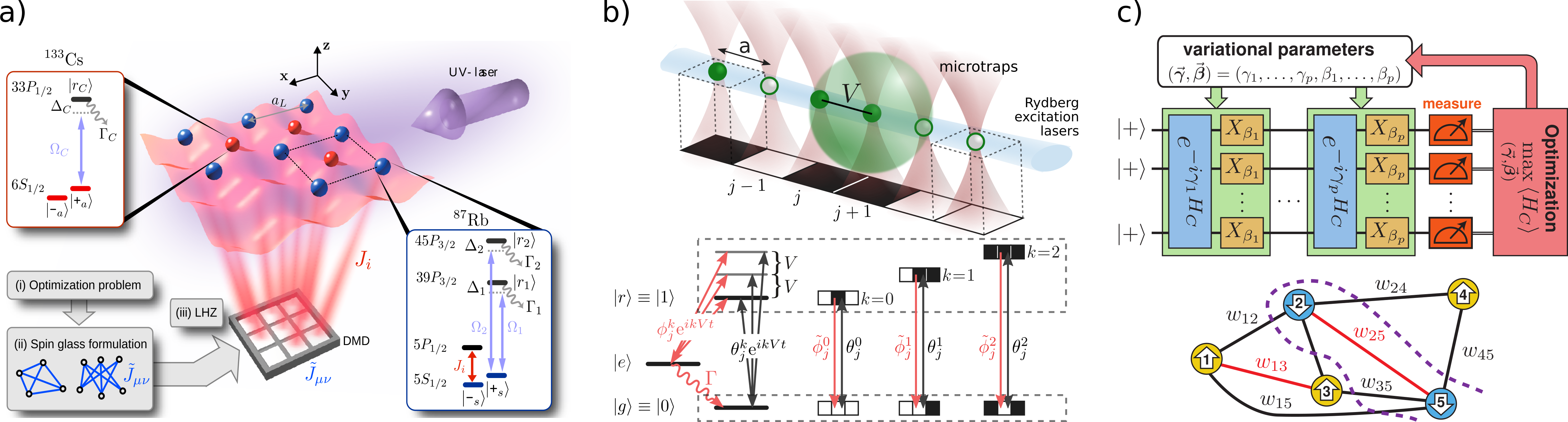}
\caption{Physically inspired models for quantum computing with Rydberg qubits. a) Sketch of a proposed setup for quantum annealing using an array of physical and ancilla qubits that realize a Ising spin-glass Hamiltonian based on the LHZ architecture. Reproduced from \textcite{glaetzle2017coherent}, Nat Commun, 8, 15813 (2017), licensed under CC-BY 4.0. b) Approach for realizing quantum cellular automata with Rydberg qubits. A multifrequency Rydberg excitation laser can be used to realize a set of programmable unitary and non-unitary dynamical rules. Reprinted figure with permission from \textcite{wintermantel2020unitary}, Phys. Rev. Lett., 124, 070503 (2020), Copyright 2020 by the American Physical Society. c) Proposed implementation of the quantum approximate optimization algorithm for solving computationally hard optimization problems such as MaxCut using Rydberg qubits. Reproduced from \textcite{zhou2020quantum}, Phys. Rev. X, 10, 021067 (2020), licensed under CC-BY 4.0.}
  \label{fig:proposals}
\end{figure*}

Seeing the rapid advances reviewed in the previous section, it is likely that Rydberg quantum processors will soon help solve important problems reaching beyond the field of many-body physics. In this section we highlight three recent theoretically proposed models for quantum computation that are uniquely suited to the Rydberg platform, even with relatively limited control over individual qubits or relatively low circuit depths typical of the current generation of experiments.

\subsection{Rydberg quantum annealing}
Quantum annealing is a technique for efficiently solving complex optimization problems of widespread practical importance\cite{hauke2020perspectives}. The basic idea is to exploit quantum evolution to prepare the ground state of a Ising spin Hamiltonian that encodes the problem of interest\cite{kadowaki1998quantum,hauke2020perspectives}. This can be achieved using a quantum simulator that implements a Hamiltonian of the form
\begin{align}\label{eq:spinglass}
\hat H(t)\! =\! A(t)\!\sum_j \!\hat X_j \!+\! B(t)\!\!\left[\sum_j \biggl(\!\Delta_j \hat Z_j +\sum_{k\neq j} V_{j,k}  \hat Z_j \hat Z_k\biggr)\right ],
\end{align}
where the optimization problem is encoded in the local fields $\Delta_j$ and two-body interactions $V_{j,k}$.

One typically starts by initializing the system in the trivial ground state that is easy to prepare with $[A(t)=1,B(t)=0]$ and then slowly transforming the system Hamiltonian such that $[A(t)=0,B(t)=1]$. The adiabatic theorem dictates that the system will follow the instantaneous ground state of the system at each moment of time, ending at the ground state of the target Hamiltonian. The solution to the problem can then be read out by measurements on the final qubit states.

One of the difficulties of this protocol however is the need for all-to-all (infinite range) interactions which are not native to the Rydberg system (or indeed most physical systems). An elegant solution to this problem was proposed by Lechner-Haucke-Zoller (LHZ) in 2015\cite{lechner2015quantum}. The basic idea is to map the infinite range spin glass Hamiltonian on a lattice spin model where each physical qubit represents the relative orientation of two logical spins in Eq.~\eqref{eq:spinglass}. A major advantage of this approach is that the necessary non-local interactions between logical spins can be implemented with local fields acting on physical qubits that are relatively simple to implement. However additional four-body constraints must be imposed to ensure that configurations with conflicting relative orientations of logical spins are energetically penalized. 

\textcite{glaetzle2017coherent}, showed that a programmable quantum annealer can be implemented with Rydberg-dressed qubits following the LHZ architecture. Their approach relies on global Rydberg dressing lasers and local AC Stark shifts applied using a digital micromirror device that encode the optimization problem (Fig.~\ref{fig:proposals}a). The necessary four body constraints are naturally realized by exploiting the Rydberg blockade effect between two species of atomic qubits, e.g., Rb and Cs atoms representing physical and ancilla qubits respectively. The number of physical spins equates to the number of connections in the original model, requiring $N(N-1)/2$ physical qubits to represent a system of $N$ logical spins with all-to-all connectivity. Through numerical calculations they demonstrate the feasibility for quantum annealing of the minimal instance consisting of 8 physical qubits and 3 ancillas (corresponding to 4 all-to-all connected logical spins) depicted in Fig.~\ref{fig:proposals}a. The results showed that the ground state can be prepared with success probability $>75\%$ within $\sim 0.5\,$ms (approaching unity for slower sweeps). This is compatible with recent experiments with Rydberg dressed qubits\cite{zeiher2017coherent} (see also Sec.~\ref{sec:demonstrations_qsim}). Recently a three-dimensional architecture for quantum annealing was proposed where distant couplings can be engineered by ferromagnetic Ising quantum wires\cite{qiu2020programmable}.

While this form of quantum annealing is not thought to be universal, a coherent quantum annealer may still provide a quantum speedup for certain computational problems with modest quantum resources. It can also be thought of as a key step toward more general purpose adiabatic quantum computers, a promising alternative to digital circuit based approaches\cite{aharonov2008adiabatic,albash2018adiabatic}.

\subsection{Quantum cellular automata}

Quantum cellular automata (QCA) is another paradigm for quantum information processing, that can be represented as a class of quantum circuits that consist of homogeneous local interactions that are periodic in space and time. 

This is in analogy with classical cellular automata: computational models involving a regular network of cells (e.g., two-state systems) that evolve according to a set of dynamical rules that update each cell depending only on the state of the cells in its neighborhood. For example, 
$$
\begin{array}{l|cccccccc}
\textrm{neighborhood} & 111 & 110 & 101 & 100 & 011 & 010 & 001 & 000\\
\hline
\textrm{update} & 1 & 1 & 0 & 0 & 1 & 0 & 0  & 1
\end{array}
$$
describes the elementary cellular automaton CA201 for 3-cell neighborhoods in 1D (the update string $11001001 =\nobreak 201$ in decimal representation). 

Quantum cellular automata replaces bits with quantum bits and update rules with (usually unitary) multi-qubit operations. Thus QCA supports quantum mechanical features such as probabilistic measurement outcomes, reversibility, superposition and entanglement. It also constitutes an intrinsically parallel model for universal quantum computing that does not require local addressing\cite{lloyd1993potentially}. In this model, quantum information is first loaded onto the array and then computations are performed by applying a sequence of global unitary rules which permute and transform the states of each qubit conditional on its neighbors\cite{lloyd1999programming,karafyllidis2004definition}. While there has been no general physical realization of QCA to date, Rydberg qubits with their versatile multiqubit interactions may be uniquely suitable. For instance, the quantum mechanical version of rule CA201 above can be understood as a bit-flip on the central qubit conditioned on both neighbors being in the $0$ state. Thus it coincides with the PXP model introduced in Sec.~\ref{sec:hamiltonians} and experimentally realized in Rydberg atom chains~\cite{bernien2017probing}. 

In 2020, \textcite{wintermantel2020unitary} proposed a physical realization of QCA using arrays of Rydberg gr-qubits. The key idea is to replace the single frequency laser coupling from $\ket{g}\leftrightarrow \ket{r}$ with a multifrequency coupling with individually controllable frequencies and amplitudes as shown in Fig.~\ref{fig:proposals}b. The relative frequency of each component should be tuned to match the interaction energy of a particular configuration, e.g., $\Delta \approx kV$, where $k$ counts the number of Rydberg excitations in the neighborhood configuration (assuming additive interactions). In this way it is possible to generalize the PXP model to a form (in 1D)
$$
\hat H = \frac{1}{2}\sum_j\sum_{\alpha,\beta} \theta^k P^\alpha_{j-1} X_j P^\beta_{j+1},
$$
where $\alpha,\beta \in (0,1)$, $\theta^k$ is the strength of the coupling for the frequency component $k=\alpha+\beta$, $j$ is the site index, and the projection operators $P^\alpha_j=\ket{\alpha}_j\!\bra{\alpha}$. This coincides with the full set of so-called totalistic QCA rules (rules that only depend on the total number of excited neighbors). It was also shown that dissipative conditional interactions could be readily implemented via multifrequency couplings to a short lived auxiliary state, which broadens the QCA set to include both unitary and non-unitary rules. These rules can be applied both in continuous fashion (in parallel to all qubits) or in discrete time using a block partitioning scheme\cite{brennen2003entanglement}, which could be implemented using two different atomic species or by structuring the excitation laser fields to address a subset of qubits. 

Using numerical simulations on small systems consisting of up to 9 qubits, they demonstrate the feasibility of the scheme to engineer highly entangled quantum states. To generate a GHZ$^N$ state via purely unitary discrete-time QCA evolution requires at least $(N-1)/2$ timesteps\cite{brennen2003entanglement}. Using a variational optimization procedure they also find combinations of QCA rules that generates the same state via continuous time evolution and using non-unitary rules. Although this specific protocol appears to be slower than the fully unitary case, the use of dissipative interactions potentially makes the protocol less sensitive to precise pulse timings and errors in the initial state.

Also in 2020, \textcite{hillberry2020entangled} theoretically studied the complexity of QCA rules and identified a set of rules which produce highly entangled and highly structured quantum states. Related work has highlighted the potential for Rydberg QCA for studying non-equilibrium universality and criticality\cite{gillman2020nonequilibrium}, simulating lattice gauge theories\cite{notarnicola2020realtime}, and non-ergodic dynamics\cite{rakovszky2020statistical,iadecola2020nonergodic}.

\subsection{Variational quantum algorithms}

Variational quantum algorithms are a hybrid quantum-classical approach to quantum computing, put forward as a way to efficiently prepare ground states of quantum Hamiltonians (e.g., molecular Hamiltonians, solutions to combinatorial optimization problems, or metrologically useful quantum states). For an overview, see Refs.\cite{mcclean2016theory,moll2018quantum} and references therein. Typically a problem of interest is encoded as a Hamiltonian and then a programmable quantum processor is used to generate trial wavefunctions. Measurements are then made and the expectation values are used to calculate the objective function of a classical optimization routine. The problem of finding good parameters is outsourced to highly efficient classical algorithms (running on classical computers) while the available quantum resources of the quantum hardware are used as efficiently and robustly as possible (depicted in Fig.~\ref{fig:proposals}c). Therefore variational quantum algorithms offer a very promising route toward solving important problems on near term NISQ devices. Some specific approaches are the quantum approximate optimization algorithm (QAOA)\cite{farhi2014quantum}, the variational quantum eigensolver (VQE) algorithm (oriented toward applications in quantum chemistry)\cite{mcclean2016theory,moll2018quantum} and variational spin-squeezing algorithms\cite{kaubruegger2019variational} applicable to quantum sensors with programmable Rydberg interactions\cite{gil2014spin,arias2019realization}.

In comparison to (analog) quantum annealing, variational quantum algorithms such as QAOA do not depend on (quasi-)adiabatic evolution and can potentially find solutions to hard optimization problems while tolerating some noise. It has been proposed as a possible approach to achieving a quantum advantage with low-depth circuits~\cite{farhi2016quantum,Bravyi308}. However relatively little is known about the applicability of QAOA to general optimization problems\cite{akshay2020reachability} (especially beyond depth $d=1$ circuits).

In 2018, \textcite{pichler2018quantum} proposed that QAOA combined with Rydberg qubits and a strong Rydberg-blockade interaction provides a natural way to solve the unit-disk maximum independent set (UD-MIS) problem. UD-MIS is a graph partition problem where the goal is to find the largest set of unconnected vertices in a 2D plane, where a pair of vertices ($v,w$) are connected if they are separated by less than a unit length. Their proposed protocol uses gr-qubits as vertices positioned in space such that two qubits are connected if they are closer than the blockade radius ($R_{v,w}<R_b$). They then consider a circuit consisting of a series of resonant Rydberg excitation pulses with Rabi frequency $\Omega$ and varying durations and phases. For small systems they find that QAOA performs similarly well to quantum annealing in terms of the required measurement count ($10^2-10^3$) when the annealing time is constrained to $T=10/\Omega$. Assuming realistic experimental parameters they estimate that it should be possible to solve computationally hard problems with $N\sim 10^2-10^3$ vertices on near term Rydberg quantum processors. \textcite{henriet2020robustness} extended the analysis using a similar protocol to include the effects of spontaneous emission, finding that the performance of Rydberg QAOA is relatively noise resilient and could be further improved through careful choice of the objective function.

\textcite{zhou2020quantum} analyze the performance of QAOA on a closely related weighted MaxCut problems, where the goal is to find two subsets of vertices that maximize the total weight of the edges connecting the two subsets (an example of a five vertex graph is shown in Fig~\ref{fig:proposals}c). This problem can be cast in the form of a quantum Ising Hamiltonian $\hat{H}$ where the interaction $V_{v,w}$ represents the weight of the edge connecting two vertices. To solve the problem one seeks a classical configuration of spins which maximizes the expectation value of $\hat H$. The proposed protocol involves an array of gg-qubits subject to alternating layers of the problem Hamiltonian and a mixing Hamiltonian $\sum_v \hat X_v$ (Fig.~\ref{fig:proposals}c) with variable durations. The mixing terms correspond to global Rabi coupling pulses with tunable durations and the interaction terms can be implemented stroboscopically using Rydberg blockade gates. Finally, measurements on each qubit in the computational basis are then used to compute the expectation value of $\hat{H}$ ($\equiv$ objective function) which is in turn maximized by a classical algorithm. They find an efficient parameter optimization procedure and show that QAOA can exploit non-adiabatic mechanisms to circumvent challenges associated with vanishing spectral gaps in quantum annealing.

\section{Future challenges and opportunities}\label{sec:outlook}

We have reviewed the state-of-the-art in the use of Rydberg qubits for quantum simulation and quantum computing and some proposed implementations that are uniquely suited to the Rydberg platform. In this final section we look forward at some of the next opportunities and challenges on the road toward fully-programmable quantum simulators and quantum computers of the future.\\ 

\noindent\textit{Scaling up the number of qubits}

As we have seen, atomic qubits have a notable advantage when it comes to scalability, with optical tweezer systems already capable of preparing more than 100 individually resolvable qubits (Fig.~\ref{fig:100atoms}). Progress is very fast, with latest demonstrations involving as many as 196 and 256 deterministically prepared Rydberg-interacting qubits\cite{scholl2020programmable,ebadi2020quantum}. This compares favorably with trapped ion systems with currently around 15-20 fully addressable qubits\cite{friis2018observation, egan2020fault} (and $\sim 50$ individually resolved qubits for quantum simulation\cite{zhang2017observation53} or hundreds of ions in Penning traps without local addressing~\cite{britton2012engineered,Bohnet1297}) and SC quantum computers with $\sim 50-70$ qubits\cite{arute2019quantum} (while special purpose quantum annealing devices can have several thousand qubits). The current methods for preparing even larger Rydberg qubit arrays (much beyond a few hundred qubits) suffer from particle loss during the increasingly complex rearrangement process. This can be improved using more deterministic initial loading mechanisms\cite{brown2019gray,wang2020preparation} combined with more efficient rearrangement protocols (e.g., using optimal sorting strategies, parallel moves and multiple rearrangement cycles\cite{mello2019defect,schymik2020enhanced}). Monte-Carlo simulations based on realistically achievable parameters indicate that it should be possible to load tweezer arrays with $N=1000$ fully occupied sites with 90\% global success probability\cite{mello2019defect,brown2019gray}. To exceed this will likely eventually require fundamentally different techniques or cryogenic environments to improve vacuum-limited trap lifetimes.\\

\noindent\textit{Increasing coherence times}

Right now, the performance of Rydberg quantum processors is limited by the fidelity of quantum operations or relatively short ground-Rydberg coherence times. This is due to a combination of many small effects, all of which can be overcome, including phase and intensity noise on the Rydberg excitation lasers, differential light shifts due to the tweezer traps and motion of the atoms (Doppler dephasing)\cite{saffman2011rydberg,leseleuc2018analysis,graham2019rydberg}. Cavity stabilized Rydberg excitation lasers with sub-kilohertz linewidths have already been developed\cite{legaie2018subkilo}, which combined with cavity filtering techniques\cite{levine2018highfidelity} should allow for the timescales associated with laser-induced dephasing to exceed the Rydberg state lifetime. Doppler and trap induced dephasing could be largely eliminated through the use of magic trapping techniques to cancel differential light shifts\cite{lundblad2010experimental,zhang2011magic,yang2016coherence,lampen2018long,sheng2018high,guo2020balanced} and cooling the atoms close to the motional ground state\cite{kaufman2012cooling,thompson2013coherence,norcia2018microscopic,wang2019preparation,lorenz2020raman}. However this is not a critical requirement, as many gate protocols can be made robust against these effects\cite{goerz2014robustness,yu2019adiabatic,theis2016high,petrosyan2017high,Shi2019fast,shi2020suppressing} (Sec.~\ref{sec:interactions}), and the use of additional echo pulses\cite{zeiher2016many,zeiher2017coherent,levine2019parallel,shi2018accurate} or dynamical decoupling protocols\cite{suter2016colloquium} could enable useful quantum computations and quantum simulations even in the presence of some technical noise.\\

\begin{figure}[!t]
  \includegraphics[width=.95\linewidth]{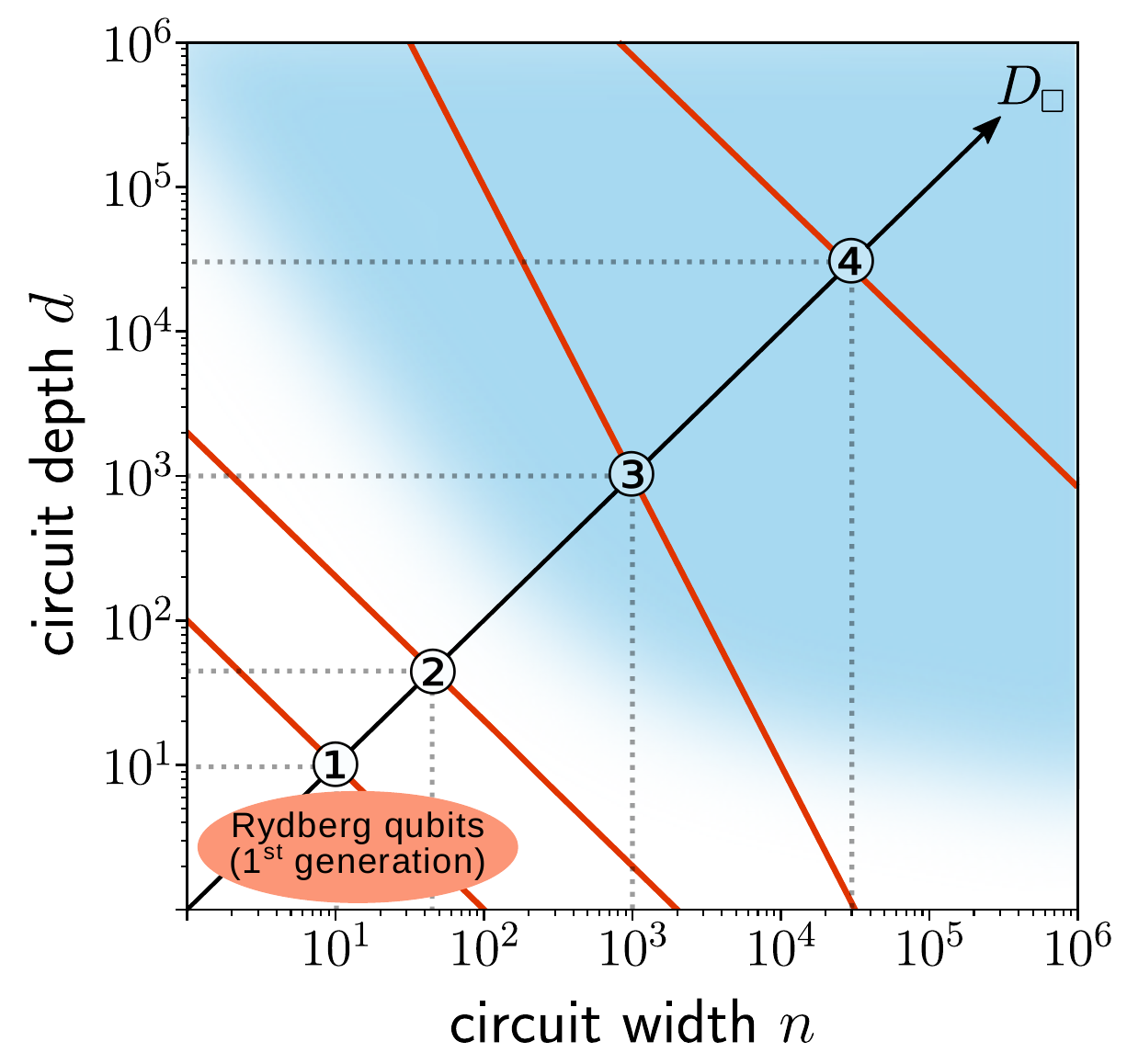}
  \caption{\small Current status and future evolution of Rydberg quantum registers in terms of the circuit depth $d$ and the number of physical qubits $n$ in the circuit (assuming $n/2$ two-qubit gates per layer). The blue shaded area represents the approximate resources required for fault tolerant quantum computing while the white area represents the NISQ regime. The red lines depict successive barriers associated with the finite lifetime $T_1$ of Rydberg qubits: \textcircled{1} finite dwell time of Rydberg states $T_1=5\,\mu$s, yielding $\Dsquare=10$ (multiqubit gate count $N_g=50$); \textcircled{2} spontaneous decay of the Rydberg state $T_1=100\,\mu$s, yielding $\Dsquare= 44$ ($N_g=1000$); \textcircled{3} trap loss due to background gas collisions $T_1=100\,$s, yielding $\Dsquare= 10^3$ ($N_g= 5\times 10^5$); \textcircled{4} parallel quantum operations and $T_1=100\,$s, yielding $\Dsquare \approx 3\times 10^4$ ($N_g\approx 5\times10^8$). The barriers are computed assuming each gate has a duration $\tau_g=50\,$ns and consists of a cumulative time in the Rydberg state $\tau=50\,$ns (counting both qubits).}
  \label{fig:future}
\end{figure}

\noindent\textit{Increasing circuit depth and breaking the lifetime barrier} 

Rydberg qubits are very attractive for their scalability and fast gate times which will facilitate the realization of wide and deep quantum circuits and quantum simulations. But for this it will be necessary to overcome some barriers concerning their finite lifetimes which will likely require new technological innovations. Figure~\ref{fig:future} depicts roughly where some of these barriers can be anticipated.

The first barrier (labelled \textcircled{1}) concerns the short dwell time $\lesssim 5\,\mu$s of Rydberg excitations that are either not trapped or anti-trapped by standard red-detuned optical tweezers. Assuming each two-qubit gate involves a cumulative time of $\sim 50\,$ns in the Rydberg state (integrated for both qubits), this imposes a limit of $\lesssim 50$ gate operations or $\Dsquare \lesssim 10$. However solutions to this problem are already being implemented using (blue detuned) optical or magnetic traps that can simultaneously trap both ground and Rydberg states\cite{zhang2011magic,mukherjee2011many,piotrowicz2013twodimensional,boetes2018trapping,barredo2020three}.

The second barrier \textcircled{2} concerns the inherent interaction of Rydberg atoms with thermal and vacuum electromagnetic fields, which limits cumulative excitation times to $\sim 100\,\mu$s. In principle this is sufficient to realize $\sim 1000$ gate operations or an effective circuit depth of $\Dsquare \lesssim 44$, which is ambitious but within reach (this implies an average gate fidelity $F\approx 0.9995$). This is a very interesting regime for solving useful problems beyond the capabilities of classical computers\cite{Villalonga_2020}, especially for e.g., quantum optimization and variational quantum algorithms.

To overcome this barrier, one can conceivably structure the electromagnetic environment and the atom-vacuum-field couplings. One possible route to reaching lifetimes in the minute range was proposed by \textcite{nguyen2018towards}, using long-lived circular Rydberg states and embedding the atoms inside cryogenic cavities to inhibit black-body induced transitions and microwave spontaneous emission.

At this point, Rydberg quantum processors will likely be limited by the trapping lifetime of the atoms themselves (usually by collisions with background gas atoms or molecules to $\sim 100\,$s). This means, for a $N=1000$ qubit processor, one qubit will be lost on average every $\sim 100\,$ms. Even if these atoms can be replaced from an extra reservoir this constitutes an error on the many-particle quantum state. If gates are applied sequentially, this imposes a maximum of $\lesssim 5\times 10^5$ operations before an error occurs ($\Dsquare \lesssim 10^3$, \textcircled{3}). This is in an interesting range for implementing quantum error correcting codes suitable for fault-tolerant quantum computing. Alternatively, with the ability to apply gates in parallel to many qubits at a time it should be possible to reach $N_g = 5\times 10^8$ or $\Dsquare \approx 3\times 10^4$ (\textcircled{4}, assuming correspondingly high numbers of qubits can be prepared), well within the interesting range for important commercial applications\cite{wecker2014gate,nam2019low,kuhn2019accuracy}.\\

\noindent\textit{Advanced multi-qubit control and benchmarking} 

Recent experiments have demonstrated the possibility to perform quantum operations acting on multiple qubits in large 2D arrays with low crosstalk\cite{graham2019rydberg}, however most demonstrations still operate with relatively limited control over individual qubits (discussed in Sec.~\ref{sec:singlequbit}). Scalable electronic control systems optimized for other multi-qubit quantum platforms are being developed commercially and can also be used for the Rydberg platform. 
 
Combined with fast and precise optical modulators and robust calibration procedures, this should allow for implementing shaped optical pulses tailored for each qubit across the register in parallel (as opposed to sequential operations). Thus we expect the realization of more complex quantum circuits involving many-qubits (e.g., in two-dimensional arrays) to be demonstrated soon. This will presumably include verification protocols and benchmarking circuits for quantifying performance that could then be meaningfully compared across different platforms. \\

\noindent\textit{All-to-all couplings and qubit routing}

For general purpose quantum computing it will eventually be necessary to entangle qubits in distant parts of the processor, thereby overcoming limitations associated to the physical geometry of the architecture. This can be achieved using SWAP networks at the cost of extra gate operations\cite{beals2013efficient}. However the availability of larger numbers of physical qubits in the Rydberg platform could also allow for clever logical qubit encodings such as the LHZ architecture, that might replace the need for physical all-to-all couplings for certain problems. Another very attractive feature of Rydberg qubits is their strong atom-photon interactions and the possibility for mediating strong photon-photon interactions (not focused on in this review, see instead Refs.\cite{pritchard2012nonlinear,Firstenberg_2016,adams2019rydberg,lauk2020perspectives,wu2020concise}). This could provide additional possibilities for mediating gates between distant qubits, interconverting stationary and flying qubits, and connecting different quantum computing nodes.\\

\noindent\textit{Non-destructive readout and quantum error correction} 

Another outstanding challenge concerns the relatively slow and destructive fluorescence based readout of Rydberg qubits. This makes it difficult to act on measurement outcomes, which is usually required for quantum error correction schemes. Quantum non-demolition measurements could be realized with high fidelity and low crosstalk, e.g., two-species architectures\cite{vasilyev2020monitoring}, possibly involving multiple readout qubits per computational qubit as sketched in Fig.\ref{fig:register}b to achieve sub-millisecond readout times. Alternatively there are promising approaches to quantum error correction that may not require intermediate readout steps. \textcite{premakumar2020measurement} recently showed that measurement free error correction\cite{pazsilva2010fault,crow2016improved} could be efficiently implemented using C$_k$NOT gates (controlled by ancilla qubits) which are native to the Rydberg platform. Another interesting approach involves tailored dissipative processes to stabilize quantum states\cite{verstraete2009quantum, reiter2017dissipative}, e.g., by coupling Rydberg qubits to additional short-lived states in a way to autonomously correct errors. Proof-of-principle demonstrations of logical qubit encoding and correction of arbitrary single-qubit errors have recently been demonstrated with trapped ions\cite{egan2020fault} and superconducting circuits\cite{gong2019experimental}. The development of these techniques for the Rydberg platform should ultimately enable the realization of quantum error correcting codes with sizable numbers of logical qubits and logical gates for fault-tolerant quantum computing\cite{kitaev2003fault,fowler2012surface}.

\subsection*{Conclusion} 
In conclusion, we have presented trapped arrays of neutral atoms with Rydberg mediated interactions which offer an extremely versatile platform for quantum simulation and quantum computing. Since the last major review of the field of Rydberg-based quantum computing (\textcite{saffman2016quantum}), we have seen breakthrough achievements, including: the realization of large qubit arrays with programmable spatial configurations with of the order of $\sim 100$ qubits; the first high-fidelity entangling operations with $F\gtrsim 0.90$ ($>0.991$ in one case\cite{madjarov2020high}); experiments on different atomic species, including alkali atoms, alkaline earth atoms and ions and dual species systems, each of which come with different advantages and new opportunities; the design and demonstration of a broad set of quantum gates specifically optimized for the Rydberg platform; optical qubit addressing in large qubit arrays and a better understanding of the dominant errors affecting gate operations; the realization of elementary quantum circuits including entangling gates with more than two qubits; large scale entanglement generation using optimized control sequences; and programmable quantum simulations with $\gtrsim 100$ interacting qubits which have already enabled the study of interesting new physics.

These achievements generally reflect a transition from few- and many-body physics experiments to more general purpose quantum simulation and quantum computing applications that has just begun\cite{henriet2020quantum}. While there are still substantial scientific and engineering challenges ahead, continuing improvements to the existing technology should allow for realizing large quantum circuits with $\sim 1000$ qubits and effective circuit depths $\Dsquare\gtrsim 40$, far beyond the current state-of-the-art in any platform. Combined with continuing algorithmic advances that take advantage of the unique features of the Rydberg platform, this will enable solutions to important problems well beyond the reach of classical computers.
We anticipate the first such applications in high-performance scientific computing and hybrid quantum-classical computing for many-body physics, quantum state engineering (for metrology), quantum chemistry, material science, mathematics, optimization, machine learning, and presumably other, yet undiscovered applications.

\vskip -1em
\begin{acknowledgements}
We thank R. Gerritsma, F. Nori, G. Pupillo, J. Schachenmayer, Y. Wang, T. M. Wintermantel and P. Zoller for careful reading of the manuscript and valuable discussions. This work is supported by the `Investissements d’Avenir’ programme through the Excellence Initiative of the University of Strasbourg (IdEx) and the University of Strasbourg Institute for Advanced Study (USIAS). M.M acknowledges the French National Research Agency  (ANR) through  the  Programme  d’Investissement  d’Avenir  under  contract  ANR-17-EURE-0024 and  QUSTEC funding from the European Union’s Horizon 2020 research  and  innovation  programme  under  the  Marie Sk\l{}odowska-Curie grant agreement  No. 847471.
\end{acknowledgements}

\vskip -1em
\appendix
\section{Achievable circuit depth $\Dsquare$}\label{app:dsquare}

\begin{figure}[!t]
  \includegraphics[width=.6\linewidth]{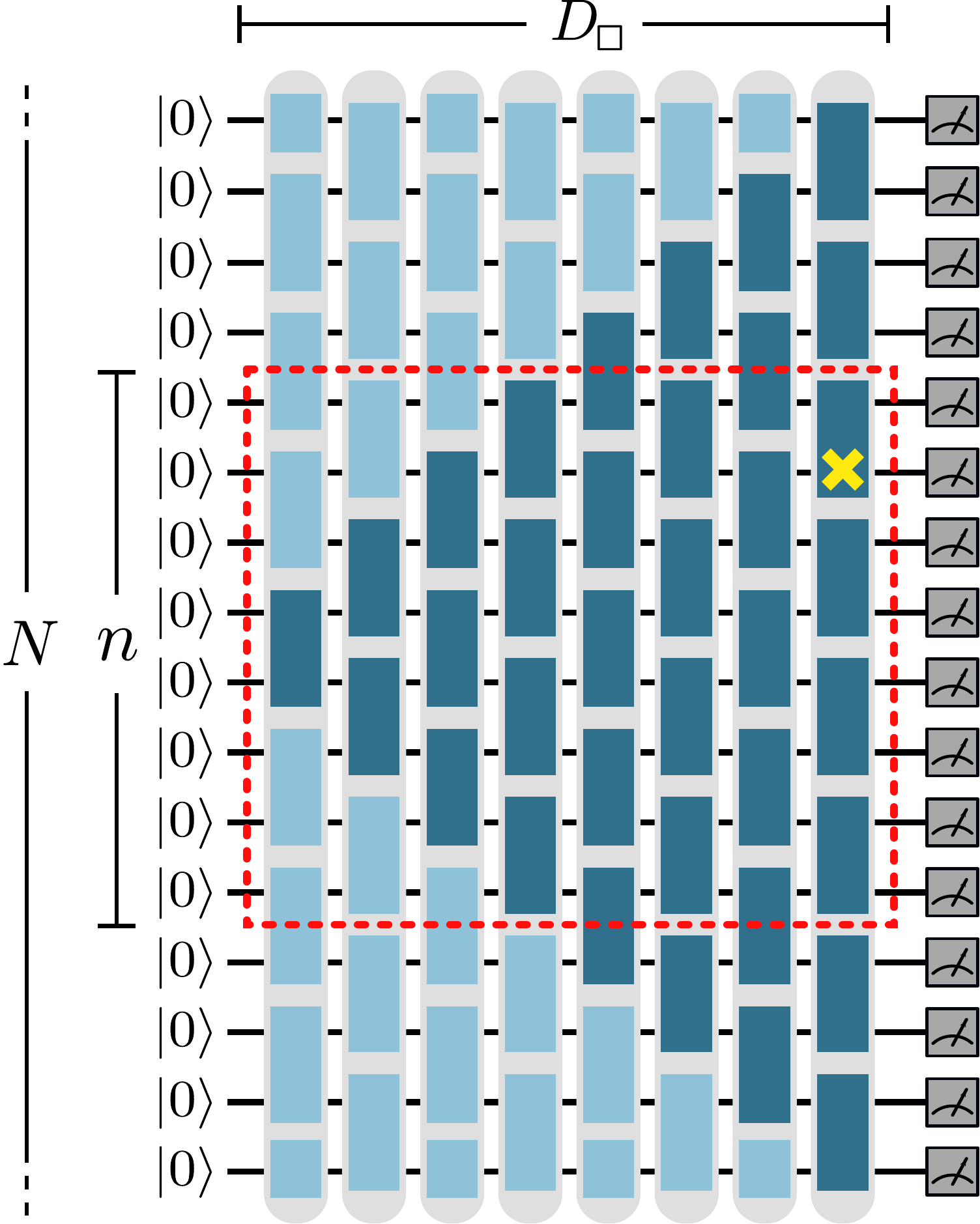}
  \caption{\small Model circuit for characterising the performance of Rydberg quantum processors consisting of the parallel application of local two-qubit operations to all qubits (blue rectangles). The effective depth $\Dsquare$ is defined as the maximum dimension of a square circuit (indicated by the red box) that can be reliably executed before an error occurs (yellow cross). The dark shaded rectangles represent the causal cone for the spreading of (quantum) correlations.}
  \label{fig:circuit}
\end{figure}

In Sec.~\ref{sec:criteria} we introduce the effective circuit depth $\Dsquare$, which can be used to compare the capabilities of different quantum simulators and quantum computers in a way that in principle takes into account all the available quantum resources.

Consider a $N$-qubit device that can run quantum circuits of variable size, i.e., involving subsets of $n$-qubits and $d$-layers (illustrated in Fig.~\ref{fig:circuit}). We search for the largest square circuit (i.e., $n=d$) that can be reliably executed before an error is statistically likely to occur (with some chosen threshold probability). By our convention we consider circuits consisting of the parallel application of two-qubit gates or Hamiltonian evolution involving pairwise interactions applied to all qubits. $\Dsquare$ is then defined by 
\begin{align}\label{eq:QV}
  \Dsquare \equiv \underset{n\leq N}{\mathrm{argmax}} \,\mathrm{min}\left[ n, d(n) \right]  .
\end{align}

This can be seen in Fig.~\ref{fig:circuit} as the area inside the red box which comprises $N_g=\Dsquare^2/2$ gates with an average error probability per gate $\epsilon = 1/(2N_g)$. In this way $\Dsquare$ captures the true performance of the device including different error channels and overheads associated to how the gates are physically implemented (e.g., errors on idle qubits for sequential application of gates). We emphasize however that $\Dsquare$ does not necessarily dictate the maximum size of a useful quantum circuit, since for example particular protocols could be used to entangle more than $\Dsquare$ qubits with the same overall number of gates (especially for circuits featuring more-than-two qubit operations and a high degree of connectivity).

For the purpose of this review we assume the error per gate is system size independent and that current Rydberg experiments are not limited by the number of available qubits. Therefore a successful circuit can be defined according to $nd\times \epsilon \leq 1$, i.e., $\Dsquare = \lfloor \epsilon^{-1/2}\rfloor$, following the convention in \textcite{moll2018quantum}. For digital circuits $\epsilon$ can be approximated by $1-F$, where $F$ is the fidelity for two-qubit operations. The average error probability $\epsilon$ is harder to define for the case of analog quantum simulations. As an estimate we take $\epsilon \sim \pi/(V_{\text{max}}T_{\text{coh}})$. This can be interpreted as the inverse of the number of half-interaction cycles that can be reliably simulated within some characteristic coherence time $T_{\text{coh}}$. An good estimate for $T_{\text{coh}}$ is the irreversible dephasing time $T_{2}$.

The achievable circuit depth defined in Eq.~\eqref{eq:QV} can also be directly determined using actual hardware (e.g. using randomized circuits and appropriate statistical tests). In that case then $\Dsquare$ can serve as a benchmark in the spirit of the quantum volume ($V_{Q} \sim 2^{\Dsquare}$) introduced by IBM as a way to benchmark the performance of quantum computers \cite{cross2019validating}. However, by their convention a benchmark circuit consists of the parallel application of random two-qubit gates acting on disjoint pairs of qubits followed by a random permutation of all the qubit indices in each step (requiring all-to-all connectivity), but this could be adapted depending on certain applications\cite{blume2019volumetric}.  

\bibliography{aipsamp}

\end{document}